\RequirePackage{ifpdf}
\documentclass[12pt]{JINST}

\usepackage{amsmath}
\usepackage{amssymb}
\usepackage{amsfonts}
\usepackage{xspace}
\usepackage{url}
\usepackage{enumerate}

\newcommand{\GeV}{\ensuremath{\mbox{GeV}}\xspace}

\newcommand{\GeVc}{\ensuremath{\mbox{GeV}/c}\xspace}

\newcommand{\cm}{\ensuremath{\mbox{cm}}\xspace}

\newcommand{\mus}{\ensuremath{\mu\mbox{s}}\xspace}

\newcommand{\rig}{\ensuremath{\mbox{~({\it right})}}\xspace}
\newcommand{\lef}{\ensuremath{\mbox{~({\it left})}}\xspace}

	\title{
         NA61/SHINE facility at the CERN SPS:\\
         beams and detector system}
	\author{
  \small
N.~Abgrall${}^{11}$,
	O.~Andreeva${}^{16}$,
	A.~Aduszkiewicz${}^{23}$,
	Y.~Ali${}^{6}$,
	T.~Anticic${}^{26}$,
	N.~Antoniou${}^{1}$,
	B.~Baatar${}^{7}$,
	F.~Bay${}^{27}$,
	A.~Blondel${}^{11}$,
	J.~Blumer${}^{13}$
% ...
\\

    \llap{$^{ }$}\\
    \llap{$^{1}$}University of Athens, Athens, Greece\\
    \llap{$^{2}$}University of Belgrade, Belgrade, Serbia\\
    \llap{$^{3}$}University of Bergen, Bergen, Norway\\
    \llap{$^{4}$}University of Bern, Bern, Switzerland\\
    \llap{$^{5}$}Wigner Research Centre for Physics of the Hungarian Academy of Sciences, Budapest, Hungary\\
    \llap{$^{6}$}Jagiellonian University, Cracow, Poland\\
    \llap{$^{7}$}Joint Institute for Nuclear Research, Dubna, Russia\\
    \llap{$^{8}$}Fachhochschule Frankfurt, Frankfurt, Germany\\
    \llap{$^{9}$}University of Frankfurt, Frankfurt, Germany\\
    \llap{$^{10}$}European Organization for Nuclear Research (CERN), Geneva, Switzerland\\
    \llap{$^{11}$}University of Geneva, Geneva, Switzerland\\
    \llap{$^{12}$}University of California, Irvine, USA\\
    \llap{$^{13}$}Karlsruhe Institute of Technology, Karlsruhe, Germany\\
    \llap{$^{14}$}University of Silesia, Katowice, Poland\\
    \llap{$^{15}$}Jan Kochanowski University in  Kielce, Poland\\
    \llap{$^{16}$}Institute for Nuclear Research, Moscow, Russia\\
     \llap{$^{17}$}Laboratory of Astroparticle Physics, University Nova Gorica, Nova Gorica, Slovenia\\
     \llap{$^{18}$}LPNHE, University of Paris VI and VII, Paris, France\\
     \llap{$^{19}$}Faculty of Physics, University of Sofia, Sofia, Bulgaria\\
     \llap{$^{20}$}St. Petersburg State University, St. Petersburg, Russia\\
     \llap{$^{21}$}Institute for Particle and Nuclear Studies, KEK, Tsukuba,  Japan\\
     \llap{$^{22}$}National Center for Nuclear Research, Warsaw, Poland\\
     \llap{$^{23}$}Faculty of Physics, University of Warsaw, Warsaw, Poland\\
	\llap{$^{24}$}Warsaw University of Technology, Warsaw, Poland\\
	\llap{$^{25}$}University of Wroc{\l}aw, Wroc{\l}aw, Poland\\
    \llap{$^{26}$}Rudjer Boskovic Institute, Zagreb, Croatia\\
    \llap{$^{27}$}ETH, Zurich, Switzerland\\
}

\date{\today}

	\abstract{
NA61/SHINE (SPS Heavy Ion and Neutrino Experiment)
is a multi-purpose
experimental facility to study hadron production in hadron-proton,
hadron-nucleus and nucleus-nucleus collisions at
the CERN Super Proton Synchrotron.
It recorded the first physics data with hadron beams
in 2009 and with ion beams (secondary $^7$Be beams) in 2011.

NA61/SHINE has greatly profited from the long development of the
CERN proton and ion sources and the accelerator chain as well as
the H2 beamline of the CERN North Area. The latter
has recently been modified to also serve as a fragment separator
as needed to produce the Be beams for NA61/SHINE.
Numerous components of the NA61/SHINE set-up were inherited from
its predecessors, in particular, the last one, the NA49 experiment.
Important new detectors and upgrades of the legacy
equipment were introduced by the NA61/SHINE Collaboration.

This paper describes the state of the NA61/SHINE  facility - the beams and the
detector system - before the CERN Long Shutdown I, which started in March 2013.
     }

\begin{document}
\section{\large Introduction}
\label{intro}

NA61/SHINE (SPS Heavy Ion and Neutrino Experiment)~\cite{proposal}
is a multi-purpose
facility to study hadron production in hadron-proton,
hadron-nucleus and nucleus-nucleus collisions at
the CERN Super Proton Synchrotron (SPS).
The NA61/SHINE physics goals include:
\begin{enumerate}[(i)]
\setlength{\itemsep}{1pt}
\item
study
the properties of the onset of deconfinement~\cite{review_onset}
and search for the critical point of strongly interacting matter
which is pursued by investigating
p+p, p+Pb and nucleus-nucleus collisions and
\item
precise hadron production measurements for improving calculations
of the initial neutrino beam flux in the long-baseline
neutrino oscillation experiments~\cite{T2K-experiment, NA61-future}
as well as for more reliable simulations of cosmic-ray
air showers~\cite{Auger, KASCADE}.
\end{enumerate}

The experiment was proposed to CERN in November 2006~\cite{proposal}.
Based on this proposal pilot data taking took place in September 2007.
The Memorandum of Understanding~\cite{mou} between
CERN and the collaborating institutions
was signed in October 2008. The first physics data with hadron beams
were recorded in 2009 and with ion beams (secondary $^7$Be beams) in 2011.
The first experimental results were published in ~\cite{abgrall2011, abgrall2012}.

NA61/SHINE has greatly profited from the long development of the
CERN proton and ion sources, the accelerator chain, as well as
the H2 beamline of the CERN North Area. The latter
has recently been modified to also serve as a fragment separator
as needed to produce the Be beams for NA61/SHINE.
Numerous components of the NA61/SHINE set-up were inherited from
its predecessors, in particular, the last one, the NA49 experiment.

\begin{figure}[t]
\centering
\includegraphics[scale=.42]{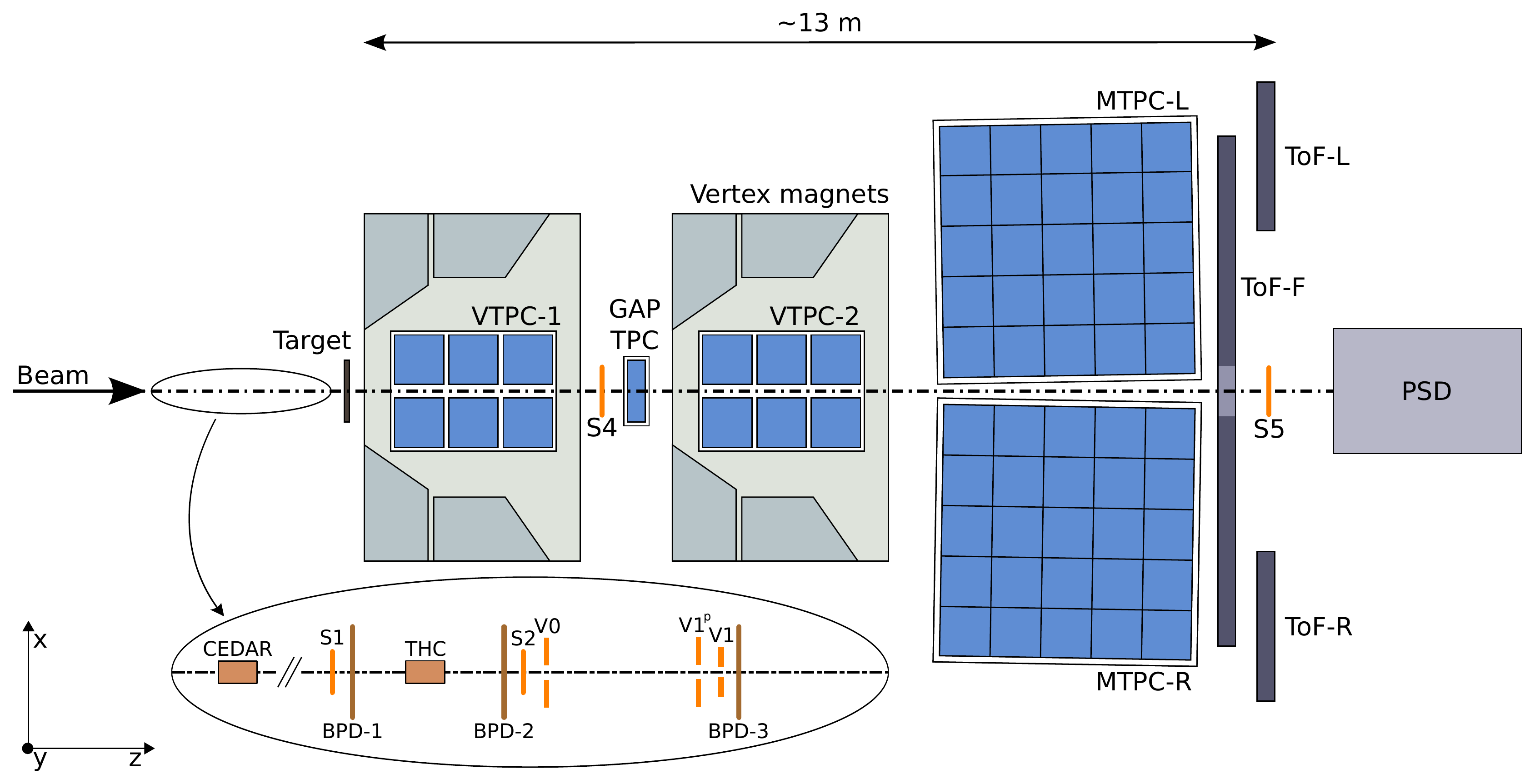}
\caption{Schematic layout of the NA61/SHINE experiment at the CERN SPS
(horizontal cut in the beam plane, not to scale).
The beam and trigger counter configuration used for data taking
on p+p interactions in 2009 is presented.
The chosen
right-handed coordinate system is shown on the plot.
The incoming beam direction is along the z axis.
The magnetic field bends charged particle trajectories in the x-z
(horizontal) plane.
The drift direction in the TPCs is along the y (vertical) axis. }
\label{fig:na61} 
\end{figure}

The layout of the NA61/SHINE detector is sketched in Fig.~\ref{fig:na61}.
It consists of a large acceptance hadron spectrometer with
excellent capabilities in charged particle momentum measurements and
identification by a set of six Time Projection Chambers as well as
Time-of-Flight detectors. The high resolution forward calorimeter,
the Projectile Spectator Detector, measures energy flow around the beam
direction, which in nucleus-nucleus reactions is primarily a measure of
the number of spectator (non-interacted) nucleons and thus
related to the centrality of the collision.
For hadron-nucleus interactions, the collision centrality is determined
by counting low momentum particles emitted from the nuclear target
with the LMPD detector (a small TPC) surrounding the target.
An array of beam detectors identifies beam particles, secondary
hadrons and ions as well as primary ions, and measures precisely
their trajectories.

This paper describes the NA61/SHINE facility, the beams
and the detector system.
Special attention is paid to the presentation of the
components which were constructed for NA61/SHINE.
The components inherited from the past experiments and
described elsewhere are presented only briefly here.
The paper is organized as follows.
In Sec.~\ref{beams} the proton and ion acceleration chains are
introduced and the North Area H2 beamline is described.
Moreover basic properties of the employed hadron and ion beams are given.
The NA61/SHINE beam and trigger detectors as well as the trigger system
are presented in Sec.~\ref{beam_det}.
Section~\ref{tpc} is devoted to the TPC tracking system which consists of
the TPC detectors with front end electronics,
two beam pipes filled with helium gas and two super-conducting magnets.
The Time of Flight system is described in Sec.~\ref{tof} and
the Projectile Spectator Detector in Sec.~\ref{psd}.
In Sec.~\ref{target} the targets and the Low Momentum Particle Detector are
presented. Finally, data acquisition and detector control systems are
described in Sec.~\ref{daq}.
Section~\ref{summary} closes the paper with summary and outlook.

\section{\large Beams}
\label{beams}
This section starts from the presentation of
the CERN proton and ion accelerator chains,
and continues with a brief description of the H2
beamline and secondary hadron and ion beams for the experiment.

\begin{figure}[htb]
\centering
\vspace*{-5cm}
\includegraphics[scale=.72]{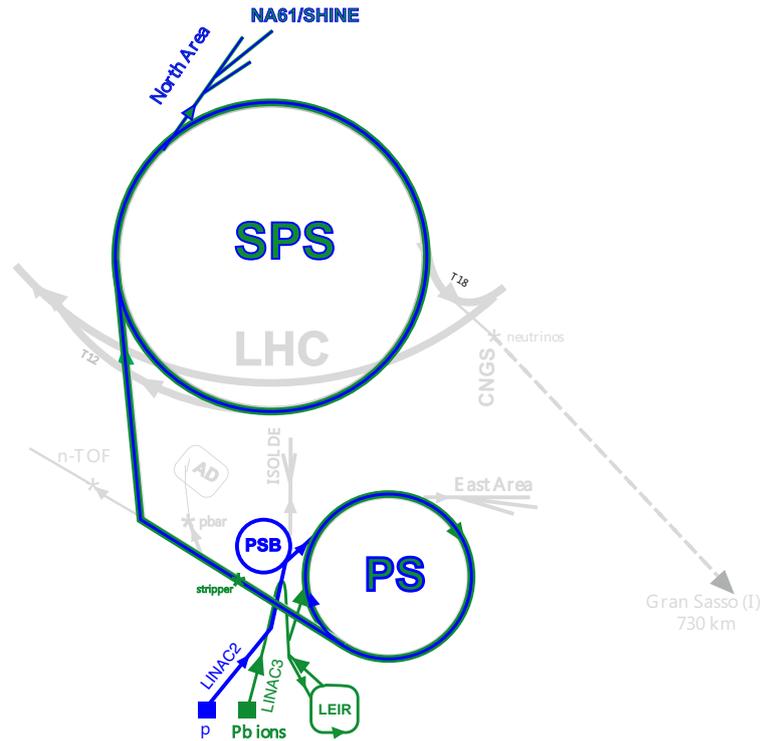}
\vspace*{-4cm}
\caption{Schematic layout of the CERN accelerator complex
relevant for the NA61/SHINE ion and proton beam operation.
(top view, not to scale).}
\label{fig:acc} 
\end{figure}

The CERN accelerator chain, with its components relevant for
NA61/SHINE beams exposed, is shown in Fig.~\ref{fig:acc}.
From the source, the beams of ions and protons pass through
a series of accelerators, before they reach the SPS for
final acceleration and subsequent extraction to
the North Area and the NA61/SHINE experiment.
The protons and ions follow a different path in
the pre-injector chain to the PS,
required to match the beam parameters for their acceleration.
\vspace*{-0.55cm}

\subsection{Proton acceleration chain}
\label{beams:p_acc}

The proton beam is generated from hydrogen gas by
a duo-plasmatron ion source, which can provide a beam current of
up to 300~mA~\cite{c1}. The Radio-Frequency Quadrupole
RFQ2~\cite{c2} focuses and bunches the beam,
and accelerates it to 750~keV for injection into LINAC2,
a three-tank Alvarez drift tube linear accelerator.
The three tanks have a total length of 33.3~m,
and the energy of the beam at the exit of the tanks is
respectively 10.3, 30.5, and 50~MeV.
With a repetition rate of 0.8~Hz, LINAC2 delivers a current of
up to 170~mA within a $90\%$ transverse emittance of
15$\pi$\ mm\ mrad, during a $120~\mus$ pulse length~\cite{c3}.
The 50~MeV proton beam from LINAC2 is then distributed in the
four rings of the PS booster (PSB) using a staircase-like
kicker magnet in the transfer line.
The multi-turn injection system of the PSB allows  to accumulate for up to 13
turns, over 10$^{13}$ protons per ring.
After acceleration to 1.4~\GeV, the beam from the four rings is
extracted and recombined in the extraction line,
to be sent to the Proton Synchrotron (PS),
CERN's oldest accelerator~\cite{c4}.
The PS has a circumference of 628~m and accelerates the beams
to 14~\GeVc for injection into the Super Proton Synchrotron (SPS).
In a typical proton cycle for the fixed-target experiments,
the PSB beam after recombination consists of a train of eight bunches
(two per ring), which are injected into the eight consecutive buckets of
the RF of the PS that operates at harmonic eight.
During the acceleration to 14~\GeVc, the beam passes from an intermediate
flat-top where the RF system splits the eight bunches into two
and changes the operation to harmonic sixteen.
At the top momentum, the beam is de-bunched and re-captured at harmonic 420~in
order to easily match the RF structure of the receiving machine,
the 6.8~km circumference SPS. The beam is extracted from the PS over five turns
using a staircase-shaped kicker pulse.
This ''continuous transfer'' multi-turn extraction is unique to the PS.
As the SPS is 11 times larger than the PS in circumference, it takes
two PS cycles to fill the SPS with the five-turn extraction.
The remaining two half-turn gaps are used for the rising and falling
edge of the SPS injection kicker. In the SPS, the 14~\GeVc beam
is accelerated to 400~\GeVc, on fixed harmonic 4620 (200~MHz).
At the top momentum, the beam is de-bunched and slowly extracted over
several seconds using a third-integer resonance.
The spill duration for the North Area experiments depends on
the overall optimisation for the SPS machine and its users,
and can be between 4.5~s and 10~s with a duty cycle of about 30\%.

\subsection{Ion accelerator chain}
\label{beams:ion_acc}

The CERN ion production complex was first designed in the 1990s
for the needs of the SPS fixed target programme~\cite{d1, d2}.
It was rejuvenated at the beginning of the 21$^{st}$ century~\cite{d3}
in order to cope with the LHC's stringent demands for
high brightness ion beams \cite{d4}. The ions are generated in
the ECR source. Here the case of the Pb beam production is considered
as an example.

A sample of isotopically pure $^{208}$Pb is inserted in a
filament-heated crucible at the rear of the ECR source,
whilst oxygen is injected as support gas.
With a 10~Hz repetition rate, a 50~ms long pulse of 14.5~GHz microwaves
accelerates electrons to form an oxygen plasma,
which in turn ionizes the lead vapor at the surface of the crucible.
The source operates in the so-called afterglow mode, i.e.
the microwave pulse is switched off, when the intensity of
the high charge state ions from the source increases dramatically.
The ions are electrostatically extracted from the source with
an energy of 2.5~keV/u. Out of the different ion species
and charge states, a 135$^o$ spectrometer situated at the exit of
the source selects those lead ions which were ionized 29 times
(Pb29+) to enter the LINAC3 accelerator.
The beam is first accelerated to 250~keV/u by the 2.66~m long RFQ
which operates at 101.28~MHz. The RFQ is followed by a four-gap RF
cavity which adapts the longitudinal bunch parameters to
the rest of the linear accelerator which is a three-cavity
interdigital H (IH) structure, that brings the beam energy
up to 4.2~MeV/u requiring about 30~MV of accelerating voltage,
for a total acceleration length of 8.13~m.
The first cavity operates at the same frequency as the RFQ (101.28~MHz),
while the second and third cavities operate at 202.56~MHz.
Finally, a 250~kV ''ramping cavity'', also operating at 101.28~MHz,
distributes the beam momentum over a range of $\pm 1\%$,
according to the time along the 200~$\mu$s pulse.
The LINAC3 currently operates at 5~Hz. A $0.3~\mu$m
thick carbon foil provides the first stripping stage at the exit of LINAC3,
followed by a spectrometer which selects the Pb54+ charge state.
A current of about $22~\mu$A of Pb54+ from LINAC3 is injected over
about 70 turns into the Low Energy Ion Ring (LEIR),
whose unique injection system fills the 6-dimensional phase space.
This is achieved by a regular multi-turn injection with a decreasing
horizontal bump, supplemented by an electrostatic septum tilted
at 45$^o$, and the time dependence of the momentum distribution
coupled with the large value (10~m) of the dispersion function
in the injection region.
The injection process is repeated up to six times, every 200~ms.
The whole process is performed under electron cooling reducing
the transverse and longitudinal emittances.
At the end of the seven injections, the beam is bunched on harmonic 2,
accelerated to 72~MeV/u by Finemet$^{TM}$ cavities,
and fast-extracted towards the PS.
At this point the total beam intensity is about 10$^9$ ions.

In the PS, the two ion bunches from LEIR are injected into
two adjacent buckets of harmonic 16 and accelerated by the 3-10~MHz
system to 5.9~GeV/u, with an intermediate flat-top for batch expansion.
This process consists of a series of harmonic changes:
h~=~16, 14, 12, 24, 21, in order to finally reach a bunch spacing of
200~ns at the top energy.
Before the fast extraction to the SPS, the bunches are finally rebucketed
into h~=~169, using one of the PS's three 80~MHz cavities.
At the exit of the PS, the beam traverses a final stripping
stage to produce Pb82+ ions, through a 1~mm thick aluminium foil.

The two bunches, now about 3$\cdot$10$^8$ ions each,
are injected into the SPS, with a bunch-to-bucket transfer into
the 200~MHz system. This process can be repeated up to 12 times every
3.6 seconds; the repetition rate is limited by the duration of the LEIR cycle,
while the total number of injections is currently limited by
the SPS controls hardware. At a fixed harmonic, the heavy mass of
the ions during acceleration would yield a too large frequency swing
(198.51 - 200.39~MHz) for the range of the travelling wave cavities of
the SPS (199.5 - 200.4~MHz). Hence, instead of using a fixed harmonic,
the ions are accelerated using the "fixed frequency" method,
in which a non-integer harmonic number is used, by
turning the RF ON at the cavity center frequency during the beam passage
and switching it OFF during its absence, to correct the RF phase
and be ready for its next beam passage through the cavities.
The phase is adjusted by an appropriate modulation of the frequency.
Aside from its complexity, one drawback of the method is that
the beam has to be constrained in a relatively short portion
$(40\%)$ of the circumference of the machine.
In the SPS, the ion beam, after acceleration to the required energy,
is left to de-bunch naturally
and is then slowly extracted to the North Area using
the third integer resonance like for protons.
The acceleration range in SPS varies between the 13~GeV/u,
which is the lowest possible operational range due to stability reasons,
and 160~GeV/u due to the limits in the power supplies
and energy in the magnets. The spill duration is preferred to be long
at about 10~s, although as for the protons the overall structure
depends on the number of users at the SPS.

\subsection{H2 beamline}
\label{beams:h2}

The extracted beam from the SPS is transported over about 1~km by
bending and focusing magnets and then split into three parts
each one directed towards a primary target where secondary particles
are created. The H2 secondary beamline emerges from the T2 primary
target and is able to transport momentum selected secondary particles
to the Experimental Hall North 1 (EHN1).
The NA61/SHINE experiment is located in the middle part of EHN1,
the NA61/SHINE production target being at 535~m distance from the T2 target.
The H2 beamline can transport charged particles in a wide range of
momenta from $\sim9$~\GeVc up to the top SPS energy of 400~\GeVc.
Alternatively the beam can transport a primary beam of protons or ions,
of low intensity to comply with the radiation safety conditions for
the experimental hall.

The North Area target cavern (TCC2) where the T2 target is located,
is about 11~m underground in order to contain the radiation produced
from the impact of the high-intensity extracted beam from the SPS.
In the proton mode the extracted intensity from the SPS is typically a
few $10^{13}$ protons per cycle at 400~\GeVc from which only a fraction
of about 40\% interacts in the targets, and the rest is dumped in
a controlled way in the TCC2 cavern. The surrounding earth acts as natural
shielding for radiation and the height difference to the experimental
hall is sufficient to reduce the muon background to the experiments.
The T2 target station hosts several beryllium (Be) plates of different
lengths. From these the target plate is chosen which best optimizes the yield
of the requested secondary particle momentum and type.
Beams for NA61/SHINE are usually produced using a target length
of 100 or 180 mm. A further optimisation can be achieved using
a set of upstream dipole magnets that can modify the incident angle
of the primary proton beam on the target and thus the production angle
of the secondary particles emitted into the H2 beamline.

\begin{figure}[ht]
   \centering
\includegraphics[scale=.30]{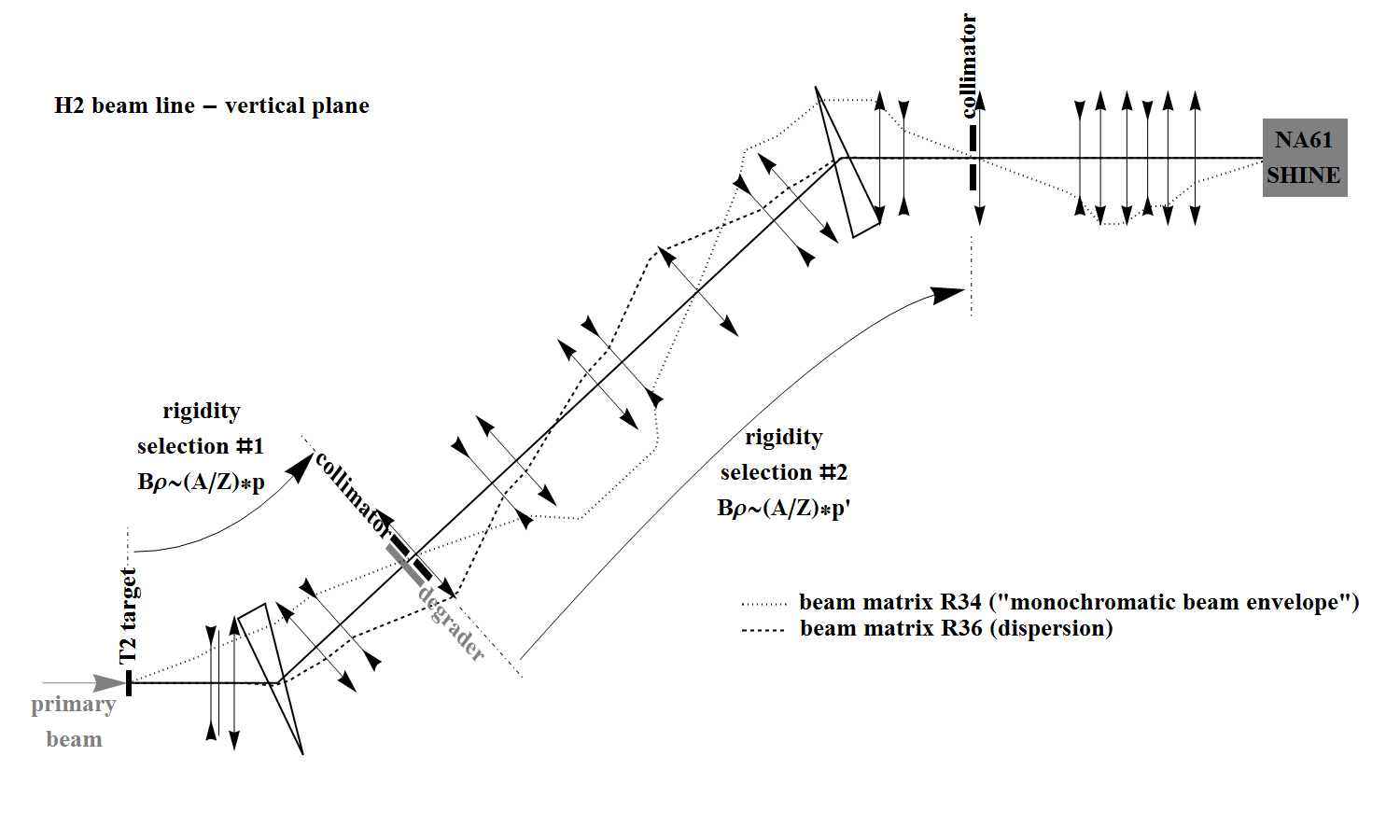}
\caption{
Schematic view of the vertical plane of the H2
beamline in the configuration used for the ion fragment separation.
The dimensions are not to scale, e.g. the beamline is more than 600~m long,
the height difference between T2 and the EHN1 is about 12~m,
the aperture of the quadrupoles is $\pm45$~mm.}
\label{fig:H2layout} 
\end{figure}

The momentum selection in the beam is done in the vertical plane,
as shown in Fig.~\ref{fig:H2layout}, where the beamline basically
consists of two large spectrometers able to select particles according
to their rigidity, i.e. the momentum to charge ratio $B\rho\approx 3.33 p_{b}/Z$, where $B\rho$
in Tesla$\cdot$meters is set by the beam optics, $p_b$ is the momentum of the beam particles in \GeVc,
and $Z$ the charge of the particle (in proton charge units).
For the energies discussed here the particle momentum can be written in
the relativistic limit as $p_b=\gamma_b\, m$, where $\gamma_b$ is the beam Lorentz factor and $m$
the particle mass, which can  be expressed in terms of the atomic mass number
$A$ times the atomic mass unit (1~amu = 0.931~\GeV).
So finally the rigidity relation for the beamline  is $B\rho\approx 3.31 \gamma_b (A/Z)$.
Each spectrometer consists of six dipole magnets that deflect the beam by a total angle of 41~mrad
and collimators defining the central trajectory. The beam spectrometer has
an intrinsic resolution of about 0.13\% and a maximum rigidity acceptance of $\pm1.7$\%.
Besides the magnetic spectrometers, the beamline is equipped with dedicated devices
which provide information on the beam position, profile and intensity
at various locations, as well as particle identification detectors like Cherenkov
or pulse height and spectrum analysis detectors to identify multiply charged particles
like heavy ions.

For the NA61/SHINE experiment secondary hadron beams in the momentum range from 13~\GeVc to 350~\GeVc
were used, as well as attenuated primary Pb82+ ion beams in the range from 13$A$~\GeVc to 158$A$~\GeVc.
For the start of the ion physics program of the experiment secondary $^7$Be ion beams
in the same momentum range were used as primary ions other than Pb82+ were not available before 2015.
The secondary $^7$Be ions were produced via fragmentation of the Pb82+ ions in the T2 target.
Starting in 2015 attenuated primary $^{40}$Ar and $^{131}$Xe ion beams in
the same momentum range will be delivered to NA61/SHINE.
The next section presents further details of the hadron and ion beams employed by the experiment
before Long Shutdown I.

%

\subsection{Hadron beams}
\label{beams:hadrons}
The H2 beamline can transport and deliver positively or negatively
charged secondary hadrons ($p$, $K^+$, $\pi^+$ and $\bar{p}$, $K^-$, $\pi^-$) to
the NA61/SHINE experiment, produced at the T2 target from the impact of
the primary proton beam from the SPS.
For a given beam tune, i.e. rigidity selection, the momentum selected hadrons
are mixed with muons, electrons and tertiary hadrons from interactions
with the collimators or the beam aperture limits.
In order to select the wanted hadrons, the beamline is equipped with a special
differential Cherenkov counter, the Cherenkov Differential Counter
with Achromatic Ring Focus (CEDAR)~\cite{CEDAR} counter.
This counter uses a gas as radiator, Helium for beam momenta higher
than 60\,\GeVc and Nitrogen for lower momenta.
The counter has a sophisticated optical system that collects and focuses
the Cherenkov photons onto the plane of a diaphragm whose opening can be tuned,
in relation to a given gas pressure, such as to allow only the photons from
the wanted species to pass through and get detected by the 8~PMTs of the counter.
By using a coincidence logic, 6-, 7- or 8- fold, a positive tagging of
the wanted particles can be achieved. For the NA61/SHINE experiment the 6-fold coincidence
signal is used for the beam trigger.

\begin{figure}[t]
\centering
\includegraphics[scale=.065]{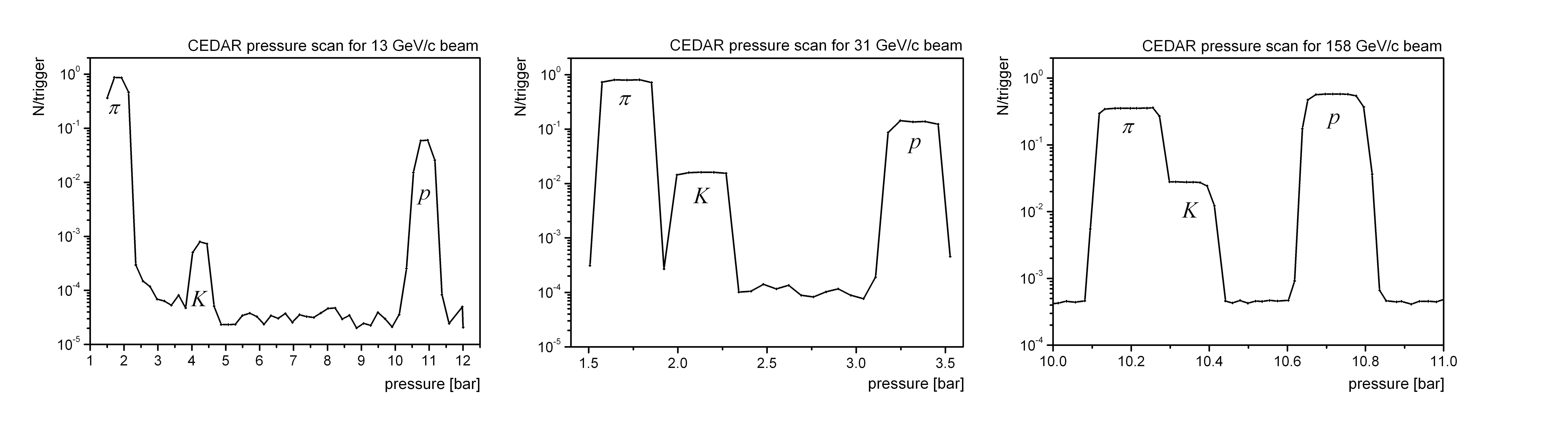}
\caption{Counts of hadrons per incident beam particle from the CEDAR counter
         as a function of the  gas pressure within the pressure range
         which covers maxima of pions, kaons and protons at 13~($left$),
         31~($middle$) and 158~\GeVc~($right$).}
\label{fig:scan} 
\end{figure}

The pressure range at which counts of a wanted hadron
($p$ ($\bar{p}$) or $\pi^+$ ($\pi^-$), or~$K^+$ ($K^-$))  per incident particle
are maximal is
found by performing  a pressure scan in the domain
where the wanted hadron peak is expected. Results of the gas
pressure scan performed in the full relevant pressure range
for beams at 13, 31 and 158~\GeVc are presented in Fig.~\ref{fig:scan}.
The final setting is the pressure corresponding to the center of the wanted hadron peak.
The actual value of the optimal pressure depends on possible admixtures in the gas,
as well as on temperature, since the production angle of Cherenkov
radiation is a function of the gas density.
The width of the signal peaks and their separation can be modified by
changing the aperture of the diaphragm.
The counter is installed in a special location of the beamline where
the beam has almost zero divergence. However to make sure the light is collected with high
symmetry, an angular alignment of the CEDAR counter has to be performed each time
the beam position at the detector changes.
Under standard operational conditions the detection efficiency of the detectors is
better than 95\%. The number of misidentified particles is lower than 0.8\%.
For beam momenta lower than 40~\GeVc, the trigger definition also requires
the signal from a carbon dioxide-filled
Threshold Cherenkov~\cite{Threshold1,Threshold2}\footnote{Ref.~\cite{Threshold1}
discusses H$_2$ filled Cherenkov detectors, similar in construction to those used
at present but operating up to 5~\GeVc.} detector in anti-coincidence.

The H2 beam optics provides a smooth focus at the target of the NA61/SHINE experiment
with an RMS width slightly larger than 2\,mm at the lowest momenta
and 1.2\,mm at 158~\GeVc. The momentum spread is typically lower than 1\%, defined by
the collimator settings in the line.

\subsection{Primary and secondary ion beams}
\label{beams:ions}

The physics program of NA61/SHINE requires beams of Beryllium (Be), Argon (Ar)
and Xenon (Xe) ions. For reasons of compatibility with the LHC program the
Be beam is obtained from the fragmentation of Pb ions,
whereas the Ar and Xe beams will be specifically produced for NA61/SHINE.

The transport of primary ions in the H2 beamline is straightforward.
The beam tune, i.e. rigidity is set to match the ion momentum extracted from the SPS,
and the detectors in the beamline are adapted to the non-standard (high)
charge of the beam particles. To respect the radiation limits
and classification in the Experimental Hall,
the intensity in the SPS machine is kept to a minimum,
typically using a single (or double) injection into the SPS resulting in
a total of $\sim 6\times10^8$ ions. The intensity is further reduced in the H2 beamline
by collimation, down to a rate of a few $10^5$ ions at the experiment.

The selection and transport of a specific ion species from a fragmented heavy (Pb82+)
ion beam for nuclear reaction experiments is not straightforward. The H2 beamline selects on rigidity, i.e. $\sim \gamma_b\,(A/Z)$, and the desired ions produced from the fragmentation of the primary Pb beam will be mixed with a variety of other nuclei with similar mass to charge ratios and slightly different rigidity values within the beam acceptance. Moreover, rigidity overlaps occur not only for ions with the same mass to charge ratio but also
for neighbouring elements
due to the nuclear Fermi motion which smears fragment momenta.
Without Fermi motion the fragments would leave the interaction region
almost undisturbed with the same momentum per nucleon as the incident Pb ions.
The Fermi motion depends on the fragment and projectile mass, and can spread the longitudinal momenta for light nuclear fragments by up to 3-5\%, much larger than the beam acceptance.

\begin{figure}[htbp]
\begin{center}
\begin{minipage}[b]{0.95\linewidth}
\includegraphics[scale=0.35]{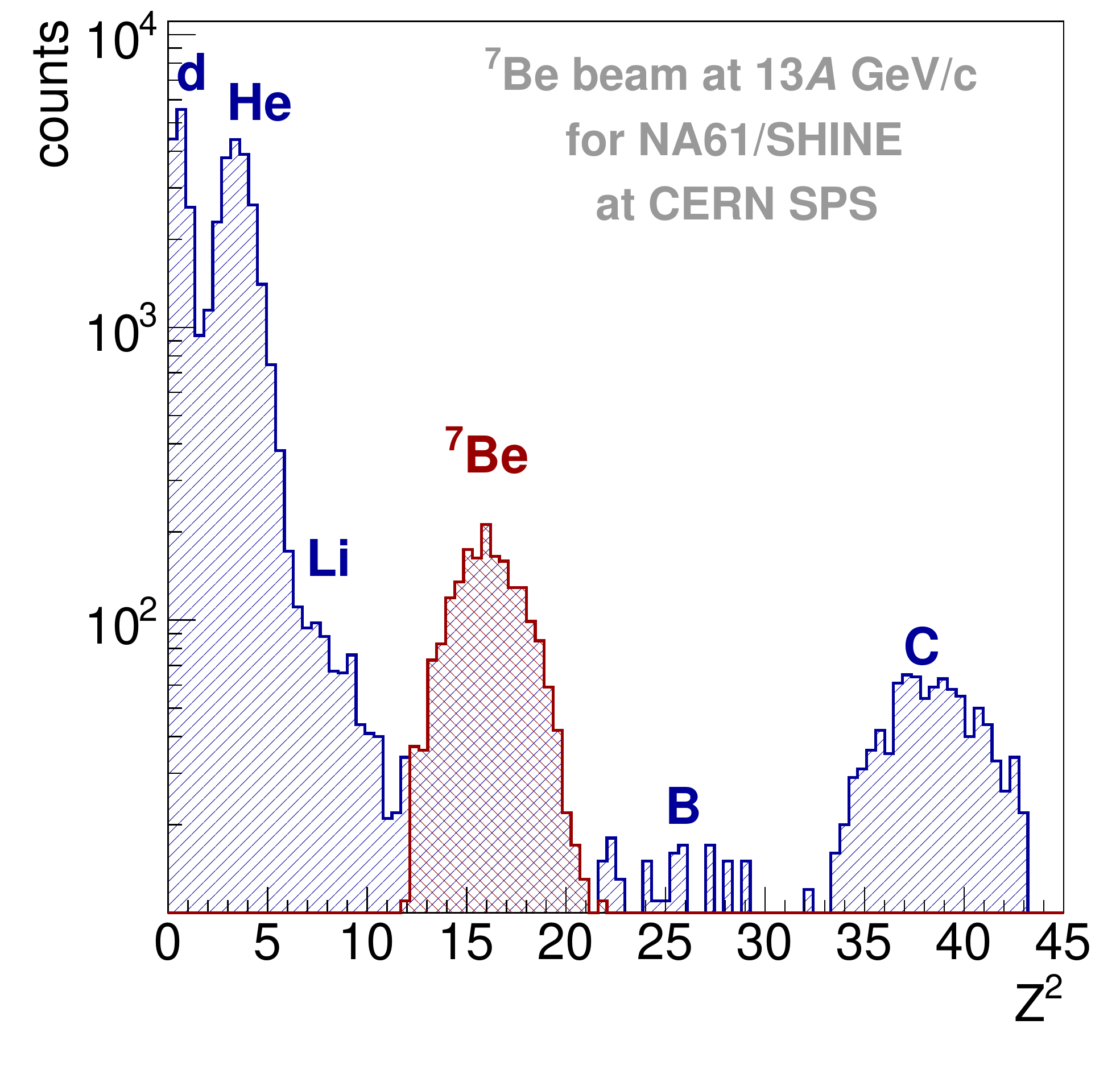}
\includegraphics[scale=0.35]{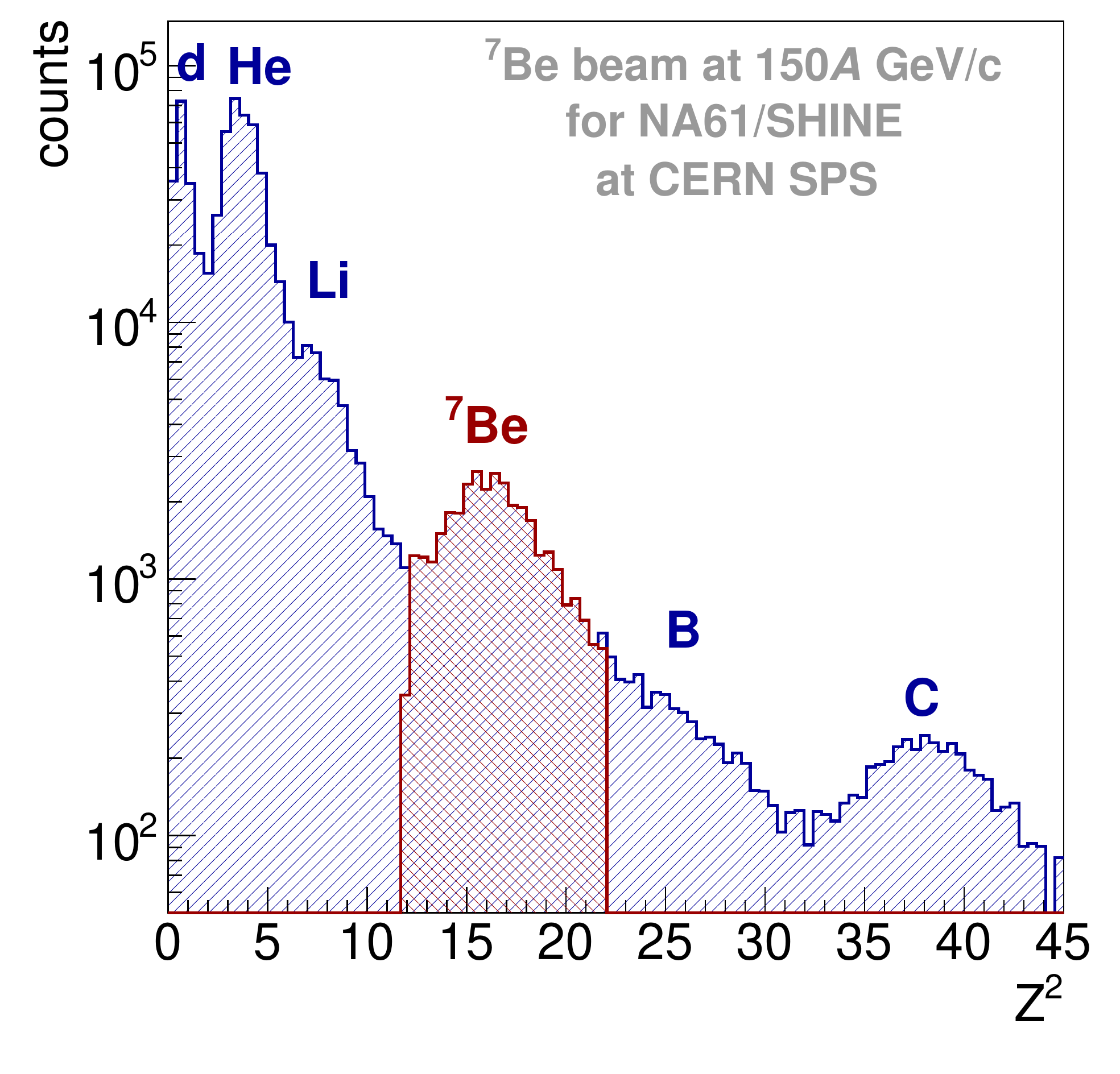}
\end{minipage}
\end{center}
\caption{
Charge spectrum measured with the Quartz Z-detector for 13$A$  and
150$A$~\GeVc
ion beams.
}
\label{fig:zdet3}
\end{figure}

The $^7$Be ion ("wanted ion") was selected for the NA61/SHINE beam, because it has
no long-lived near neighbors and thus allows to make a light ion beam with a large ratio
of wanted to all ions. The near neighbors to $^7$Be are isotopes
$^6$Be and $^8$Be and the nuclei with a charge difference
of one and a similar mass to charge ratio (e.g. $^5$Li, $^9$B).
Furthermore $^7$Be has more protons ($Z=4$) than neutrons ($N=3$).
Such nuclear configurations are disfavored with increasing nuclear mass,
since a surplus of protons causes a Coulomb repulsion which cannot be balanced by
the attractive potential of the fewer neutrons. As can be seen in
Fig.~\ref{fig:zdet3}, $^7$Be fragments are accompanied
mainly by $^2$D and He ions whose rigidity overlaps with the one of the
wanted ions due to Fermi motion.
A problematic choice of wanted ion species would be a nucleus
with mass to charge ratio of two, which would be accompanied by stable or
long lived nuclei from $^2$D up to $^{56}$Ni.

At low energies, a better selectivity of the wanted ions is achieved with
the "fragment separation" method~\cite{FragSep}, profiting from the double
spectrometer of the H2 beamline (see Fig.~\ref{fig:H2layout}).
In the first spectrometer, ions are selected
within a rigidity value that maximizes the wanted to all ratio for the
fragments as produced by the primary fragmentation target. At the focal
point of the first spectrometer, a piece of material (called degrader) is
introduced, in which ions lose energy in dependence of their charge.
Then in the second spectrometer the different ions will be spatially
separated according to their charge state, which then can be selected
using a thin slit (collimator). The drawback
of this method is a loss of beam intensity due to nuclear interactions
and the beam blow-up due to multiple scattering in the material of the
degrader which rises with increasing thickness.  Thus a high
separation power of the fragment separator spectrometer goes along with a
high loss of intensity. Furthermore, for a given degrader thickness
both the nuclear cross section and the energy loss are to a large extent
energy independent. This means that the separation power ($\Delta E/E$)
increases with decreasing energy.

Under typical running conditions a beam of several 10$^8$ Pb ions per spill
from the SPS is focused onto
the 180~mm long fragmentation target made of Be. During its passage through
the target
the Pb beam undergoes (mostly peripheral) collisions with the (light) target
nuclei. Part of the resulting mix of nuclear
fragments is captured by the H2 beamline, tuned to a rigidity such
that the fraction of the created $^7$Be to all ions is maximized (see Fig.~\ref{fig:zdet3}).
Ion fluxes at the NA61/SHINE apparatus were 5000 to 10~000 $^7$Be particles with
10 to 20 times more unwanted ions.

\begin{figure}[h!]
 \begin{center}
 \includegraphics[width=0.45\linewidth]{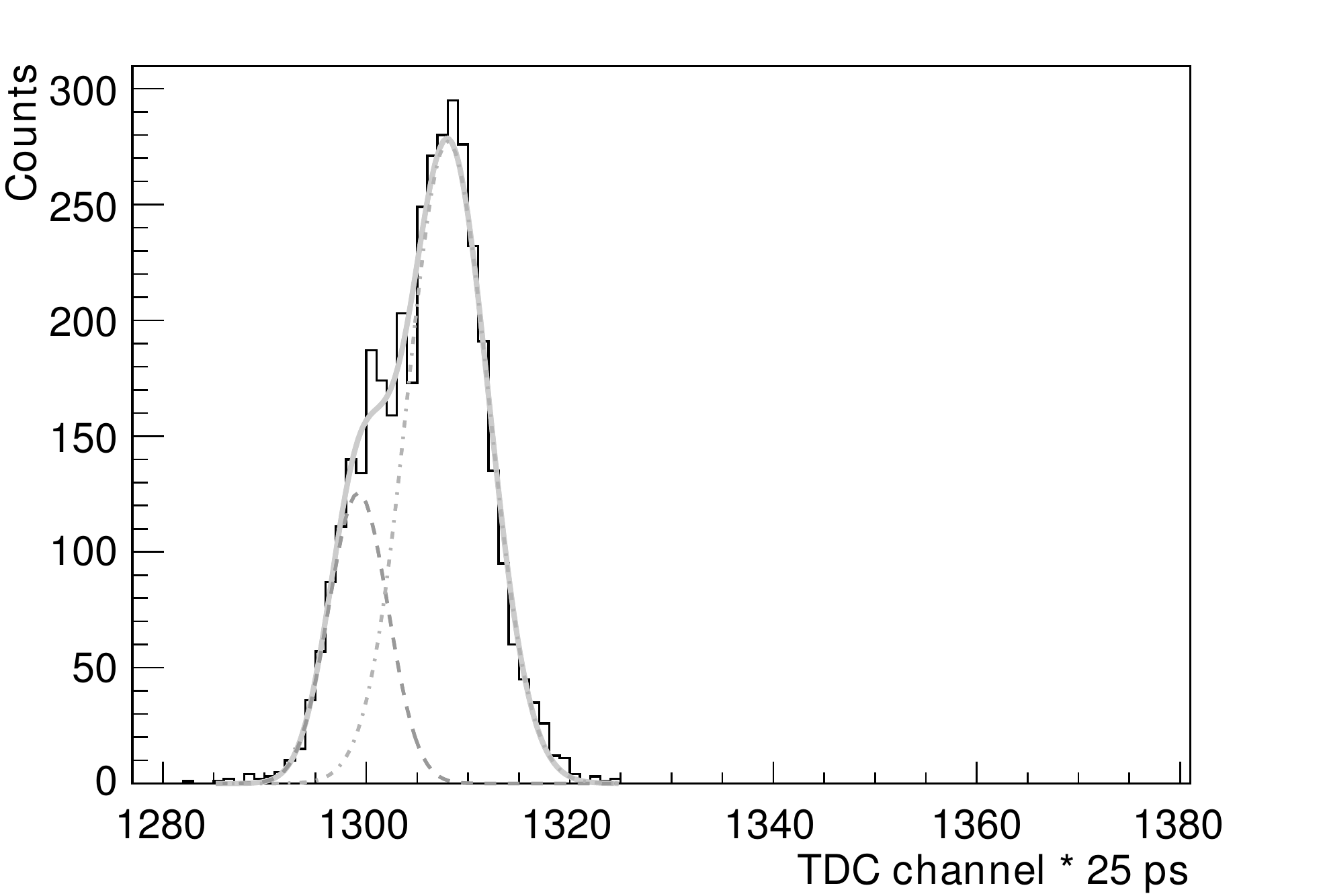}
 \includegraphics[width=0.45\linewidth]{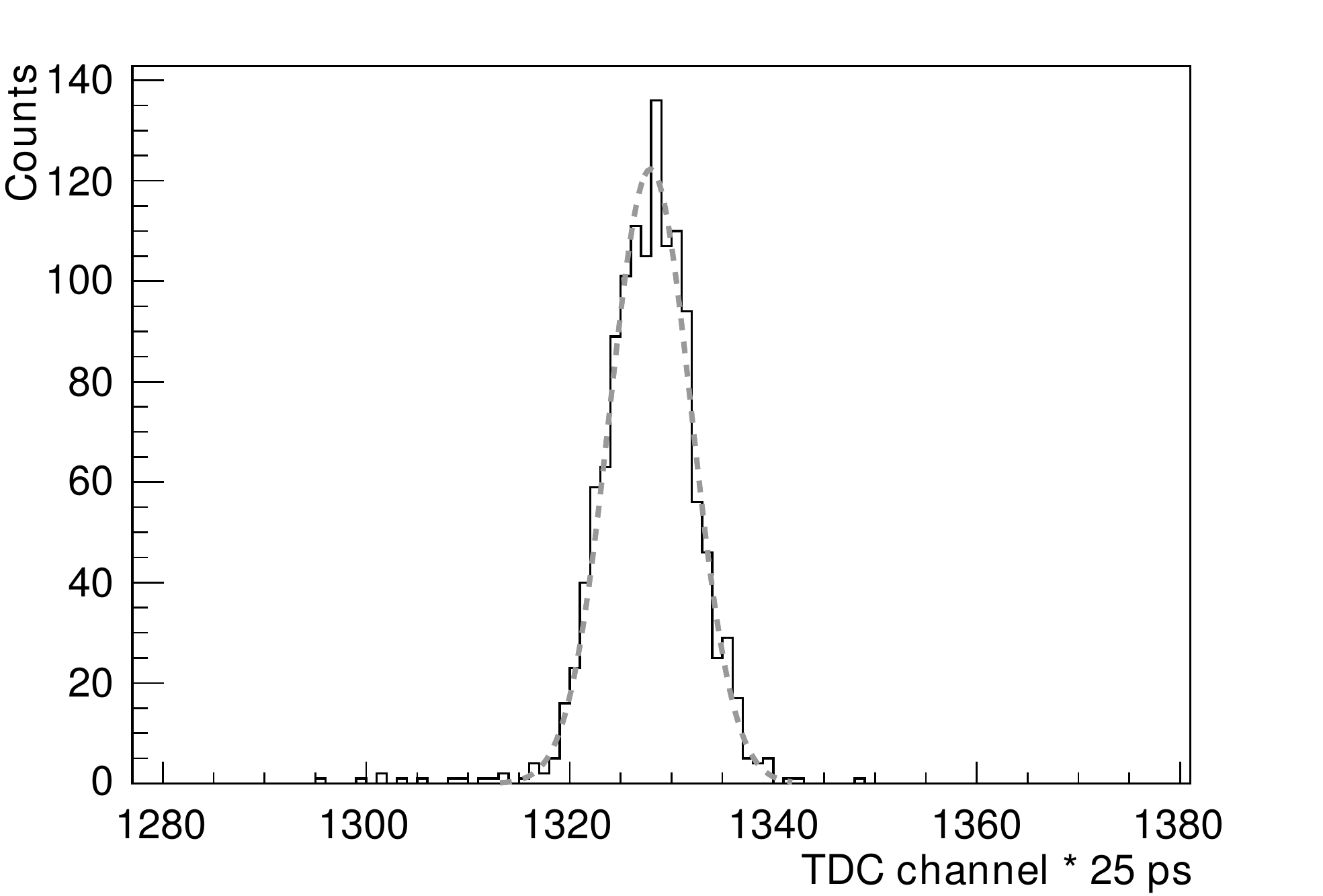}
 \includegraphics[width=0.45\linewidth]{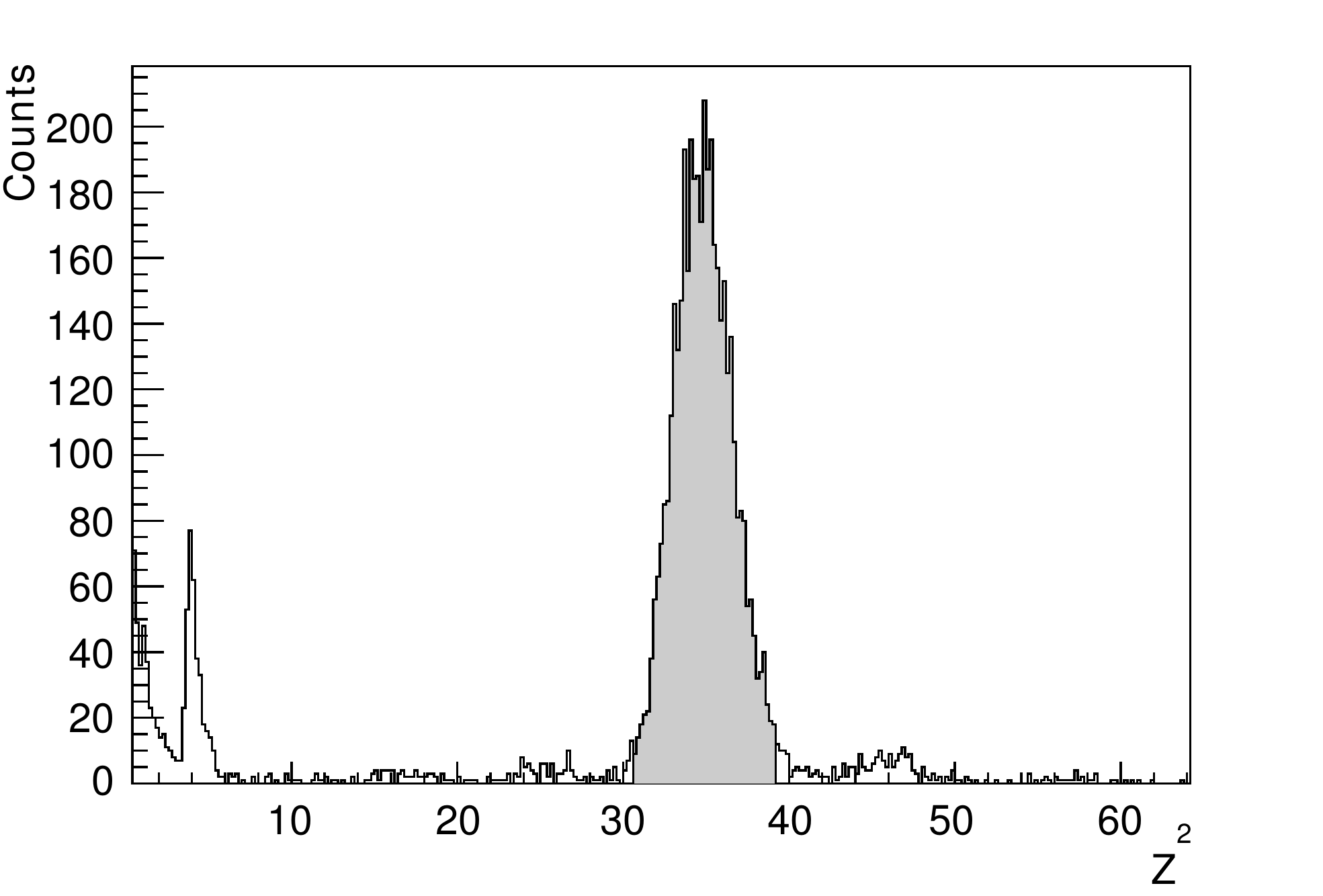}
 \includegraphics[width=0.45\linewidth]{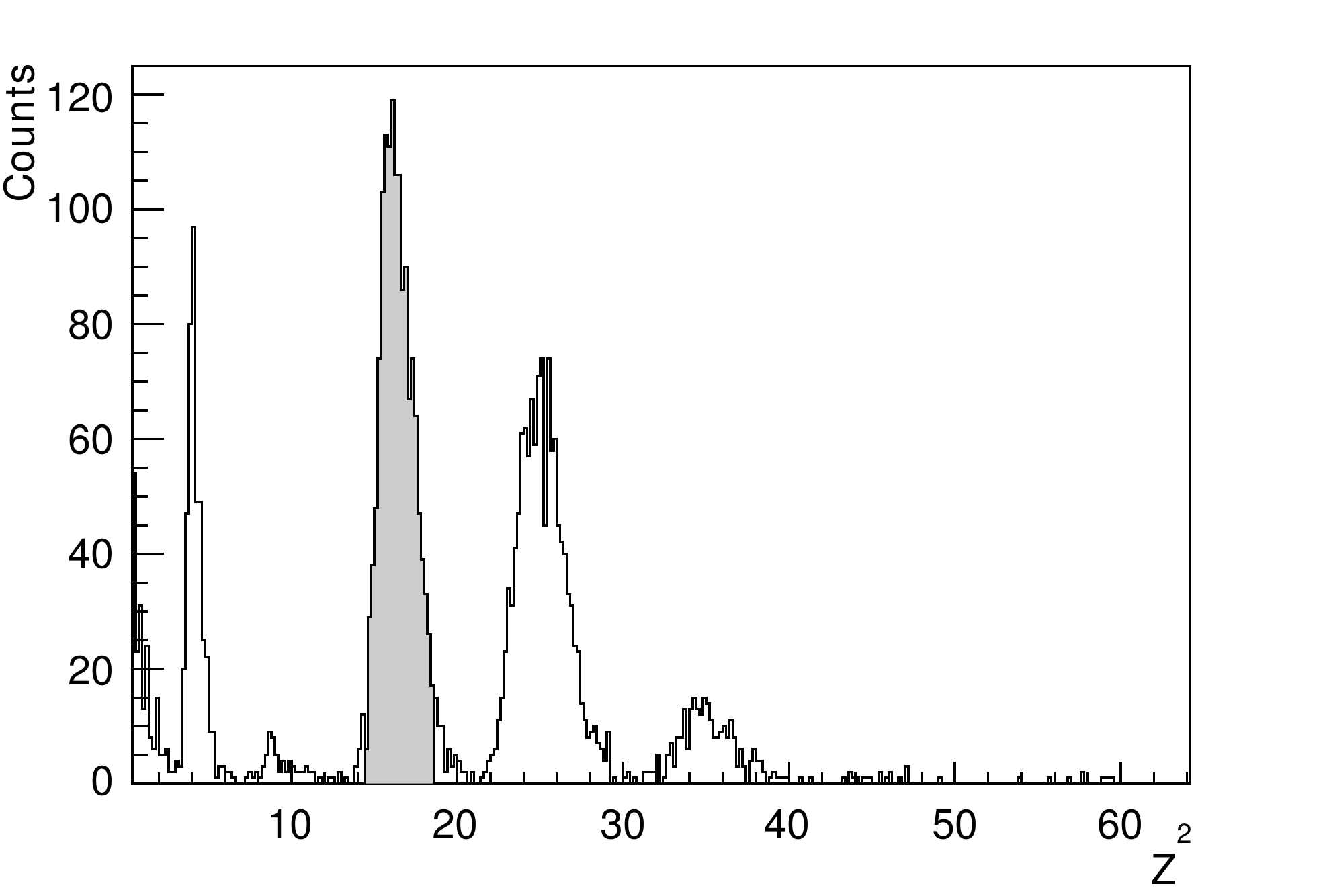}
 \end{center}
 \caption{
 $Top:$ The time of flight spectra of ions for the carbon ($left$) and
 beryllium ($right$)  peaks resulting from fragmentation
 of the primary Pb beam at 13.9$A$~\GeVc.
 The spectrum of carbon is fitted by a sum of two Gauss functions, whereas
 that of beryllium by a single Gaussian.
 $Bottom:$ The $Z^2$ spectra with the selected (gray area) peaks
 of carbon ($left$) and beryllium ($right$).
 The spectra were measured by the A-detector. }
 \label{fig:atof}
 \end{figure}

The optional degrader (1 or 4\,cm thick Cu plate) is located between the two
spectrometer sections (see Fig.~\ref{fig:H2layout}).
The composition of the ion beam can be monitored by
scintillator counters which measure the charge ($Z^2$) and time of flight ($tof$)
of the ions. The latter allows mass ($A$) determination
of the ions for momenta lower than 20$A$~\GeVc and thus
a check of the purity of the $^7$Be beam.
This is illustrated in Fig.~\ref{fig:atof} where the $tof$
spectra of ions for the carbon ($left$) and
beryllium ($right$)  peaks resulting from fragmentation
of the primary Pb beam at 13.9$A$~\GeVc are shown.
The carbon spectrum shows clear evidence for two isotopes, whereas
only the single Be isotope ($^7$Be) is seen in the Be spectrum.
We therefore decided to select $^7$Be for data taking.

\begin{figure}[htb]
   \centering
    \includegraphics[scale=.42]{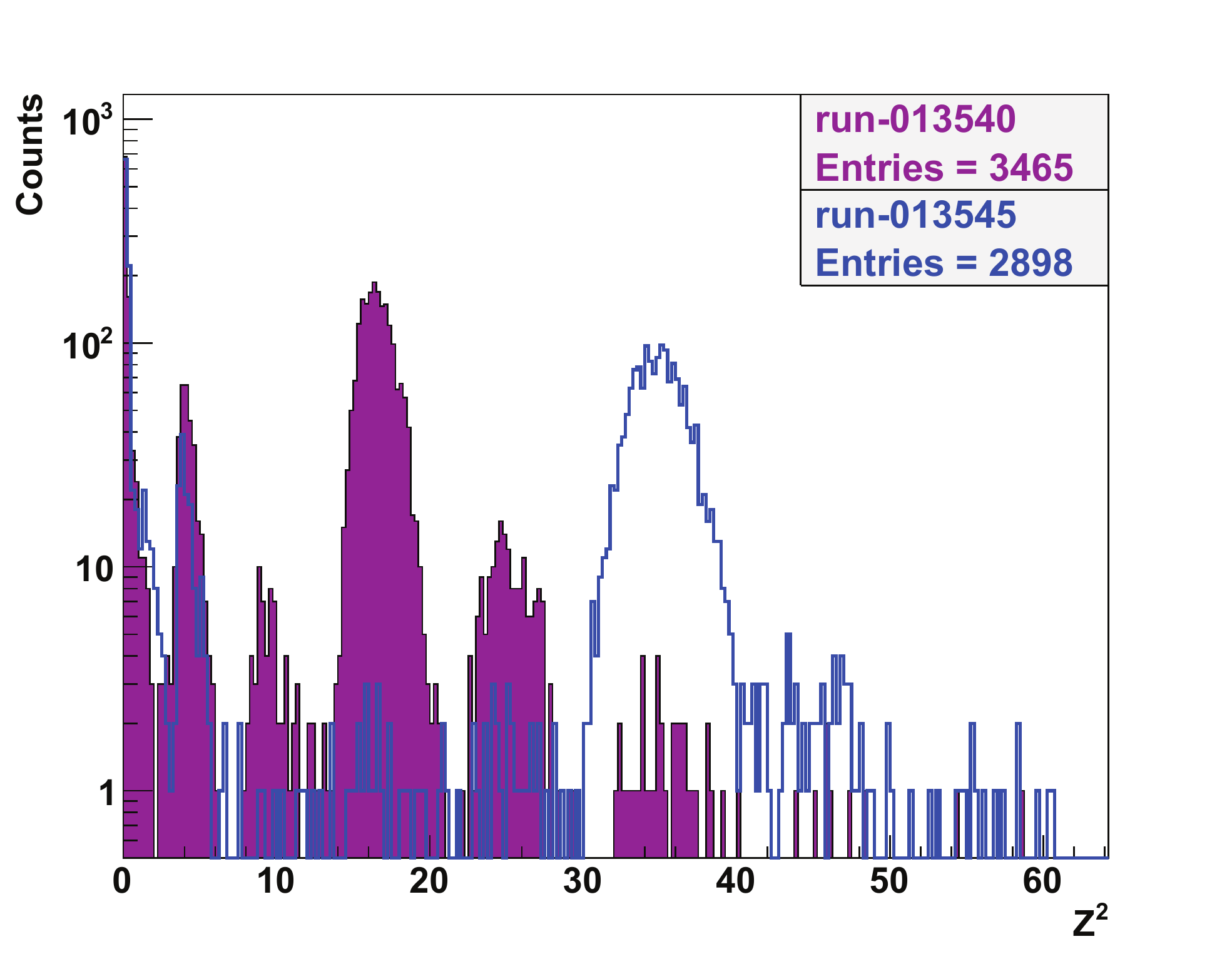}
   \caption{Fragment charge distributions obtained with the H2 beamline in
    fragment separator mode and collimator settings
    optimized for $^7$Be (filled purple) and $^{11}C$ (open blue) fragments.
    The primary Pb beam momentum is 13.9$A$~\GeVc.}
   \label{fig:coll_frag}
\end{figure}

The fragment separation method was tested
in 2010 with a 13.9$A$~\GeVc Pb beam incident on the primary target of the
H2 beamline with the 4~cm thick degrader in place. Figure~\ref{fig:coll_frag}
shows for a given rigidity setting
the charge distributions resulting from two collimator settings which optimize
either the $^7$Be (filled purple histogram) or the $^{11}$C (open blue
histogram) content. Comparison with Fig.~\ref{fig:zdet3} demonstrates that
by using the degrader the ratio of wanted to all ions is improved by nearly
an order of magnitude.

During the 2011 running period the NA61/SHINE collaboration used the beamline
configuration without degrader at beam momenta of 150$A$, 75$A$, and 40$A$~\GeVc
with settings optimised for $^7$Be ions. This choice allowed tagging pure
$^7$Be ions without incurring the penalty on intensity
from the degrader while still resulting in an acceptable ratio
of wanted to all ions.
In 2012 and early 2013 fragmented ion beams at 30$A$, 19$A$, and at 13$A$~\GeVc
(the lower momentum limit of the accelerator capabilities)
were provided to NA61/SHINE.

\section{\large Beam detectors and trigger system}
\label{beam_det}

A set of scintillation and Cherenkov counters as well as the beam position
detectors located upstream of the target provide precise timing reference,
along with charge and position measurement of the incoming beam particles.
Interaction counters located downstream of the target allows to trigger on
interactions in the target. Typical parameters of these detectors are
summarized in Table~\ref{tab:trigger}.
\par
\begin{figure}[htb]
\centering
\includegraphics[width=1.00\textwidth]{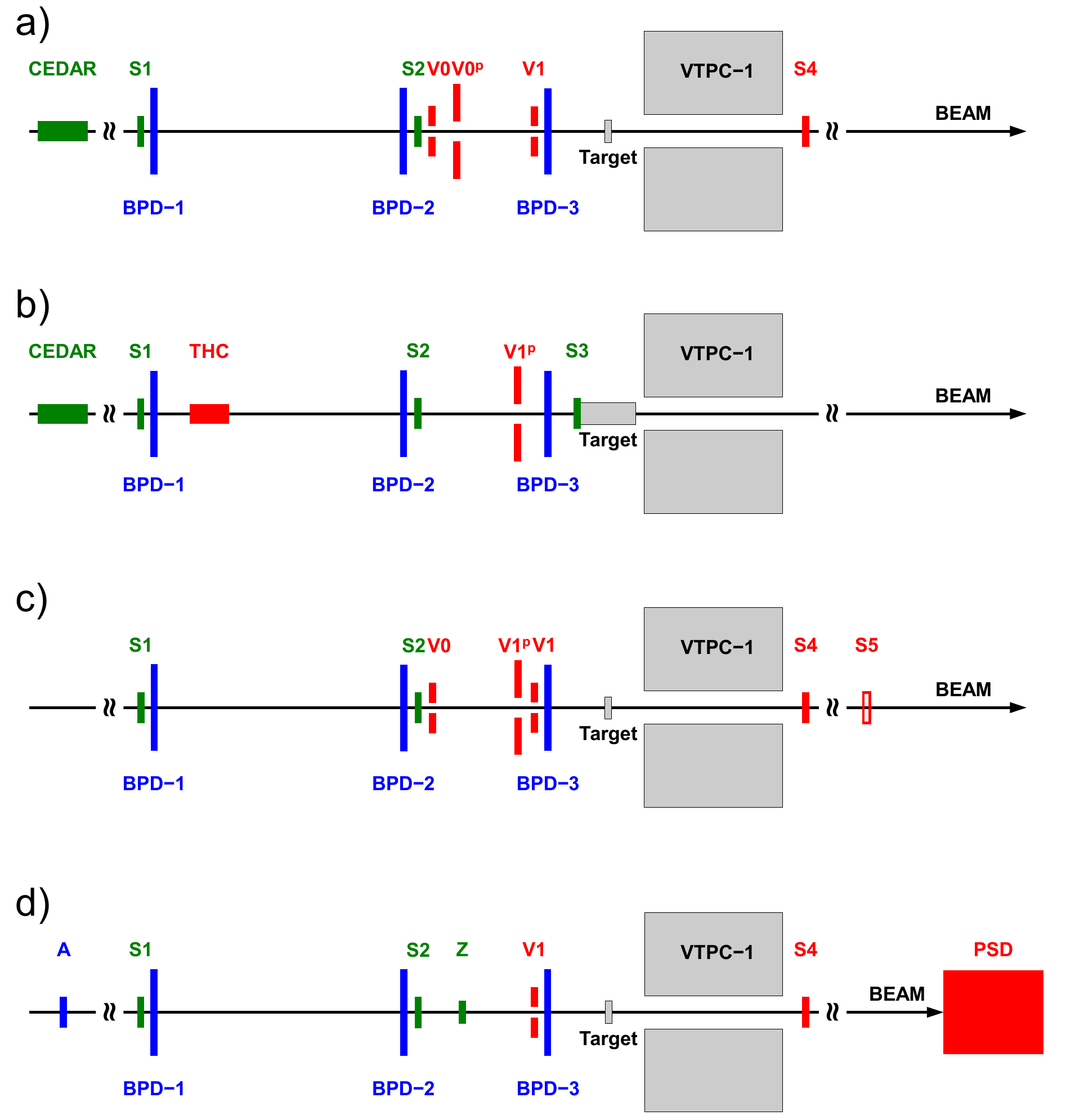}
\caption
{
  Schematic layout (horizontal cut in the beam plane, not to scale) of
  the beam detectors in four configurations of the beamline,
  see text for details.
}
\label{fig:beamLines}
\end{figure}

The locations of the beam detectors in four exemplary configurations of the beamline are
indicated in Fig.~\ref{fig:beamLines}.
These were used for data taking on (from top to bottom):
\begin{enumerate}[(a)]
\setlength{\itemsep}{1pt}
\item
p+p interactions at 13 and 158\,\GeVc in 2011,
\item
p+(T2K replica target) interactions in 2009 and 2010
(the S3 detector was glued to the surface of the replica
target),
\item
h$^-$+C interactions at 158 and 350\,\GeVc in 2009
(an additional interaction trigger was used
with the S5 counter instead of S4 to estimate the trigger bias),
\item
$^7$Be+$^9$Be interactions at 13$A$, 19$A$ and 30$A$\,\GeVc in 2013.
\end{enumerate}
Colour indicates the detector function
in the NA61/SHINE trigger.
Namely signals of detectors in green  are used in coincidence,
in red in anti-coincidence and detectors in blue are not used in the trigger logic.

\begin{table}
\centering
\begin{tabular}{|c|c|c|c|c|c|}
\hline
detector & dimensions & hole & position & \multicolumn{2}{|c|}{material budget} \\ \cline{5-6}
        & [mm]       & [mm] & [m]      & [$\%\lambda_I$] & [$\%X_0$] \\
\hline
S1         &   $60 \times 60 \times 5$   &   &  -36.42 & 0.635 & 1.175 \\
S2         &   $\phi =28 \times 2$         &   &  -14.42 & 0.254 & 0.470 \\
S3         &   $\phi = 26 \times 5$        &   &  -6.58 &  0.635 & 1.175 \\
S4         &    $\phi = 20 \times 5$     &   &  -2.11 & 0.635 & 1.175 \\
S5         &   $\phi = 20 \times 5$        &   &  9.80  & 0.635 & 1.175 \\
\hline
V0         &  $\phi = 80 \times 10$       &   $\phi = 10$     & -14.16& & \\
V0$^p$        &  $300 \times 300 \times 10$  &   $\phi = 20 $ & $\approx$ -14 & & \\
V1         &  $100 \times 100 \times 10$  &   $\phi = 8 $  & -6.72 & & \\
V1$^p$        &  $300 \times 300 \times 10$  &   $\phi = 20 $ & -6.74 & & \\
\hline
A          &  $150 \times 5 \times 15$    & & $\approx$ -146 & 1.904 & 3.526 \\
Z          &  $160 \times 40 \times 2.5$    & & -13.81 & 0.562 & 2.034 \\
\hline
BPD-1      & $48 \times 48 \times 32.6 $ & & -36.20 & 0.025 & 0.070 \\
BPD-2      & $48 \times 48 \times 32.6 $ & & -14.90 & 0.025 & 0.070 \\
BPD-3      & $48 \times 48 \times 32.6 $ & & -6.70 & 0.025 & 0.070 \\
\hline
\multicolumn{3}{|c|}{Typical thin target position} & -5.81 & & \\
\hline
\end{tabular}
\caption{
  Summary of typical beam detector parameters: dimensions, positions along the
  beamline ($z$~coordinates) and their material budget (in terms of the
  nuclear interaction length~$\lambda_I$ and radiation length~$X_0$). Positions
  of most of these detectors varied in time by a few cm due to dismounting and
  remounting in subsequent runs. Exceptions are BPD-3, V1 and V1$^p$, for which
  positions varied up to 50\,cm depending on the employed target. E.g. for the
  liquid hydrogen target the positions were: BPD-3 -6.93\,m, V1 -7.20\,m and V1$^p$
  -7.23\,m.
  Positions of the A and V0$^p$ detectors were not surveyed,
  therefore are known only approximately.
}
\label{tab:trigger}
\end{table}

\subsection{Beam counters}
\label{beam_det:count}

Minimization of the total detector material in the beamline is a major
concern,
especially with ion beams.
A minimal set of plastic scintillator (BC-408) beam counters
(see Fig.~\ref{fig:beamLines}) is therefore used.
The first detector S1 is located upstream of the target at position
$z$~=~-36.42~m.
As the time of flight resolution in low-multiplicity events has
to rely on a precise reference time, the S1 counter (0.5~cm thick)
is equipped with four photomultipliers directly coupled to the scintillator.
The second beam counter S2 (0.2~cm thick) is located just behind the BPD-2 detector
at $z$~=~-14.42~m.
In the case of primary heavy ion beams the S1 and S2 scintillator detectors will be
replaced by quartz detectors of 200~$\mu$m thickness
yielding sufficient timing and pulse height resolution from the Cherenkov effect.
Downstream of
the S2 detector two 1~cm thick veto scintillator detectors V0 and V1
are positioned at $z$~=~-14.16~m
and $z$~=~-6.72~m, respectively.
The round V0 detector has outer diameter of 8~cm and in the center a hole of 1~cm
diameter.
The square (10$\times$10~cm$^2$) V1 detector also has a 1~cm central hole.
Additionally two veto detectors $V0^{'}$ and $V1^{'}$ are installed for configurations of beam line a), b)
and c), respectively. For configuration b) the Threshold Cherenkov, THC, detector is also positioned.
The S4 and S5 detectors (2~cm diameter, 0.5~cm thickness) are located
downstream of the target at
$z$~=~-2.11~m and 9.80~m, respectively.
They are used to select interactions of beam particles in the target.
Interactions in the target are signaled
by the absence of the beam particle signal in the counter.
The beam counter parameters are summarized in Table~\ref{tab:trigger}.
In the plane transverse to the beam direction (the $x-y$ plane) the detectors
are centered at the maximum of the beam profile.

\subsection{A-detector}
\label{beam_det:a}

The A-detector was constructed to verify whether the secondary Be beam consists only of
the single isotope $^7$Be.
It was located about 140~m upstream of the target and measured
the time of flight ($tof$) of beam ions  between A- and S1 detectors.
It consists of a plastic scintillator (BC-408) bar (15x0.5x1.5~cm${}^{3}$)
with light readout from both sides of the scintillator bar
by the two fast PMTs, EMI 9133.
The time resolution of the A-detector was measured to be about 80~psec
during a test in the T10 beam of the CERN PS using pions.


The time of flight spectra measured by the A-detector at 13.6$A$~\GeVc
are shown in Fig.~\ref{fig:atof} ($top$) for carbon and beryllium ions
selected by the Z$^2$ proportional signal amplitude simultaneously
measured in the A-detector ($bottom$).
The single isotope beryllium (${}^{7}$Be) spectrum demonstrates a $tof$
resolution of 60~psec.
The expected difference of  $tof$ between ${}^{7}$Be ions and
its nearest isotopes is about 170~ps at the beam momentum of 13.6$A$~\GeVc
and distance of 140~m.


\subsection{Beam Position Detectors}
\label{beam_det:bpd}

\begin{figure}[ht]
\begin{center}
\begin{minipage}[b]{0.95\linewidth}
\raisebox{-0.1\height}{\includegraphics[width=0.53\linewidth]{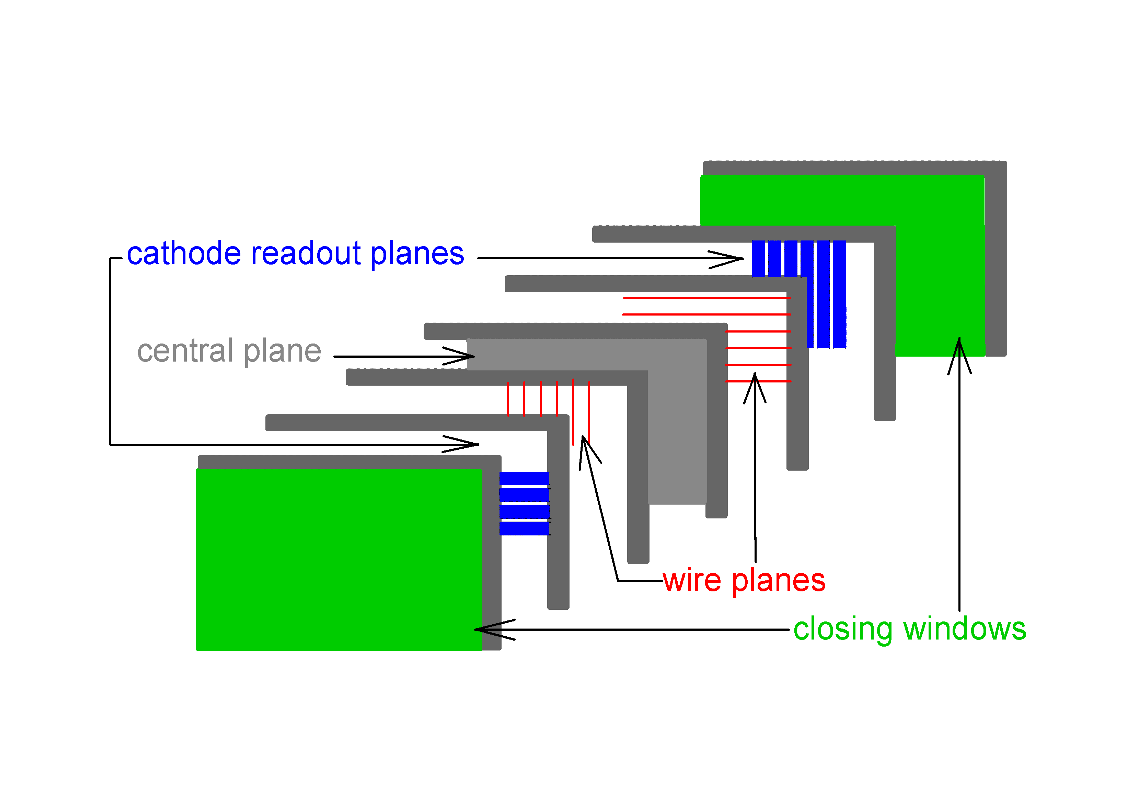}}
\qquad
\includegraphics[width=0.41\linewidth]{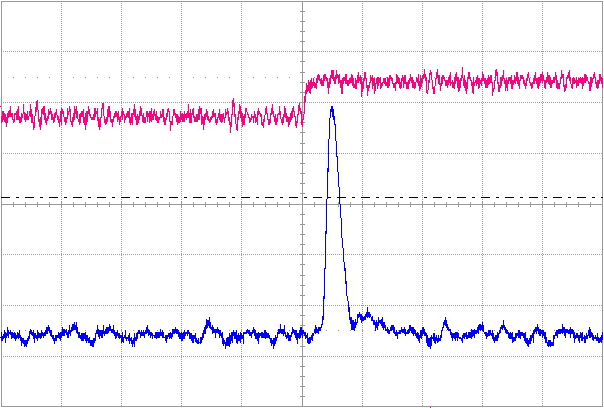}
\end{minipage}
\end{center}
\caption{
{\it Left:} The schematic layout of the BPD detector.
	{\it Right:}
The signal generated by a BPD pre-amplifier
(upper curve) and the corresponding signal at the output of a~shaper
(lower curve). The time scale of the $x$ axis is 500 ns per division.
	 }
	\label{fig:bpd1}
	\end{figure}

The positions of the incoming beam particles in the transverse plane
are measured by a~telescope of three Beam Position Detectors (BPDs)
placed along the beamline upstream of the target. The NA61/SHINE
BPDs (see Fig.~\ref{fig:bpd1}\lef) with active areas
of $48\!\times\!48\,\text{mm}^2$ were constructed  in 2009.
These detectors are proportional chambers
operated with Ar/CO$_{2}$ 85/15 gas mixture. Two orthogonal sense wire planes (15\,$\mu$m
tungsten wires with 2\,mm pitch) are sandwiched between three cathode
planes made of 25\,$\mu$m aluminized Mylar.  The outer cathode planes of these
detectors are sliced into strips of 2\,mm pitch which are connected to the
readout electronics.  In order to detect beam particles at high intensities
(about $10^{5}$ particles/second)  shapers based on AD817 integrated
amplifiers were constructed. The width of their output signal is about 350 ns.
Negative undershoots of output signals observed with the NA49 shapers are
practically eliminated (see Fig.~\ref{fig:bpd1}\rig).

Each BPD measures the~position of the trigger-selected beam particle in two orthogonal
directions independently using two planes of orthogonal strips. On each strip
plane a~charge distribution is induced with a~width of about 5~strips on
average. The reconstruction algorithm first searches for a~cluster in each
plane. The cluster is defined as a~set of adjacent strips with signal amplitudes above
a threshold value (to remove signals from pedestal fluctuations). Then, an
average of the strip positions weighted with the signal amplitudes on the strips is calculated for
the cluster to estimate the position of the beam particle (the so-called
centroid method). A~3-dimensional point measured by a~given BPD is built from
two transverse coordinates measured by the two strip planes and the position of the BPD
along the beamline.

\begin{figure}[ht]
\begin{center}
\includegraphics[scale=.55]{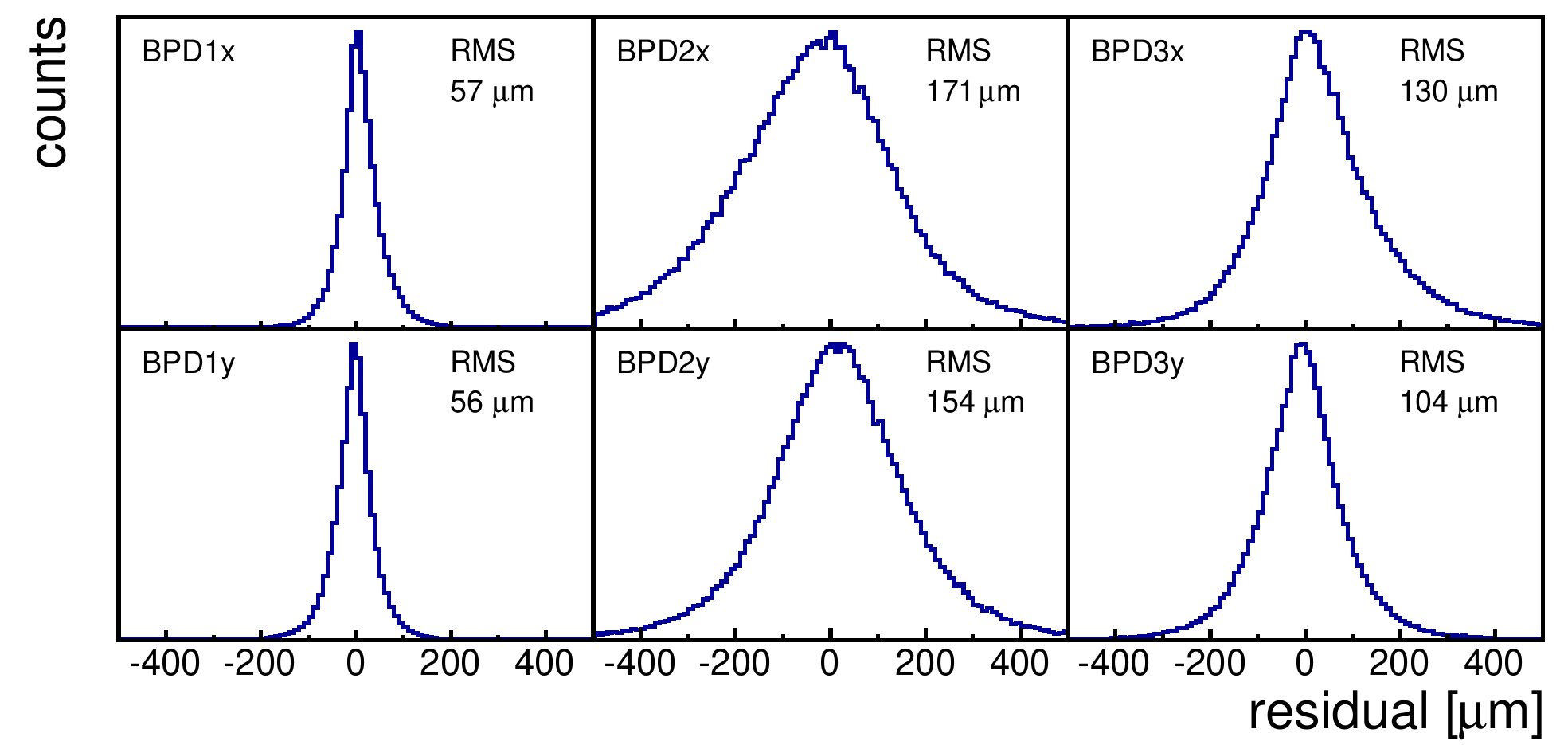}
\end{center}
\caption{
Distributions of residuals associated with beam track fits in each of the BPDs
in the $x-z$ (top row) and $y-z$ (bottom row) planes. Differences in RMS widths are mainly due to
non-equal distances between the detectors in the $z$ direction.
Plots are for the negatively charged hadron (mainly $\pi^-$)
beam at 158\,\GeVc.
}
\label{fig:bpd2} 
\end{figure}

In order to reconstruct a~beam particle track, least squares fits of straight
lines are performed to the positions measured by the three~BPDs in $x-z$ and
$y-z$ planes independently.
Distributions of residuals
associated with those fits are shown in Fig.~\ref{fig:bpd2}. The RMS
widths of the distributions indicate the order of magnitude of the accuracy of
the BPD measurements, which is in agreement with the design value of
$\sim$100\,$\mu$m. Differences of the RMS widths are mainly due to
non-equal distances
between the detectors in the $z$ direction ($d$(BPD-1, BPD-2)~=~21\,m,
$d$(BPD-2, BPD-3)~=~8\,m).

\subsection{Z-detectors}
\label{beam_det:z}

In order to select secondary ions by their charge at the
trigger level, two Cherenkov counters (Z-detectors) were constructed.
These are:
\begin{enumerate}[(i)]
\setlength{\itemsep}{1pt}
\item The Quartz Detector (QD) with a quartz (GE214P) plate
  $160\!\times\!40\!\times\!2.5\,\text{mm}^3$ as the
radiator is equipped with two photomultipliers (X2020Q) attached
to both sides of the quartz plate.
Optical grease (BC630) is applied between the PMTs and the quartz plate in
order to achieve good light transmission.
\item The Gas Detector (GD) uses a 60~cm long C$_{4}$F$_{10}$ gas radiator. The
construction of this
detector is based on the standard CERN threshold Cherenkov unit.
Cherenkov photons are reflected by a 25~$\mu$m thick aluminized
Mylar foil and focused by a parabolic
mirror onto a single photomultiplier, Hamamatsu R2059.
\end{enumerate}

The large refraction index of the quartz radiator of about 1.458 leads to a
low Cherenkov threshold momentum of $\approx$~0.88~\GeVc per nucleon.
Thus the detector performance is largely independent of the beam
momentum in the momentum range relevant for the NA61/SHINE energy
scan program.
The refraction index of the
C$_{4}$F$_{10}$ gas at normal conditions is about 1.00142 leading to a Cherenkov
threshold momentum of about 17.6~\GeVc per nucleon. For the momentum of
13$A$~\GeVc the lowest acceptable value of the refraction index is
$\approx1.0023$
requiring a minimum gas pressure of 1.63~bar.

Both detectors were tested during the 2011 run using the secondary ion beam
at 150$A$~\GeVc. The ADC signals measured in the Quartz and Gas Detectors
showed well
separated peaks of proton and He beam particles.
Both detectors performed well at this momentum, namely the measured
width to
mean value ratio for the $^4$He peak was 18-19\%. However, it was estimated
that at the lowest beam momentum the performance of the GD will
deteriorate both in terms of the charge resolution and the material budget
(at the sufficiently high gas pressure it is about a factor of
2.5 larger than for the QD).
Therefore the QD was used as the Z-detector for the physics data taking
with secondary $^7$Be beams in 2011-2013.

Figure~\ref{fig:zdet3} presents $Z^2$ distributions measured
by the QD for secondary ion beams of
13$A$,  and 150$A$~\GeVc momentum used in the data taking
on $^7$Be+$^9$Be interactions.

\subsection{Trigger system}
\label{beam_det:trig}

In designing the NA61/SHINE trigger system,
particular attention was paid to developing a flexible
and robust system
capable of handling and selecting  different reactions
using a variety of beams (pions, kaons, protons, ions) and targets
as required by the NA61/SHINE physics programme~\cite{proposal}.

The trigger is formed using several of the beam counters listed in Table~\ref{tab:trigger},
the Cherenkov detectors for beam particle (hadrons or ions)
identification,
and the PSD calorimeter, as illustrated in Fig~\ref{fig:beamLines}.

The core of the trigger logic is an FPGA (Xilinx XC3s1500) running at
120~MHz embedded in a CAMAC Universal Logic Module,
the CMC206~\cite{CMC206}.
CAMAC is used for backward compatibility with the legacy NA49 electronics.
The trigger logic is divided into three main blocks:
\begin{enumerate}[(i)]
\setlength{\itemsep}{1pt}
\item beam logic
\item beam particle identification
\item interaction logic.
\end{enumerate}
Up to four different triggers can be run simultaneously
with a selectable 12~bit pre-scaler for each trigger.
Different trigger configurations are recorded in a pattern
unit on an event-by-event basis for off-line selection.

Analog signals from the beam counters are first discriminated with constant fraction
and leading edge discriminators
before entering a second discriminator, whose role is
to shape the logic signals (12~ns width) and convert them to
ECL levels, as required by the FPGA trigger logic.
These logic signals are also recorded in pattern units on
an event-by-event basis for verifying the trigger logic in the analysis
of trigger data.
The combined use of two discriminators with different output
widths prevents also the pile-up in the trigger logic
(the length of the output of the first discriminator is around
100~ns, while the length of the second discriminator is 12~ns).
Correspondingly, the dead time of the trigger system is around 100~ns,
small compared to the dead time gated trigger rate of $\approx$100~Hz.

The simultaneous use of the beam ($T_{BEAM}$) and
interaction triggers ($T_{INT}$) allows for the direct determination
of the interaction probability, $P_{INT}$:
\[
P_{INT} =
\frac{N(T_{BEAM} \wedge T_{INT})}{N(T_{BEAM})} \; ,
\]
where  $N(T_{BEAM})$ is the number
of events which satisfy the beam trigger condition and
$N(T_{BEAM} \wedge T_{INT})$ is the number of events which satisfy
both the beam trigger and interaction trigger conditions.

\section{\large TPC tracking system}
\label{tpc}

The main tracking devices of the NA61/SHINE experiment are
four large volume Time Projection Chambers (TPC). Two of them ({\it Vertex}
TPCs: VTPC-1 and VTPC-2) are located in
the magnetic field, two others ({\it Main} TPCs: MTPC-L and MTPC-R) are
positioned downstream of the magnets symmetrically to the beamline.
In addition a smaller TPC (GAP-TPC)  is mounted between the two
VTPCs. It is centered on the beamline for measuring particles with the smallest
production angles.
Also the Low Momentum Particle Detector presented in Sec.~\ref{target:lmpd}
consist of two small TPC chambers.
The TPCs allow reconstruction of over 1000 tracks in a single Pb+Pb
interaction.
Up to 234 clusters and samples of energy loss per particle trajectory provide high
statistics for precise measurements.

The TPCs consist of a large gas volume in which
particles leave a trail of ionization electrons.
A uniform vertical electric field is established
by a surrounding field cage made of aluminized Mylar strips
that are kept at the appropriate electric potential by a voltage divider chain.
The electrons drift with constant velocity under the influence of
the field towards the top plate where their position,
arrival time, and total number are measured with proportional wire chambers.
In order
to achieve high spatial resolution the chamber top plates are subdivided
into pads of about one square centimeter area, a total of
\mbox{about 180~000}
for all TPC's.
From the recorded arrival times of the track signals and the known pixel
positions one gets a sequence of 3-dimensional measured points along the
particle trajectories.

An overview of the main parameters of the different TPCs is given in
Table~\ref{table:TPC_parameters}.

\begin{table}
\footnotesize
\begin{center}
\vglue0.2cm
\begin{tabular}{|l||c|c|c|c|}
\hline
				& VTPC-1 & VTPC-2 & MTPC-L/R & GAP-TPC\\
\hline
\hline
size (L$\times$W$\times$H) [cm] & 250 $\times$ 200 $\times$ 98 & 250 $\times$ 200 $\times$ 98 & 390 $\times$
390 $\times$ 180 & 30 $\times$ 81.5 $\times$ 70 \\
\hline
No. of pads/TPC & 26 886 & 27 648 & 63 360 & 672 \\
\hline
Pad size [mm] & 3.5 $\times$ 28(16) & 3.5 $\times$ 28 & 3.6 $\times$ 40, 5.5 $\times$ 40 & 4 $\times$ 28 \\
\hline
Drift length [cm]& 66.60 & 66.60 & 111.74 & 58.97 \\
\hline
Drift velocity [cm/$\mu$s] & 1.4 & 1.4 & 2.3 & 1.3 \\
\hline
Drift field [V/cm] & 195 & 195 & 170 & 173 \\
\hline
Drift voltage [kV] & 13 & 13 & 19 & 10.2 \\
\hline
gas mixture & Ar/CO$_2$ (90/10) & Ar/CO$_2$ (90/10) & Ar/CO$_2$ (95/5) &
Ar/CO$_2$ (90/10) \\
\hline
\# of sectors & 2 $\times$ 3 &2 $\times$ 3 & 5 $\times$ 5 & 1 \\
\hline
\# of padrows & 72 & 72 & 90 & 7 \\
\hline
\# of pads/padrow & 192 & 192 & 192, 128 & 96 \\
\hline
\end{tabular}
\caption{ Parameters of the VTPCs and MTPCs. The pad length in
the VTPC-1 equals
16 mm only in the two upstream sectors.
In the MTPCs the 5 sectors closest to
the beam have narrower pads and
correspondingly more pads per padrow.}
\label{table:TPC_parameters}
\end{center}
\end{table}

\subsection{VTPC, MTPC and GAP-TPC}
\label{tpc:det}

Each Main TPC has a readout surface at the top of
$3.9\!\times\!3.9\,\text{m}^2$
and a height of the field cage of about 1.1 m. It is filled with
a gas mixture of Ar/CO$_{2}$ in the proportion 95/5.
The track signals are read out by 25 proportional chambers providing
up to 90 measured points and ionization samples on each particle trajectory.
The accuracy of the measurement of the average ionization energy loss
for a particle is about 4\%.

Each Vertex TPC consists of a gas box with $2.0\!\times\!2.5\,\text{m}^2$
top surface area and 0.67~m depth.
The inserted field-cage structures exclude the region of 0.12~m
on either side of the beamline in which the particle density in Pb+Pb reactions
is so high that trajectories cannot be resolved.
A gas mixture of Ar/CO$_{2}$ in the proportion 90/10 is employed.
The readout is performed by 6 proportional
chambers on the top which provide up to 72 measurements
and ionization samples on the particle trajectories.
More details about the Vertex and Main TPCs can be found
in Ref.~\cite{NA49:spectrometer}.

\begin{figure}[ht]
\includegraphics[scale=.9]{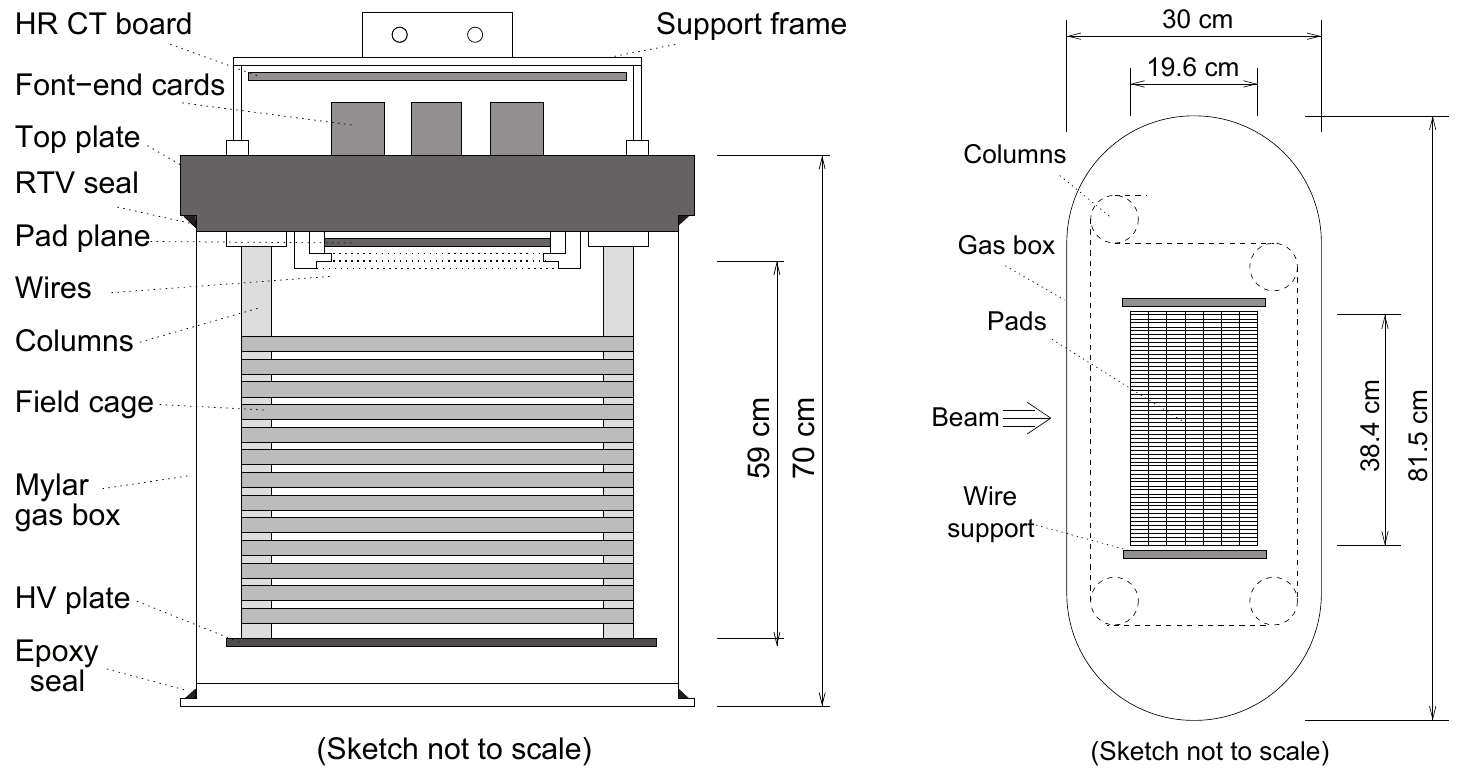}
\caption{Schematic layout of the GAP TPC: front view\lef
and top view\rig.}
\label{fig:GAPTPC_1} 
\end{figure}

Between VTPC-1 and VTPC-2 an additional tracking device,
the GAP-TPC~\cite{Varga:GAPTPC}, is located directly on the beamline.
It covers the gap left for the beam between
the sensitive volumes of the VTPCs and MTPCs. High
momentum tracks can be better extrapolated back to the primary vertex
using the additional points measured in the GAP-TPC. Particles originating from the primary vertex
but measured only in the MTPCs can be better distinguished from conversion
electrons faking high momentum tracks outside the magnetic field.
Since the beam passes through this detector its material budget
was minimized to 0.15\% of a radiation length and 0.05\% of an interaction
length.
The design follows that of the other TPCs described above (the schematic
layout is shown in Fig.~\ref{fig:GAPTPC_1}).
The electric field in the drift volume is generated
 by a field cage made from aluminized Mylar strips connected by a resistor
 chain.
 The electric field is 10.2~kV over 58.97~cm (173~V/cm) with a resulting drift time
 of about $50~\mu$s. The support of the Mylar strips is provided by tubes of
 glass-epoxy with a wall thickness of $100~\mu$m. The drift volume is enclosed
 by a gas box made of a single layer of $125~\mu$m Mylar. The readout
 plane consists
 of 7 padrows with 96 pads each, the pad dimensions are 28~mm by 4~mm. The
 gas composition
 is Ar/CO$_{2}$ (90/10) like in the VTPCs. For the readout three standard TPC
 front end cards are used
 together with one concentrator board.

\subsection{He beam pipes}
\label{tpc:he}

He filled beam pipes were installed in April 2011 in the gas volume of
the
VTPCs in order to reduce the number of $\delta$-electrons by a factor
of about 10. This
is needed in order to decrease significantly event-by-event fluctuations of
the track density
in the TPCs and thus reduce systematic uncertainties of fluctuation
measurements in
nucleus-nucleus collisions. Installation of the pipes necessitated cutting
openings in the
double wall Mylar envelopes of VTPC-1 and VTPC-2. The openings were drilled
using
specially developed tools which prevented dust and debris of material from
getting into the
inner volume of the TPCs. These openings into the gas volume were precisely
positioned
around the beam axis and were used for gluing the lightweight interface units
($200~\mu$m
wall thickness carbon fiber rings) providing support for the He beam pipes
and ensuring
hermetic VTPC volumes.
The lightest possible He beam pipe structure was produced that is feasible
with present
technology. The close-to-normal gas pressure conditions allow to use thin
($30~\mu$m) gas-leakage-tight Tedlar polyvinyl fluoride (PVF) film~\cite{Tedlar} to produce
2.5~m length cylinder
envelopes (pipes of 75 and 96~mm diameters), which form the He gas and
protective gas
volumes. The pipes were made by gluing the Tedlar film to special rigid,
light-weight carbon fiber endcaps strengthened with Airex foam~\cite{Airex}. This
provides low-mass
beam pipe fixation, separation of gases and hermetic sealing. Use of
light-weight low-$Z$
materials minimizes the probability of secondary interaction
processes. The used materials~\cite{Tedlar,Toray,Airex} include:
\begin{enumerate}[(i)]
\setlength{\itemsep}{1pt}
\item Tedlar polyvinyl fluoride (PVF) film
\item Carbon Fiber M55J, $\rho = 1.92$~g/cm$^{3}$, $X_{0} = 25$~cm
\item Toray and
\item Airex foam, $\rho$~=~0.03~g/cm$^{3}$, $X_{0}$~=~1390~cm.
\end{enumerate}

\begin{figure}[ht]
\begin{center}
\includegraphics[width=0.76\linewidth]{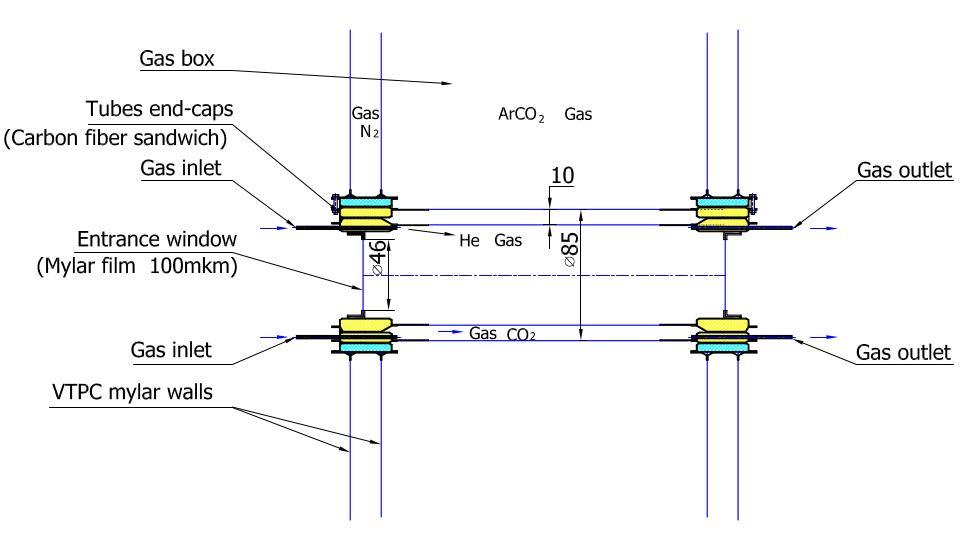}
\includegraphics[width=0.54\linewidth]{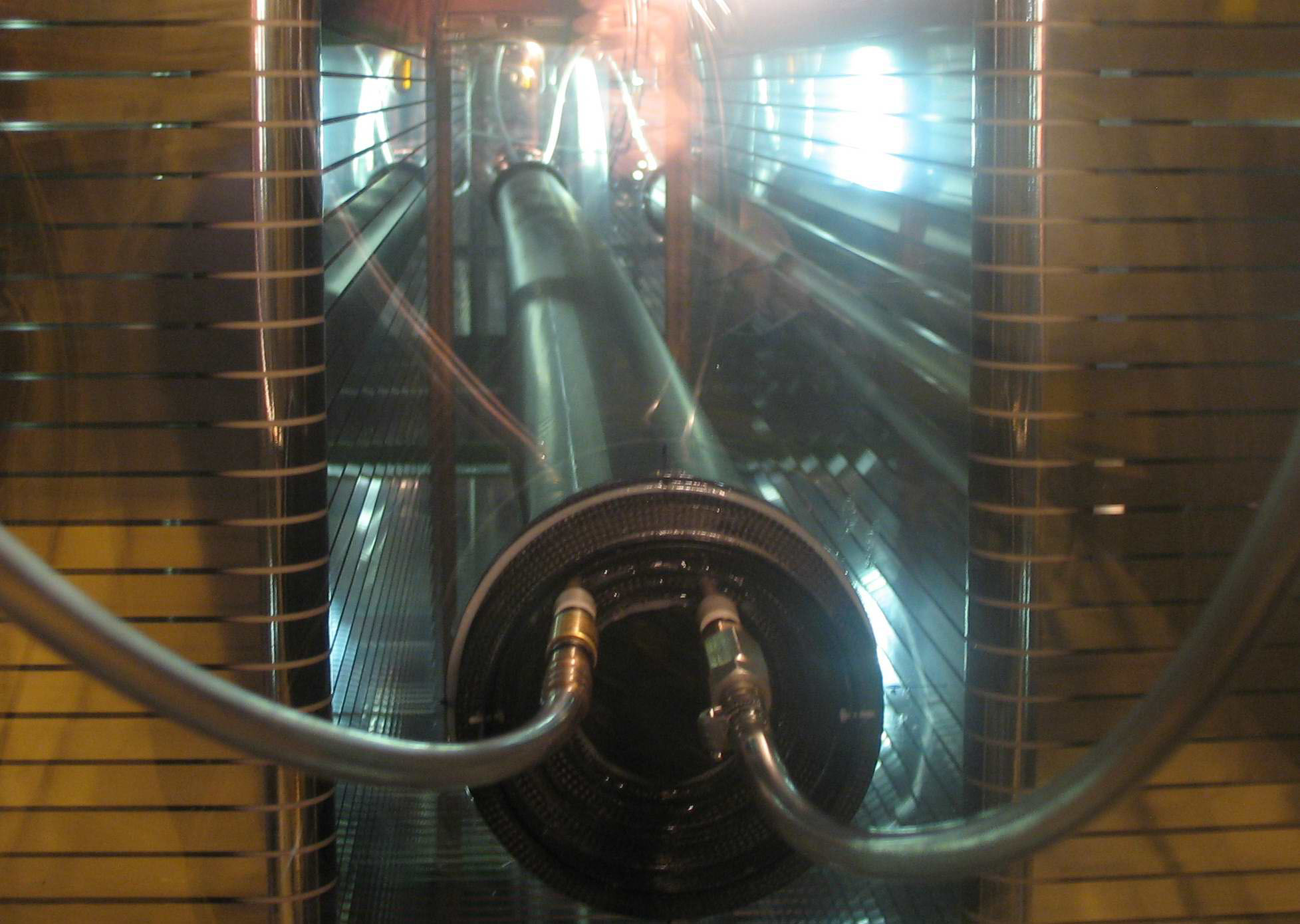}
\end{center}
\caption{
{\it Top:}
Schematic layout of the He beam pipe installed in the VTPC-1 gas
volume between the two
field cages.
{\it Bottom}:
The He beam pipe photo.}
\label{fig:he_b_p1}
\end{figure}

\begin{figure}[ht]
\centering
\includegraphics[width=0.6\linewidth]{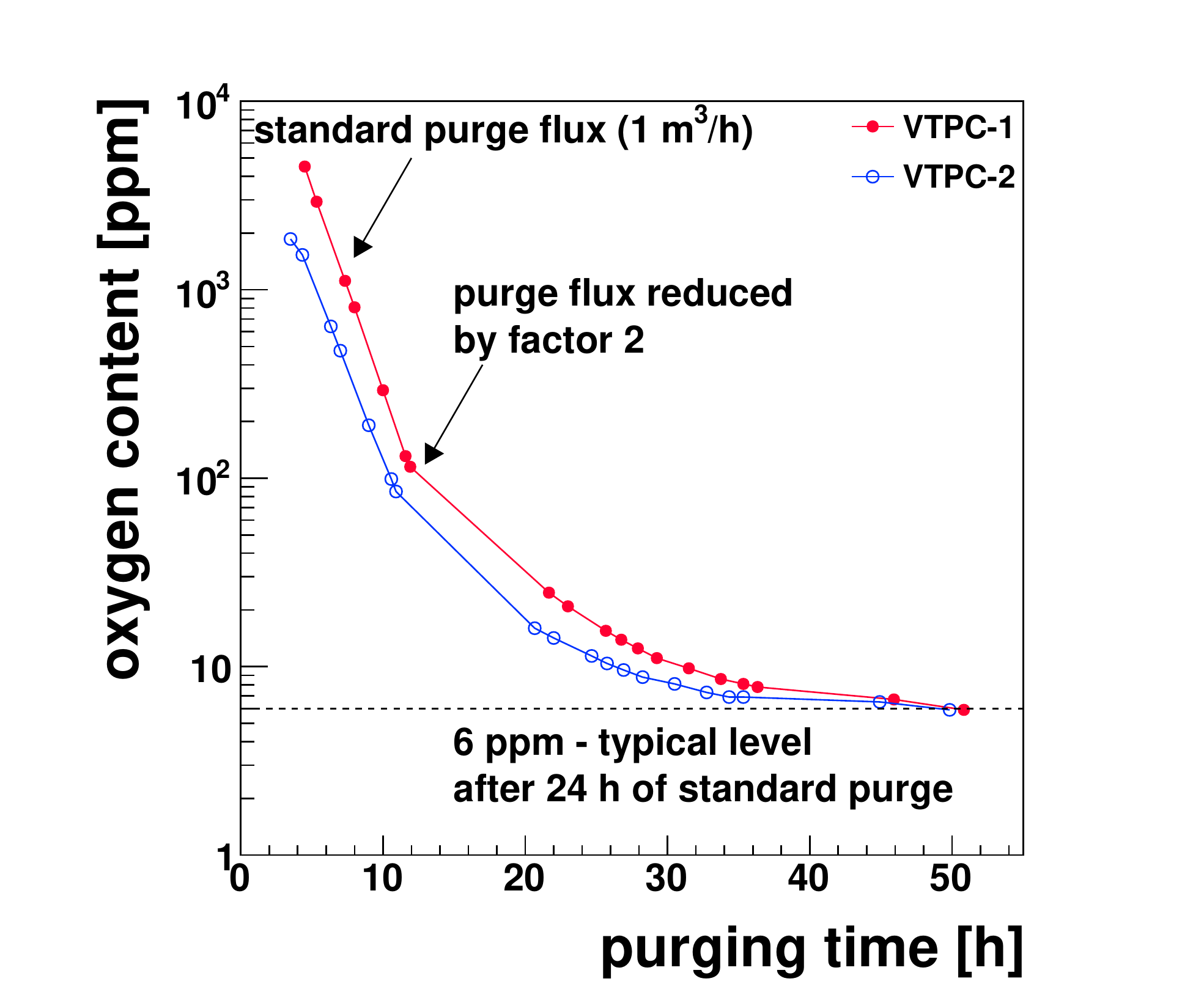}
\caption{
Oxygen level in the VTPC volumes during the gas purging period.}
\label{fig:oxygen_level}
\end{figure}

The construction is gas tight allowing  separation of the working
gas of the VTPCs from the helium used in the central pipe. In addition,
during the operation an inert gas (CO$_{2}$)
is flushed through the
outer envelope of the pipe (the protective gas volume). The He
beam pipe installed in the VTPC-1, with gas supply lines connected, is
shown in Fig.~\ref{fig:he_b_p1}. A special gas supply system was
constructed
to maintain the overpressure gradients in the two gas volumes of the pipes
with respect
to the pressure in the VTPCs. This is mandatory to ensure the
mechanical stability of the pipes. The He pipes were tested under working
conditions of the VTPCs. They showed good mechanical stability when
operating
different modes of gas circulation in the VTPCs. No leakage of helium from
the central pipes
to the outer envelope was observed. The surface of the pipes did not show
any excessive
charging-up which could distort the drift field in the active VTPC volume. Gas
tightness of the fixations of the pipes to the Mylar foils closing the VTPC
field cage
was tested by monitoring of the oxygen content in the VTPC gas. The oxygen
contamination
of the working gas decreased to a few ppms during the first 48 hours of purge
(see Fig.~\ref{fig:oxygen_level}) and remained constant
at this level. The measured level of oxygen impurity after gas
stabilization in the VTPCs was the same as that observed during the
operation before the
installation of the He beam pipes.

\begin{figure}[ht]
\centering
\includegraphics[width=0.9\textwidth]{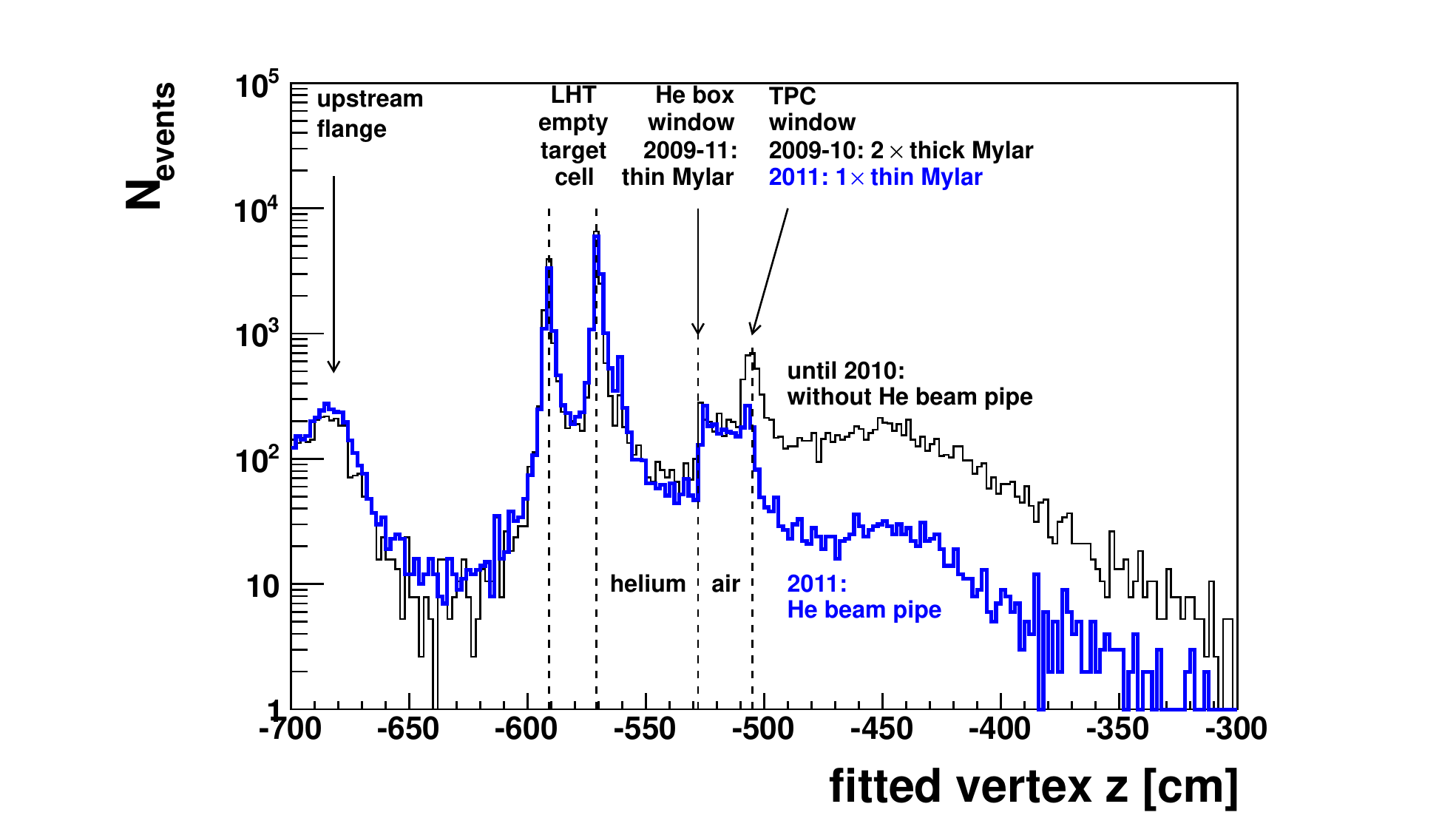}
\caption{Interaction vertex distribution along the beam axis (data from
p+p interactions
at 158~\GeVc) before the He beam pipe installation (2009 and 2010 data)
and with the He
beam pipes installed and filled with helium gas (2011 data).}
\label{fig:he_b_p2} 
\end{figure}

Figure~\ref{fig:he_b_p2} shows the interaction vertex distribution along
the beam axis (data for p+p
interactions at 158~\GeVc) before the He beam pipe installation (2009 and
2010 data) and
with the He beam pipes installed and filled with helium gas (2011
data). Clearly, the He
beam pipes reduce the number of background interactions in the volume of
the VTPCs by
about a factor of 10.

\subsection{Magnets}
\label{tpc:mag}

The two identical super-conducting dipole magnets with a maximum total
bending power
of 9~Tm at currents of 5000~A
have a width of 5700~mm and a length of 3600~mm. Their centers are
approximately 2000~mm
and 5800~mm downstream of the target. The shape of the magnet yokes is such
that the opening in the bending plane is maximized at the
downstream end. Inside the magnets a gap of 1000~mm between the upper and lower
coils leaves room for the VTPCs.
The coils have an iron-free central bore of 2~m diameter. This causes large
field inhomogeneities
where the minor components reach up to 60\% of the central
field at the extremities of the active TPC volumes. The magnetic field
inside the sensitive volumes of the Vertex TPCs was precisely measured
by Hall probes.

The standard
configuration for data taking at beam momentum per nucleon of
150~\GeVc and higher is nominally 1.5~T, in
the first and 1.1~T in the second magnet. At lower beam momenta the fields are
reduced proportional to the beam momentum keeping the ratio of the two fields constant.
More details about the magnets can found in Ref.~\cite{NA49:spectrometer}.

\subsection{Gas system and monitoring}
\label{tpc:gas}

\begin{figure}
\begin{center}
\includegraphics[width=0.99\textwidth]{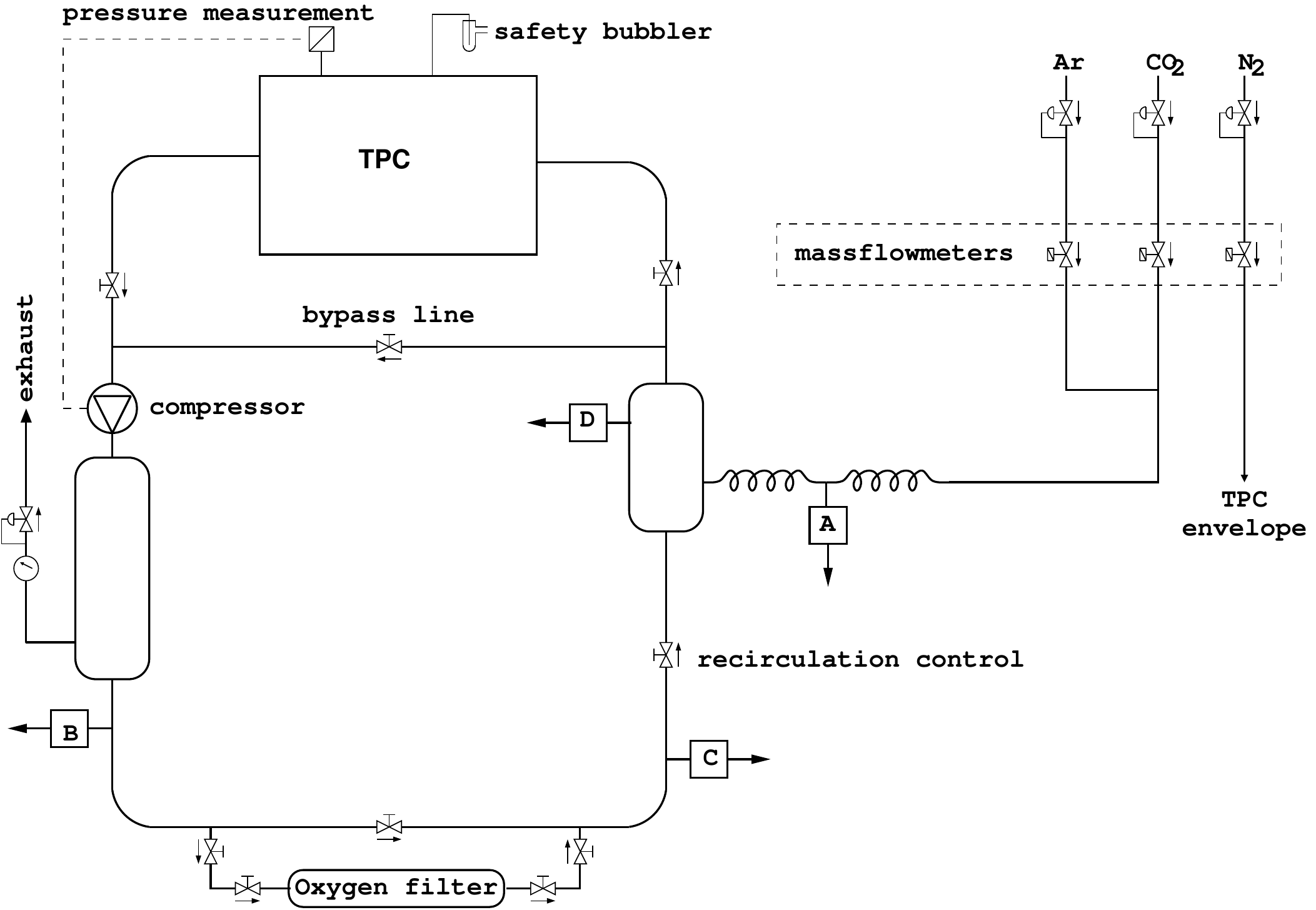}
\end{center}
\caption{
Schematic
layout of one of the four TPC gas recirculation systems. The gas flow direction
is marked by arrows next to the valves. The letters A, B, C and D mark
points where small amounts of the gas can be directed to the measuring devices.
}
\label{fig:gas1}
\end{figure}

The gas in the VTPCs and MTPCs is supplied by four independent gas systems.
A schematic drawing of one gas system is presented in Fig.~\ref{fig:gas1}.

Each system recirculates the gas with a compressor at a rate of about $20\%$
of the detector volume per hour, i.e.\ 0.9 and 3~m$^{3}$/h for VTPCs and MTPCs,
respectively.
Fresh gas is mixed through mass flow controllers from pure Ar and CO$_2$.
In normal operation, the fresh gas is supplied at only $3\%$ detector volume
per hour; in purge mode, the full recirculation rate is used.
The recirculation flow is controlled by regulating the TPC overpressure to
$0.50 \pm 0.01$~mbar via frequency modulation of the compressors.

Oxygen is cleaned from the detector gas by filter columns containing active
Cu-granules chosen for use with the CO$_{2}$ gas mixtures.
The filters are regenerated after typically 4-6 months operation periods,
using Ar/H$_{2}$ (93/7) mixture at $200^\circ$C.
Fresh filters absorb water contamination in the gas for about 2 weeks,
such that the water content may vary by 10 to 20~ppm.

Two bypass lines allow to start recirculation with the TPC isolated, and to
replace the oxygen filter without stopping the system. In case of compressor
failure a safety bubbler protects the TPC from overpressure.

The VTPC and MTPC  walls are made of two layers of 125~$\mu$m Mylar.
Nitrogen is flushed
between the layers, to prevent air from contaminating the TPC gas by diffusion
through the walls, or a potential leak.

The monitoring of gas quality is one of the major tasks of the NA61/SHINE TPC gas system.
Small amounts of gas can be directed to measuring units.
Non-linearities and calibration drift of the flow controllers are followed
by online measurements of drift velocity and gas amplification, as the required
setting accuracy is beyond the specifications of the flow regulators.
It is possible to measure the fresh gas mixture (point A in
Fig.~\ref{fig:gas1}), and the gas from the TPC (point B).
Each of the gas systems has a system of a drift velocity monitor and four
gas amplitude monitors.
Drift velocity is measured in a drift detector using the drift time difference
from a pair of $^{241}$Am $\alpha$ sources at 10~cm distance.
The gas amplification is checked in the amplitude monitors.
The amplitude monitors measure the signal from $^{55}$Fe photons in a
proportional tube. The mixing and monitoring equipment is temperature stabilized
to better than $0.1^\circ$C.

Oxygen and water contamination in the TPC gas (point B), filtered gas
(point C), and the input gas mixture (point D) can be
measured with two pairs of O$_{2}$/H$_{2}$O sensors.
Gas purities of 2--5~ppm oxygen and about 20--100~ppm water are typically
achieved.

\begin{figure}
\begin{center}
\includegraphics[width=0.98\linewidth]{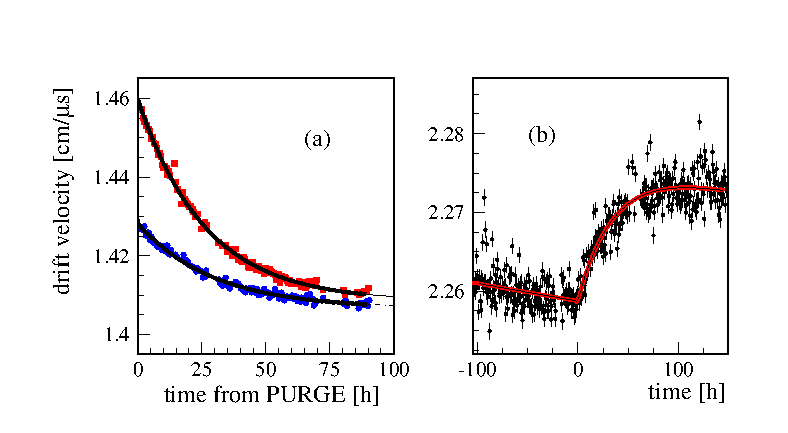}
\vspace{-0.2 cm}
\end{center}
\caption{
$Left:$ Drift velocity after purge measured for VTPC-1 (circles) and VTPC-2
(squares). The curves show the exponential dependence fitted to the data.
$Right:$ Change of the drift velocity in the MTPCs after the reduction of CO$_2$
contents by 0.1\%
at time t=0. Note that the vertical scale is expanded by a factor 2 with
respect to the left plot.}
\label{FigureDrift}
\end{figure}

The gas circulation is typically started 1--2 weeks before the beginning of
detector operation.
First, the detector is flushed with fresh gas for 2 days in the purge mode.
Due to limited precision of the input flow meters the gas mixtures needs several
days to stabilize after purge.
The drift velocity approaches a stable value exponentially with a
time constant of approximately 28 hours for the VTPCs and 35 hours
for the MTPCs (see Fig.~\ref{FigureDrift}).

The drift velocity decreases by about 1\% after purge which is attributed
to out-gassing of the detector material.
This is compensated by a slight increase of the argon content.
Also, when the water contamination increases, the drift velocity decreases; the
effect is of order of 1\% for the typical water content of several tens of ppm.
The change is slow (about 0.0002~cm/\mus per day) and is taken into account
in the drift velocity calibration based on the online drift velocity measurements
and more precise off-line drift velocity determinations.

The GAP-TPC with a volume of only about 150~l uses a much simpler gas system.
About 20~l/h of fresh gas mixture is flushed through this detector.
The gas coming out from the GAP-TPC is also passed through a drift velocity monitor.

The gas composition is Ar/CO$_2$ 90/10 in the VTPCs and GAP-TPC, and 95/5
in the MTPCs.
The higher argon content in the MTPCs is required to obtain higher drift velocity
necessary to read out the longer drift length (the safety limit on the drift
high voltage is about 20~kV).
Typical drift velocities are 1.4~\cm/\mus in the VTPCs, 2.3~\cm/\mus in the
MTPCs and 1.3~\cm/\mus in the GAP-TPC.

\subsection{TPC Front End Electronics}
\label{tpc:fee}

The electron trace of the track ionization in the TPC gas drifts in the
electric field to the amplification planes of the TPCs, which are operated in
the proportional amplification range. After passing through
the gating grid and the cathode grid, the drifted electrons get
amplified by gas electron multiplication on the sense wires of the field and
sense wire plane by about $\approx5\cdot10^{4}$. The disappearance of this
amplified electron signal on the sense wires capacitively induces an opposite-sign
signal on the two-dimensionally segmented pad plane just behind the
field and sense wire plane. Readout of the charge signal of the pads
in consecutive short time intervals provides 3 dimensional information on the particle
trajectories traversing the the TPCs. The electronic readout of the
pads is performed by the TPC Front End Electronics (FEE).

One FEE channel is dedicated to each readout pad, pre-amplifies the signal
and stores the analog charge of a given time sample in a capacitor array.
256 time slices with 200~ns time bins are used.
It is also possible to use
512 time slices with 100~ns
time bins. The time sampling is driven by a global clock for the full
TPC system in order to eliminate relative phase shifts. 32 channels
are handled by each FEE card, thus the full TPC system comprises about
6000 TPC FEE cards. After the analog charges are stored in the capacitor arrays
their digitization is performed via a Wilkinson ADC on the card, and the
digitized charge values are forwarded to the readout electronics.

The TPC FEE system~\cite{Kleinfelder:1990,Bieser:1997} was inherited from
NA49 along with the TPC system.

\subsection{Physics performance}
\label{tpc:per}

Several examples of the physics performance of the TPC system
are presented in this section, which were obtained from calibrated data.
Details on the calibration procedure can be found in Ref.~\cite{na61-calib}.

\begin{figure}[htbp]
\centering
\includegraphics[width=\linewidth]{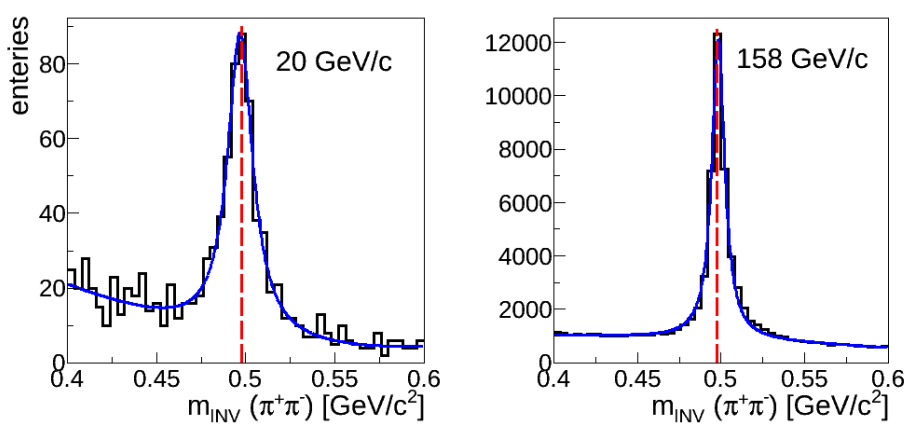}
\caption{
Invariant mass distribution of reconstructed $K^0_S$ candidates
in p+p interactions at 20 (\emph{left}) and 158~\GeVc (\emph{right}).
The curves show the result of a fit with the sum of a Lorentzian function
for the signal and a second order polynomial for the background.
The fitted peak positions are 496.8$\pm$0.6 and 498.3$\pm$0.1 MeV/$c^{2}$, respectively. The FWHM values
are 17$\pm$2 and 9.3$\pm$0.1 MeV/$c^{2}$, respectively.
The red dashed vertical line marks the PDG value of 497.6 MeV/$c^{2}$.
The magnetic fields were set for a total bending power of 1.1 and 9~Tm
at 20 and 158~\GeVc, respectively.
}
\label{fig:V0_study}
\end{figure}

The quality of measurements was studied by reconstructing
masses of $K^0_S$
particles from their $V^0$ decay topology.
As an example the invariant mass distributions of $K^0_S$
candidates found in p+p interactions at 20 and 158~\GeVc are
plotted in Fig.~\ref{fig:V0_study}.
The measured peak positions 496.8$\pm$0.6 and 498.3$\pm$0.1 MeV/$c^{2}$
are in reasonable agreement with the PDG value $m_{K^0_S} = 497.6$~MeV/$c^{2}$.

\begin{figure}[htbp]
\centering
\includegraphics[width=\linewidth]{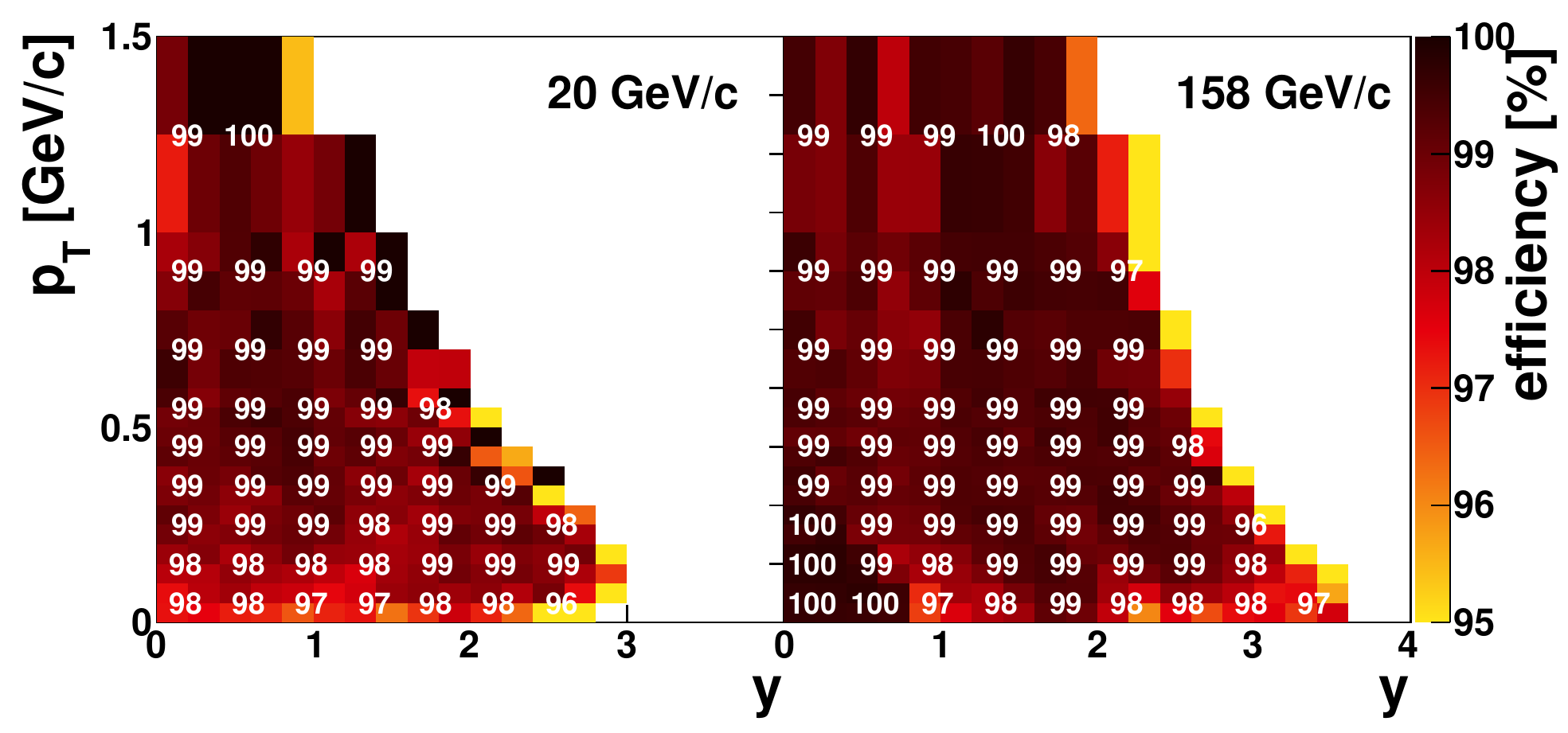}
\caption{
The reconstruction efficiency of negatively charged pions produced
in p+p interactions at 20 (\emph{left}) and 158~\GeVc (\emph{right}) as a
function of pion
rapidity and transverse momentum.
The magnetic fields were set for a total bending power of 1.1 and 9~Tm
at 20 and 158~\GeVc, respectively.
}
\label{fig:track_eff}
\end{figure}

\begin{figure}[htbp]
\centering
\includegraphics[width=0.99\linewidth]{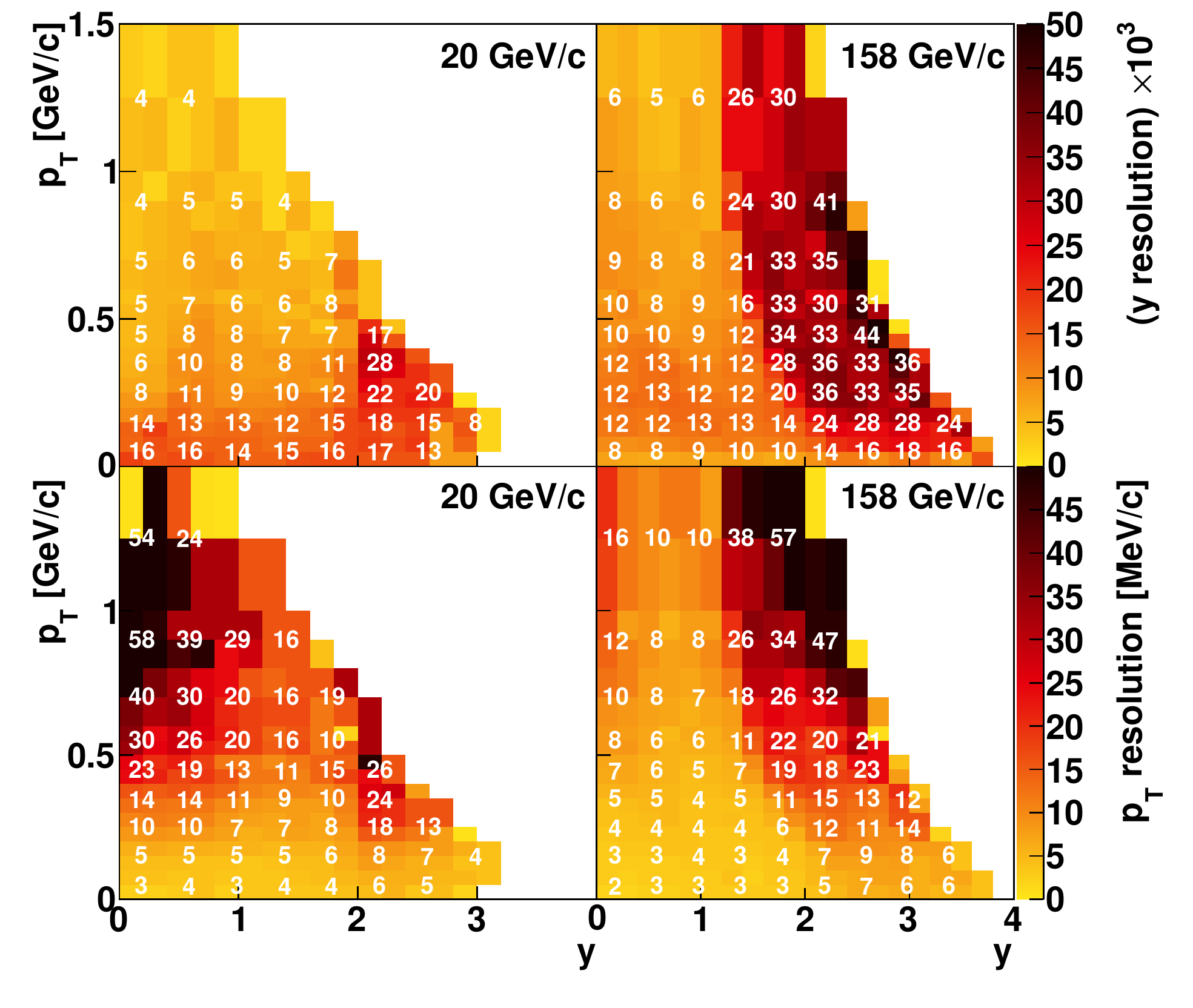}
\caption{
Resolution of rapidity (\emph{top}, scaled by $10^3$) and transverse
momentum (\emph{bottom}) measurements for negatively charged pions produced in
p+p interactions at 20
(\emph{left}) and 158~\GeVc (\emph{right}) as a function of pion rapidity
and transverse momentum.
The magnetic fields were set for a total bending power of 1.1 and 9~Tm
at 20 and 158~\GeVc, respectively.
}
\label{fig:resolution}
\end{figure}

The track reconstruction efficiency and resolution of kinematic quantities
were studied using a simulation of the detector. Estimates were obtained
by matching of generated tracks to their reconstructed partners.
As examples, the reconstruction efficiency as a function of rapidity $y$ and
transverse momentum $p_T$ for
negatively charged pions produced in p+p interactions at 20 and 158~\GeVc is
shown in Fig.~\ref{fig:track_eff}.
The measuring resolution of pion rapidity $y$ and transverse momentum $p_T$
is illustrated
in Fig.~\ref{fig:resolution}. The resolution was calculated as
the FWHM
of the distribution of the difference between the
generated and reconstructed $y$ and $p_T$.
Results presented in Figs.~\ref{fig:track_eff} and~\ref{fig:resolution} were
obtained for negatively charged pions passing the event and track selection
criteria used in the data analysis~\cite{na61-pp}, including
rejection of the azimuthal angle regions where the reconstruction efficiency
drops below 90\%.

\begin{figure}[htbp]
\centering
\includegraphics[width=0.9\linewidth]{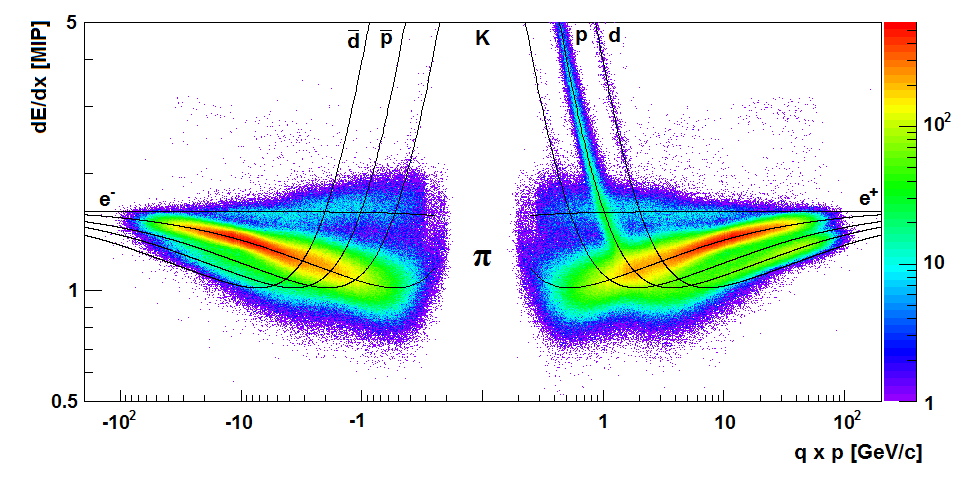}
\caption{
Specific energy loss in the TPCs for
positively (\emph{right}) and negatively (\emph{left}) charged particles as
a function of momentum measured for p+p interactions
at 80~\GeVc. Curves show parametrizations of
the mean $dE/dx$ calculated for different particle species.
}
\label{fig:dedx}
\end{figure}

The specific energy loss in the TPCs for
positively (\emph{right}) and negatively (\emph{left}) charged particles as
a function of momentum measured for p+p interactions
at 80~\GeVc is shown in Fig.~\ref{fig:dedx}. Curves show parametrizations of
the mean $dE/dx$ calculated for different particle species.

\section{\large Time of Flight systems}
\label{tof}

Since particle identification based only on energy loss measurement can not
be performed in the crossover region of the Bethe-Bloch curves,
NA61/SHINE
also uses additional and independent particle identification by Time of Flight (ToF)
detectors.
The ToF-L and ToF-R detectors were inherited from NA49.
In order to extend the identification of NA61/SHINE to
satisfy neutrino physics needs, a new forward detector (ToF-F) was
constructed. It is placed
between the ToF-L/R, just behind the MTPCs.

The particle's mass squared is obtained by combining the
information from the particle's time of flight, $tof$,
with the track length, $l$, and momentum, $p$, measured in the TPCs:
\begin{equation}
m^2=p^2
\left(\frac{c^2~tof^2}{l^2}-1 \right)~.
\label{eq:m2}
\end{equation}

\subsection{ToF-L, ToF-R}
\label{tof:lr}

\begin{figure}[htb]
  \centering
      \includegraphics[width=0.85\textwidth]{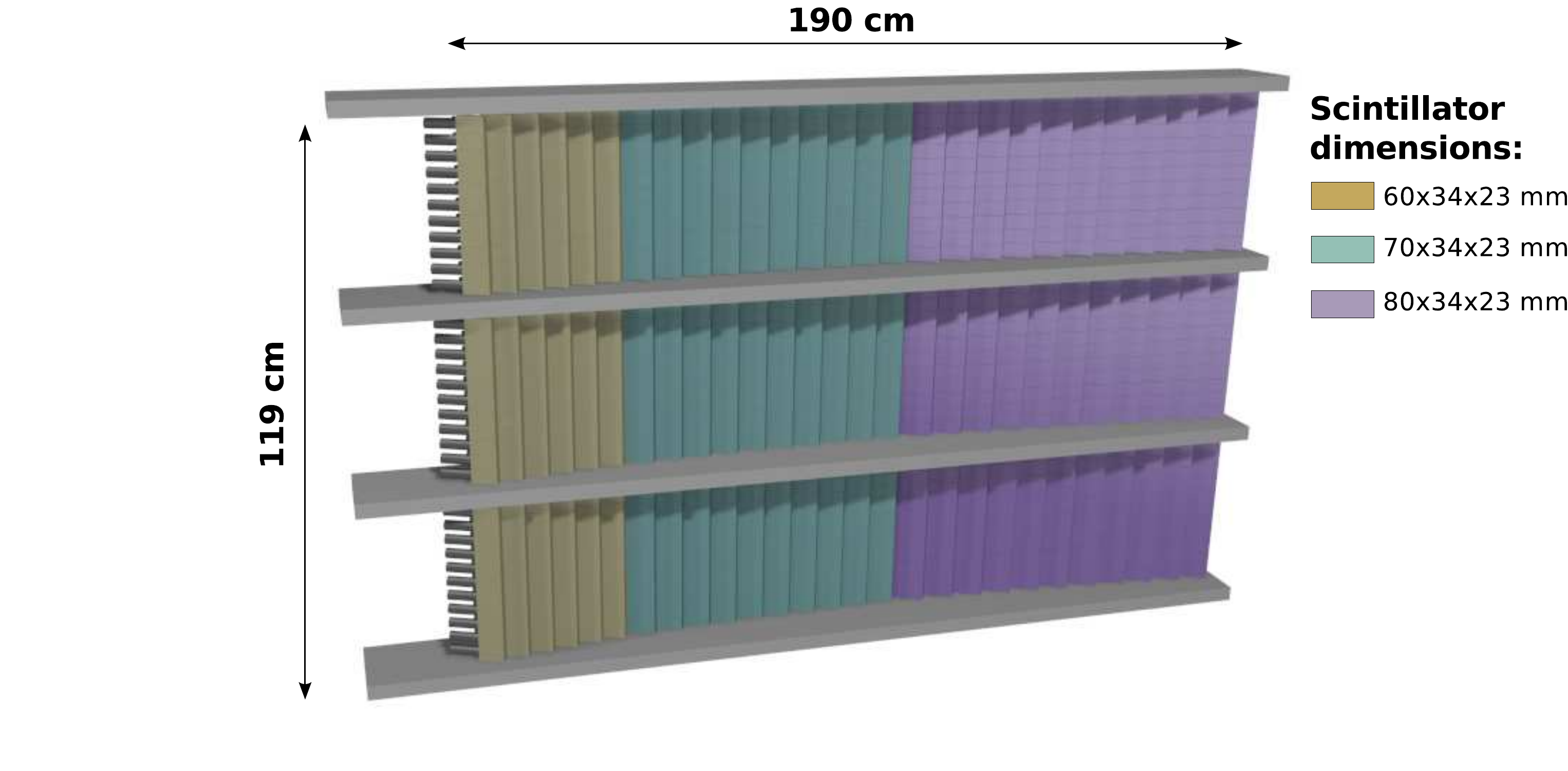}
\vspace{1.cm}
\caption{
Schematic layout of scintillators in the ToF-R detector.
}
\label{fig:toflr1}
\end{figure}

Two walls, ToF-L(eft) and ToF-R(ight), of 4.4~m${}^{2}$ total surface (see
Fig.~\ref{fig:toflr1}) are placed behind the MTPCs.
The track length of particles produced in the target is about 14 meters.
Each wall contains 891 individual
scintillation detectors
with rectangular dimensions, each having a single photomultiplier tube
glued to the short side. The scintillators
have a thickness of 23~mm matched to the photocathode diameter, a height
of 34~mm and horizontal
width of 60, 70 or 80~mm, with the shortest scintillators positioned closest
to the beamline and the longest on the far end.
The operating voltage for the PMTs in the ToF-L is
about 1600~V and in the ToF-R is around 1300~V. For the ToF-L, photomultiplier signals
enter constant-fraction discriminator
modules (CFD, KFKI custom made), housed in VME crates (WIENER) where
the analog signals are split before being passed through the actual discriminator
units. One output signal is directly sent
to FASTBUS analog-to-digital converters (96 channel ADC LeCroy 1885F)
while the other signal passes first through the discriminator unit and is then sent
to time-to-digital converters (64 channel FASTBUS TDC LeCroy
1775A). For ToF-R the PMT signals
are first split, then one line is sent to an ADC (96 channel ADC LeCroy 1882F)
while the other line goes
into a FASTBUS CFD (Struck DIS Str138) followed by a TDC (64 channel FASTBUS
TDC LeCroy 1772A). The start signal
for the TDCs is provided by the S11 PMT of the upstream S1 beam counter (scintillator)
while the stop signal comes from the
CFDs if their input is above a threshold.

The calibration procedure is based on extrapolation of the tracks reconstructed
in the MTPCs into the
area of the ToF detectors, identifying scintillators they extrapolate to and
checking for appropriate
signals in the corresponding TDC and ADC channels. Afterwards detailed corrections
are performed depending on charge
deposition and relative position of the incident particle in each scintillator.
Corrections for the position of the main interaction (in case of thick targets)
and position of the beam particle in the S1 counter are also done.

The overall time resolution of the ToF-L/R system is estimated
based on the distribution of the differences
between the measured time of flight for particles identified as pions and
that predicted from the
measured momentum and trajectory assuming the pion mass.
The distribution can be described by a Gaussian where, in p+p and Be+Be collisions,
a standard deviation of 95~ps was observed for ToF-L, while a standard deviation of
80~ps was observed for ToF-R. This value for the time resolution includes all contributions
to the $tof$ measurement (the intrinsic ToF detector resolution as well as
the start detector resolution,  uncertainties in tracking, etc.).

\begin{figure}[htbp]
\begin{center}
\begin{minipage}[b]{0.95\linewidth}
\includegraphics[scale=.30]{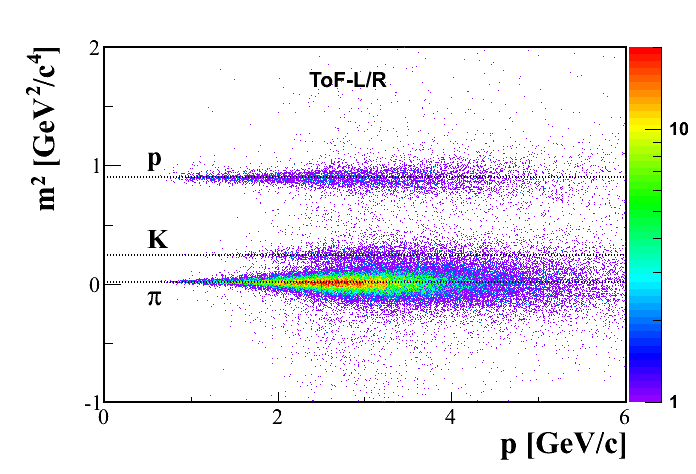}
\includegraphics[scale=.30]{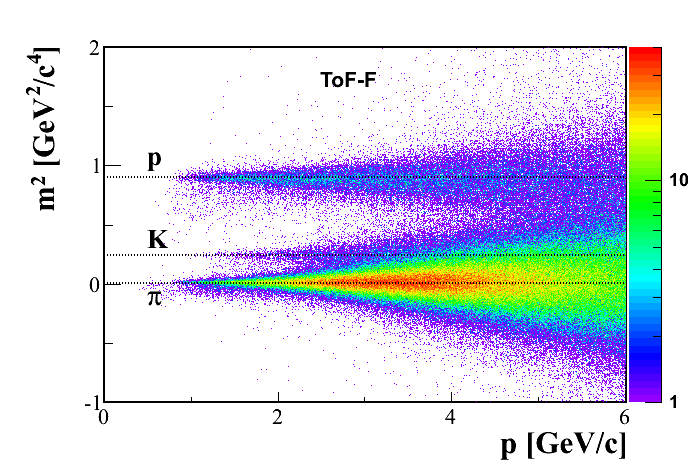}
\includegraphics[scale=.30]{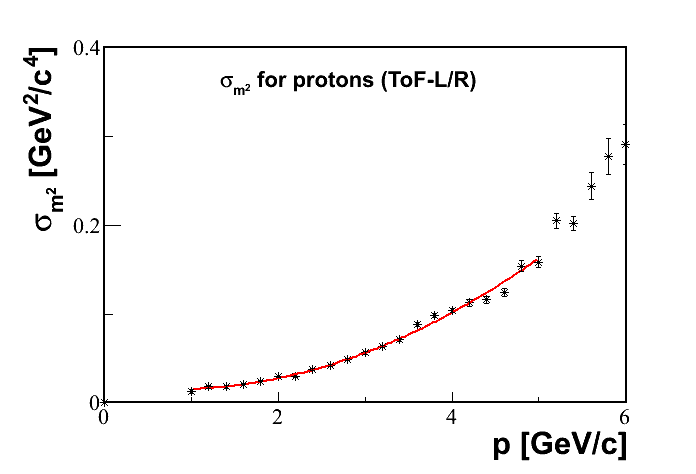}
\includegraphics[scale=.30]{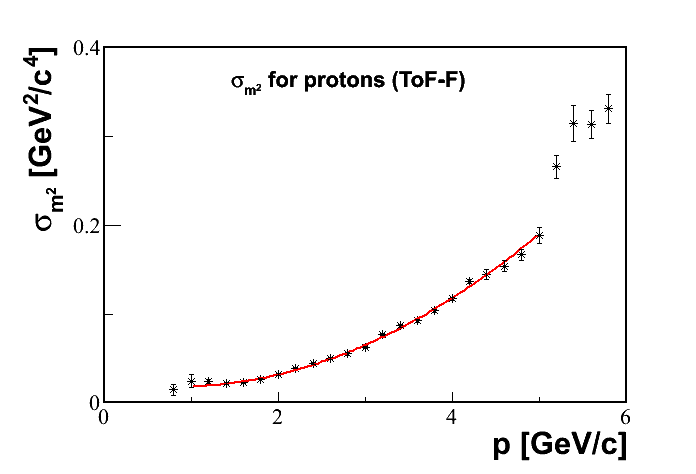}
\end{minipage}
\end{center}
\caption{
$Top$:
Mass squared versus momentum measured by ToF-L/R ($left$) and
ToF-F ($right$) detectors for particles produced in p+p
interactions at 80~\GeVc. The lines show the expected mass squared values
for different hadrons.
$Bottom$:
The resolution of the $m^2$ measurement as a function of
particle momentum for the ToF-L/R ($left$) and
the ToF-F ($right$).
}
\label{fig:toflr3}
\end{figure}

This resolution allows to separate pions and kaons  for momenta up
to 3~\GeVc (up to 5~\GeVc if $tof$
information is used along with $dE/dx$ information) and pions and protons
for even higher momenta.
Figure~\ref{fig:toflr3} ($top,~left$) shows mass squared as a function
of particle momentum for p+p interactions at  80~\GeVc.
The resulting resolution in $m^2$ is plotted as a function of
momentum in Fig.~\ref{fig:toflr3} ($bottom,~left$).

Due to aging of the electronics two upgrades are
foreseen. The HV supply will be
partially replaced and an upgrade of the readout electronics is
planned. This should,
modernize the whole system, ensure its long-term functioning, and also improve
the quality of the obtained $tof$ information, primarily the resolution.

\subsection{ToF-F}
\label{tof:f}

The NA61/SHINE data taking for the T2K neutrino oscillation experiment
requires particle identification in a phase space region which is not
covered by the  ToF-L/R. When operating at the
T2K proton beam at 31~\GeVc, a large fraction of the tracks are
produced by low-momentum particles which exit the spectrometer
between the ToF-L and ToF-R. An additional time of flight detector, the
forward ToF (ToF-F), was therefore constructed to provide full time of
flight coverage of the downstream end of the MTPCs (see
Fig.~\ref{fig:na61}).

\begin{figure}[htb]
  \centering
     \includegraphics[width=0.8\textwidth]{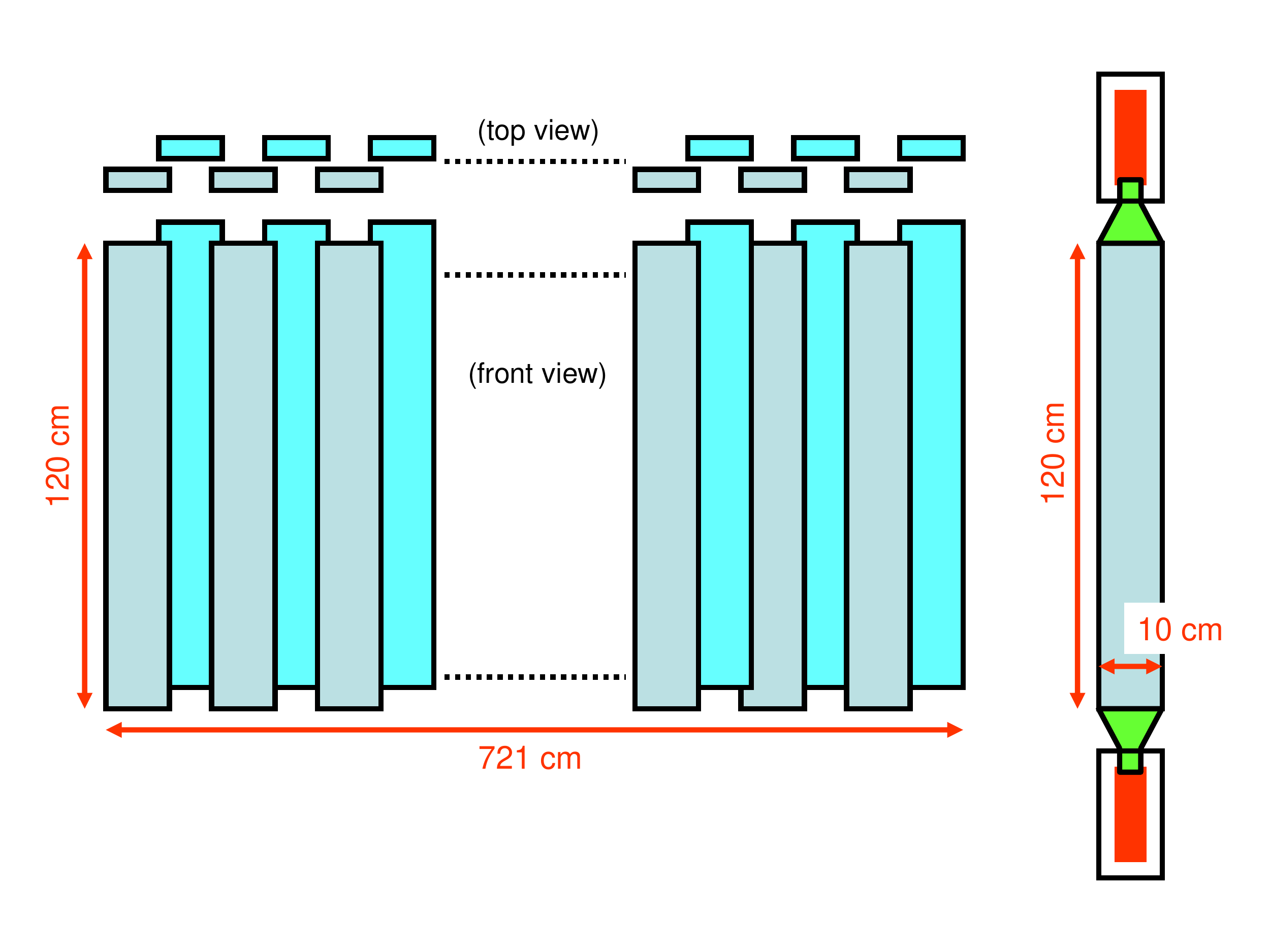}
\caption{
The schematic layout of scintillators in the ToF-F detector.
}
\label{fig:TOF-F}
\end{figure}

The ToF-F  consists of 80 scintillator bars oriented vertically
(see Fig.~\ref{fig:TOF-F}). The bars were tested and mounted in groups of 10
on independent frames (modules) out of which half were placed on the left
side and half on the right side of the detector.
The size of each scintillator is
120$\times$10$\times$2.5 cm$^3$. They are staggered with 1~cm overlap to
ensure full coverage in the ToF-F geometrical acceptance.
This configuration provides a total
active area of 720$\times$120 cm$^2$. Each scintillator bar is read
out on both sides with $2''$ photomultipliers  Fast-Hamamatsu
R1828, for a total of 160 readout channels. The scintillators are
plastic scintillator ({\em Bicron BC-408}) with a scintillation rise
time of 0.9~ns, a decay time of 2.1~ns and attenuation length of
210~cm. Their maximal emission wavelength is about
400~nm perfectly matching the PMT spectral response. Fish tail PMMA
(Poly methyl methacrylate) light-guides were glued on both ends for
the readout. The bars and light-guides were wrapped in aluminium
foils to ensure light reflection towards the light-guide and covered
with black plastic foils and tape. In order to ensure proper optical contact
between the PMTs and the light-guides a 3~mm thick silicone cylinder
matching the diameter of the PMT and light-guide is inserted at the
interface.

Most of the electronics for the ToF-F were inherited from the two NA49
Grid ToFs~\cite{Palla:2000tx} which were not used since their
acceptance coverage is marginal. Each PMT channel is
operated near 1700~V supplied by LeCroy1461 independent 12-channel high voltage
(HV) cards. The analog signals are transported from the ToF to the
counting house by 26~m RG58 50$\Omega$ coaxial cables. In order to obtain fast
logic signals and not be influenced by the variations in amplitude of
the PMT response the cables are plugged into Constant Fraction
Discriminators (16-channel KFKI CFD5.05 VME module). At the input these
include an internal passive divider of 1:3 to provide the signals for
the integrated charge and time measurements, respectively, and the
necessary delay lines at their output. In order to minimize the
crosstalk of the neighboring channels an appropriate order is chosen
between the PMT outputs and the CFD inputs. The output signals of the
CFDs of the PMT-channels serve as ``stop'' signals for the time of
flight measurements. The start signal is provided by the fast beam
counter S1 of the central trigger system.  The time measurement is
carried out by LeCroy FASTBUS Time-to-Digital Converter (TDC) units
digitizing the time in 12 bits dynamic range with a sampling time of
25~ps. The analog signals of the PMTs are converted by LeCroy
Analog-to-Digital Converter (ADC) units into 12 bits.  The time
measurement, $t$, has an offset, $t_0$, which is specific to each
channel as it depends on cable length, PMT gain or CFD response. $t_0$
was therefore carefully adjusted on a channel by channel basis by
first assuming that all produced particles are pions and shifting the
mean value of the $tof~=~t-t_\pi$ distribution accordingly. In a first
iteration, this method allows to discriminate pions from protons and
was then repeated by selecting only pions.

\begin{figure}[htb]
  \centering
     \includegraphics[width=0.6\textwidth]{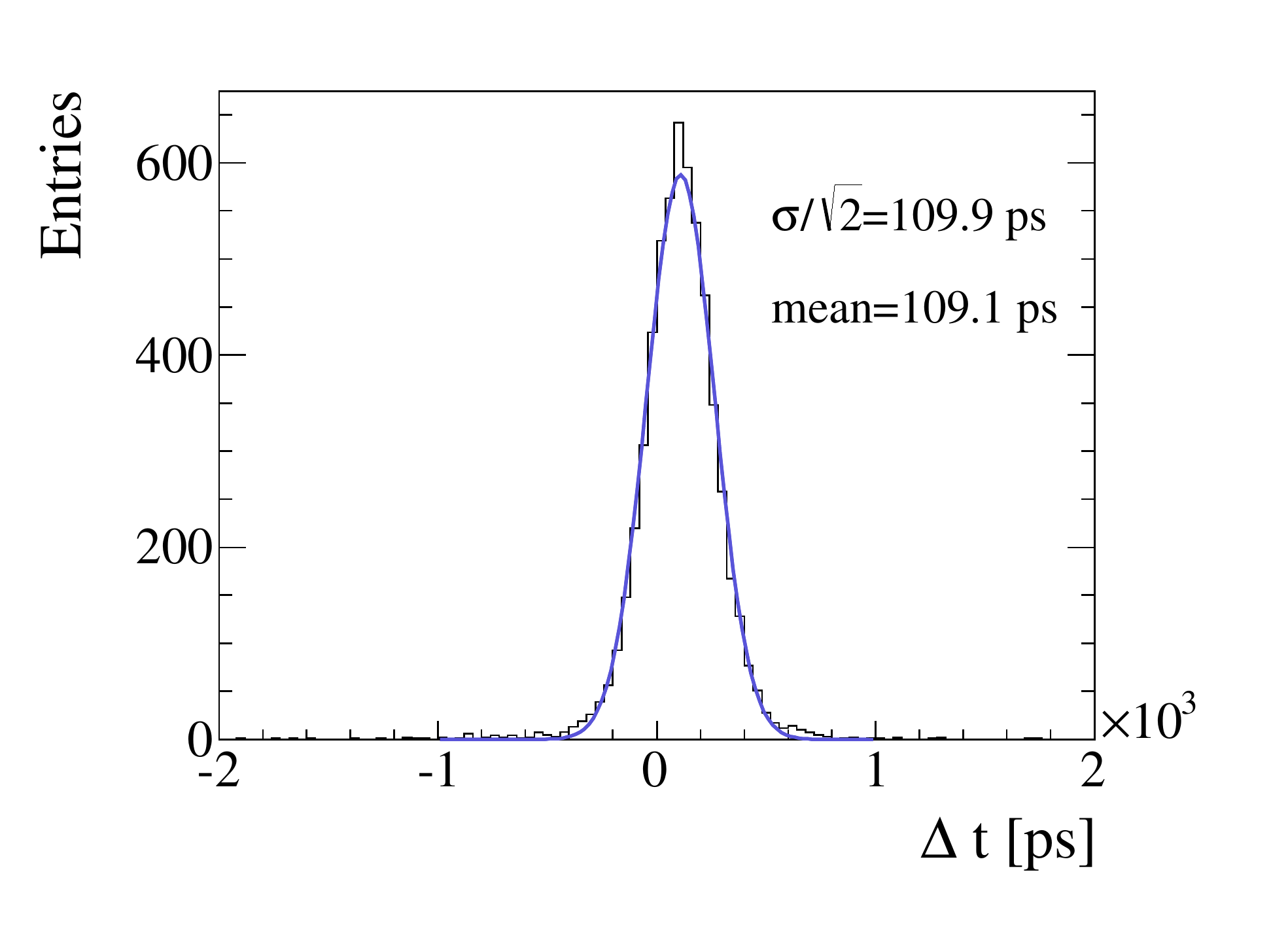}
\vspace{1.cm}
\caption{Distribution of the
	  difference between a particle's time of flight measured
	  independently by the overlapping scintillator bars of the
	  ToF-F detector. The width of the distribution is about
	  155~ps, indicating a $tof$ resolution of about 110~ps for a
	  single measurement.
	  }
\label{fig:ftof_res}
\end{figure}

The mass-squared distribution and the resolution of its measurement
as a function of momentum are presented in
Fig.~\ref{fig:toflr3} ($right$). The intrinsic resolution of the ToF-F was
also determined (see Fig.~\ref{fig:ftof_res}). This was achieved by selecting particles that
hit the region where the scintillators overlap and plotting the time
difference between the two signals. The Gaussian fit to the distribution gives a
resolution $\sigma_{tof}= \frac{155}{\sqrt{2}}\approx110$~ps.

\section{\large Projectile Spectator Detector}
\label{psd}

The development and construction of the forward hadron calorimeter was
one of the most important upgrades of the NA61/SHINE experimental setup. This
calorimeter is called the Projectile Spectator Detector (PSD). The purpose
of the calorimeter is the measurement of projectile spectator energy in
nucleus-nucleus collisions.
The PSD is used to select central (with a small number of projectile spectators)
collisions at the trigger level.
Moreover, the precise event-by-event
measurement of the energy carried by projectile spectators
enables the extraction of the number of interacting nucleons
from the projectile with the precision
of one nucleon. The high energy resolution of the PSD is important for the study
of fluctuations in nucleus-nucleus collisions which are expected to be sensitive to
properties of the phase transition between the quark-gluon plasma and
hadron-resonance matter.
Namely, the PSD  provides the precise control over
fluctuations caused by the variation of the number of interacting nucleons and
thus excludes the "trivial" fluctuations caused by variation of
the collision geometry. Basic
design requirements of the PSD are good energy resolution,${\sigma_{E} / E} <
{60\% / \sqrt{E(\rm{GeV})}}$,
and good transverse uniformity of this resolution. The PSD is a fully
compensating modular lead/scintillator hadron
calorimeter~\cite{Alekseev:2001,Fujii:2000} and meets these
requirements.

\subsection{Calorimeter design}
\label{psd:design}

\begin{figure}[htb]
  \centering
     \includegraphics[width=0.9\textwidth]{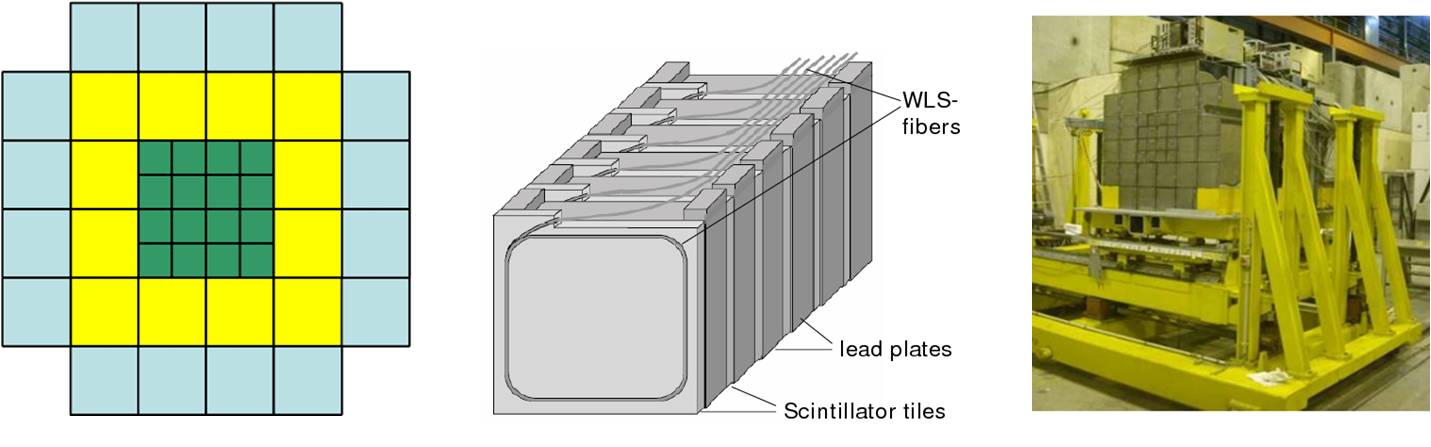}
\vspace{1.cm}
\caption{ The PSD: schematic front view\lef, schematic view of single module
($center$) and the fully assembled detector\rig.}
\label{fig:psd1}
\end{figure}

The PSD calorimeter consists of 44 modules which cover a transverse area
of $~$120x120~cm$^{2}$.
A schematic front view of the PSD is shown in Fig.~\ref{fig:psd1}\lef. The
central part of the PSD consists of 16 small modules with transverse
dimension of 10x10~cm$^{2}$ and weight of 120~kg each. Such fine transverse
segmentation decreases the spectator occupancy in one module and improves
the reconstruction of the reaction plane. The outer part of the PSD consists of
28 large 20x20~cm$^{2}$ modules with a weight of 500~kg each.

Each module, schematically shown in Fig.~\ref{fig:psd1} ($center$), consists of
60 pairs of alternating lead plates and scintillator tiles with 16~mm and 4~mm thickness, respectively. The
stack of plates is tied together with 0.5~mm thick steel tape and
placed in a box made of 0.5~mm thick steel. Steel tape and box are spot-welded
together providing appropriate mechanical rigidity. The full length of the
modules corresponds to 5.7 nuclear interaction lengths.

Light readout is provided by Kyraray Y11 WLS-fibers embedded in round
grooves in the scintillator plates. The WLS-fibers from each longitudinal section of 6 consecutive
scintillator tiles are collected together in a single optical connector at
the end of the module. Each of the 10 optical connectors at the downstream
face of the module is read out by a single photodiode. The longitudinal
segmentation into 10 sections ensures good uniformity of light collection
along the module and delivers information on the type of particle which caused
the observed particle shower. Ten photodetectors per module are placed at the
rear side of the module together with the front end electronics. A photograph
of the fully assembled calorimeter is shown in Fig.~\ref{fig:psd1}\rig.
In order to
fit the PSD transverse dimensions to the region populated by spectators the
distance between the NA61/SHINE target and the calorimeter is increased
from 17~m to 23~m with increasing collision energy.
Interactions of spectators upstream of the PSD were minimized
by the installation of a helium tube
of length 5.5~m and diameter 125~cm between the upstream
PSD face and the hut housing the MTPCs.
The entrance and exit Mylar windows of the tube had a thickness of about
125~$\mu$m. This tube was inserted
for data taking with ion beams of momenta larger than 40$A$~\GeVc
when the distance between the target and the PSD was 23~m.

\subsection{PSD photodetectors}
\label{psd:photo}

The longitudinal segmentation of the calorimeter modules requires 10 individual
photodetectors per module for the signal readout. Silicon photomultipliers
SiPMs or micro-pixel avalanche photodiodes, MAPDs~\cite{Sadygov:2006,site:zecotek}
are an optimum choice due to their remarkable properties
such as high internal gain, compactness, low cost and immunity to the
nuclear counter effect. Moreover, forward hadron calorimeter applications
have some specific requirements such as large dynamic range and linearity
of the photodetector response to intense light pulses. However, the dynamic range
and linearity of MAPDs are limited by the finite number of pixels. Most of the
existing types of MAPD with individual surface resistors have a pixel density
of $~10^3$~pixels/mm$^2$. Such a limited number leads to serious restrictions
of MAPD applications in calorimetry, where the number of detected photons is
comparable and even larger than the pixel number. The effect of saturation,
when a few photons hit the same pixel, leads to significant non-linear MAPD
response to light pulses with high intensity. Evidently, the MAPD has linear
response only if the number of pixels is much larger than the number of
incident photons. This feature represents a disadvantage of MAPDs compared
to the traditional PMTs. However, this drawback is essentially reduced for
MAPDs with individual micro-well structure~\cite{Sadygov:2006}, for which a pixel
density of 10$^4$/mm$^2$ and higher is achievable. The above considerations
motivated the choice of photodiodes of type MAPD-3A produced
by Zecotek Photonics Inc. (Singapore)~\cite{site:zecotek} for
the readout of the PSD hadron calorimeter. These MAPDs have a pixel density
of 15000/mm$^2$. Their 3x3~mm$^2$ active area fits well the size of the
WLS-fiber bunch from one longitudinal section of the PSD modules and provides
a total number of pixels of more than 10$^5$ in a single photodetector.

\begin{figure}[htb]
  \centering
     \includegraphics[width=1.\textwidth]{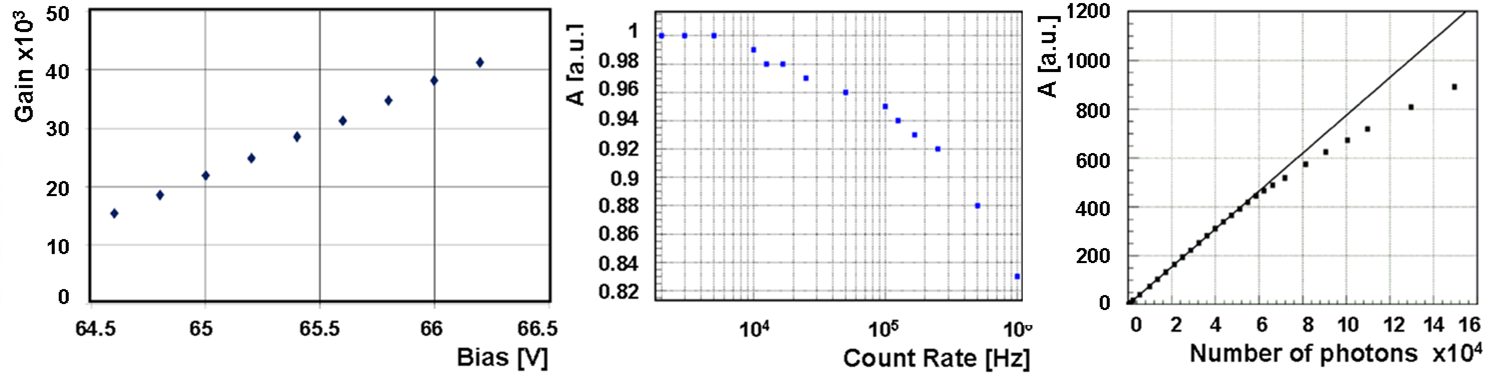}
\caption{Performance of MAPD-3A photodiodes:
Dependence of gain on the bias voltage\lef, count rate
capability ($center$) and dependence of amplitude (in arbitrary
units) on the number of incident photons\rig.
}
\label{fig:psd3}
\end{figure}

The MAPD-3A photon detection efficiency (PDE) for the Y11 WLS-fiber emission
spectrum reaches 15\% at 510~nm and is similar to the performance of PMTs. The
operation voltage for different samples of MAPD-3A detectors ranges from 65~V to 68~V.
The maximum achieved gain is 4$\times$10$^4$, see Fig.~\ref{fig:psd3}\lef.

Since the PSD calorimeter has no beam hole to ensure maximum acceptance for
the spectators, the central part of the PSD is exposed to  ion beams
with intensities up to 10$^5$~Hz.
The average amplitude in one longitudinal section of
a PSD module is expected to be about 1500 photoelectrons.
Therefore, an important requirement
for the calorimeter readout is a high count rate capability, at least in the
central region.
 In particular,
the recovery time of the selected MAPD-3A photodiode must be fast enough to
deliver stable amplitudes at signal frequencies up to 10$^5$~Hz. To check the
count rate capability of the MAPD-3A the dependence of its amplitude on the
frequency of light pulses was measured. The stability of the amplitude of
the pulses at different operation frequencies of the light emitting diode was checked
by a normal PMT. The obtained behavior of the amplitude produced by the MAPD-3A
is presented in Fig.~\ref{fig:psd3}~($center$). As seen, the MAPD-3A amplitude
would drop about 5\% for the maximum
beam intensity foreseen.

In order
to check the photodetector linearity, measurements of MAPD amplitudes were
performed with light pulses of different intensity. The number of incident
photons was determined by a reference photodiode with known quantum efficiency
and gain equal to one. The dependence of the MAPD amplitude on the number
of incident photons is shown in Fig.~\ref{fig:psd3}\rig. As seen,
the MAPD linearity is preserved for light pulses with number of photons up
to 6$\cdot$10$^4$. Taking into account the different photon detection efficiencies of
about 25\% for the tested MAPD (with the same pixel density as the MAPD-3A type)
and 15\% for the MAPD-3A, linear response is expected for
amplitudes up to 15000 photoelectrons.

The reported comprehensive studies confirm that the selected MAPD-3A
photodetectors satisfy the requirements of the NA61/SHINE experiment. At
present, 440 MAPD-3A detectors are installed in the 44 PSD modules and show
stable operation during data taking for calibration and physics with
beryllium and proton beams.

\subsection{Performance of the PSD calorimeter}
\label{psd:perf}

In order
to check the performance of the calorimeter, several tests were performed
with hadron beams of various momenta. During the first stage of the R\&D (in 2007)
a PSD module array of nine small modules (3$\times$3 array) was
assembled and tested in the H2 beamline using hadron beams
of 20-158~\GeVc momentum. The calibration of all readout channels was
done with a muon beam. The energy resolution of the tested array was
estimated from data taken with the hadron beams.

\begin{figure}[htb]
  \centering
     \includegraphics[width=0.9\textwidth]{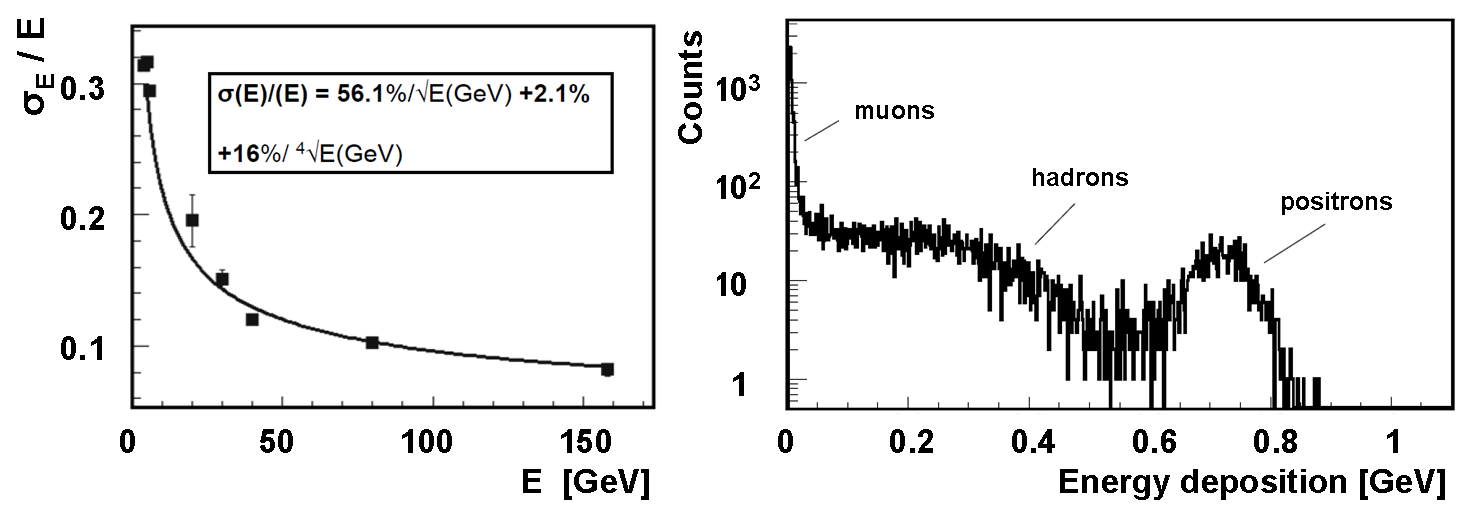}
\caption{$Left$: Energy resolution of the tested PSD prototype as a function of
hadron beam energy. The solid line is a fit to the  experimental data points
by the function shown in the inset. $Right$: Energy spectrum measured in the first
section of the central module for a 30~\GeV hadron beam containing a fraction
of muons and positrons.
}
\label{fig:psd6}
\end{figure}

The dependence of the measured energy resolution on the pion energy is shown in
Fig.~\ref{fig:psd6}\lef.

The tested prototype with 30$\times$30~cm$^{2}$ transverse size is too small
to contain the entire hadron shower. Therefore, a non-negligible lateral
shower leakage is expected. Monte Carlo simulations confirm that about 16\%
of the hadron shower energy escapes from the tested array. The influence of
shower leakage on the energy resolution was studied in
Refs.~\cite{Acosta:1991,Badier:1994},
where a third term in addition to the stochastic and constant
terms was added in the parameterization of the resolution. The fit of the
experimental data with the three-term formula,
assuming a fixed leakage term of 16\%, gives the coefficient of the
stochastic term equal to 56.1\% and of the constant term equal to 2.1\%.
The non-zero constant term might be an indication that
the selected lead/scintillator sampling does not provide full compensation.

The spectrum of deposited energy in the first section of the central module exposed
to the 30~\GeVc positively charged H2 beam is shown in Fig.~\ref{fig:psd6}\rig. The right-side
peak in the spectrum corresponds to full positron energy absorption in the
first longitudinal section which can be regarded as an electromagnetic
calorimeter with coarse sampling. The energy resolution for positrons at
30~\GeVc is about 6.5\%.

\begin{figure}[htb]
  \centering
  \includegraphics[width=0.9\textwidth]{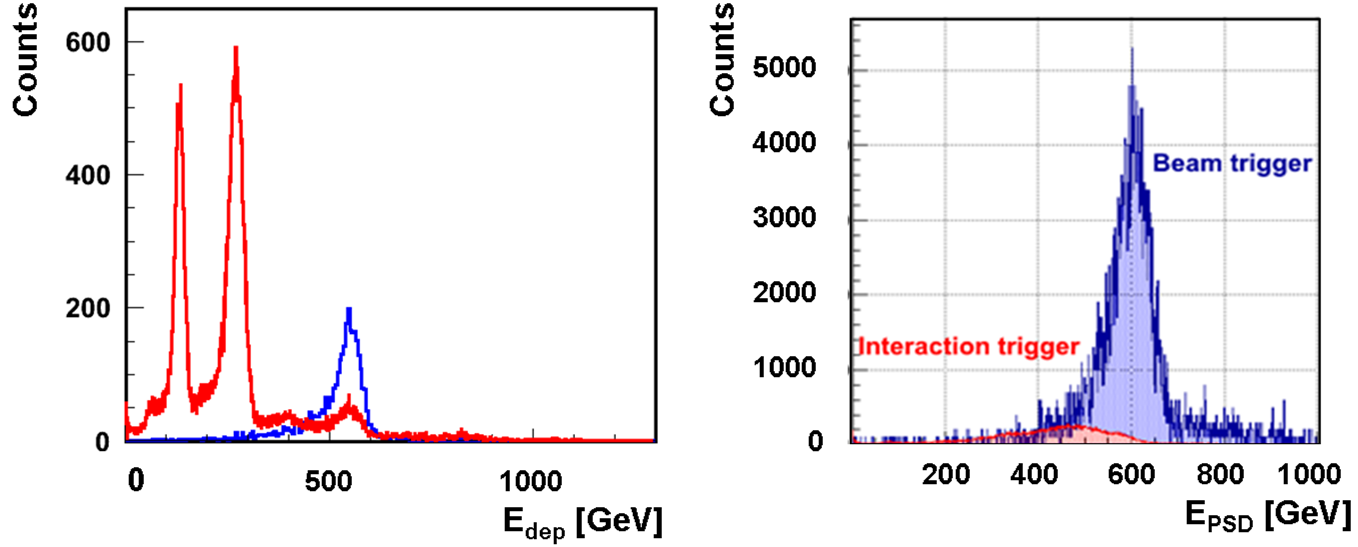}
\caption{$Left$: PSD energy distributions for the secondary ion beam at
75$A$~\GeVc with and without selection of the $^{7}$Be ions by the Z-detector.
$Right$: PSD energy distributions recorded during the data taking for $^{7}$Be+ $^{9}$Be
interactions at 75$A$~\GeVc. Spectra are shown for the beam trigger
(blue) and for the interaction trigger (red) events.
}
\label{fig:psd8}
\end{figure}

During the 2011-2013 data taking for $^7$Be + $^9$Be collisions
the PSD was also used in the
trigger for online rejection of the most peripheral events.
As a secondary $^7$Be beam was used other ions were also
present in the beam.
The distribution of the energy deposited in
the calorimeter by 75$A$~\GeVc beam ions is shown in
Fig.~\ref{fig:psd8}\lef without cut (red) and with cut (blue) on the $^7$Be
peak in the amplitude distribution of the  Z-detector. Clear identification of
$^7$Be
ions in the fragmented beam is seen here as well as contamination of deuteron
and helium ions.
Figure~\ref{fig:psd8}\rig shows the PSD energy spectra recorded during the
data taking for  $^{7}$Be+$^{9}$Be collisions at 75$A$~\GeVc. Spectra are
shown for the beam trigger (blue) and interaction trigger (red).

\begin{figure}[htb]
  \centering
  \includegraphics[width=0.75\textwidth]{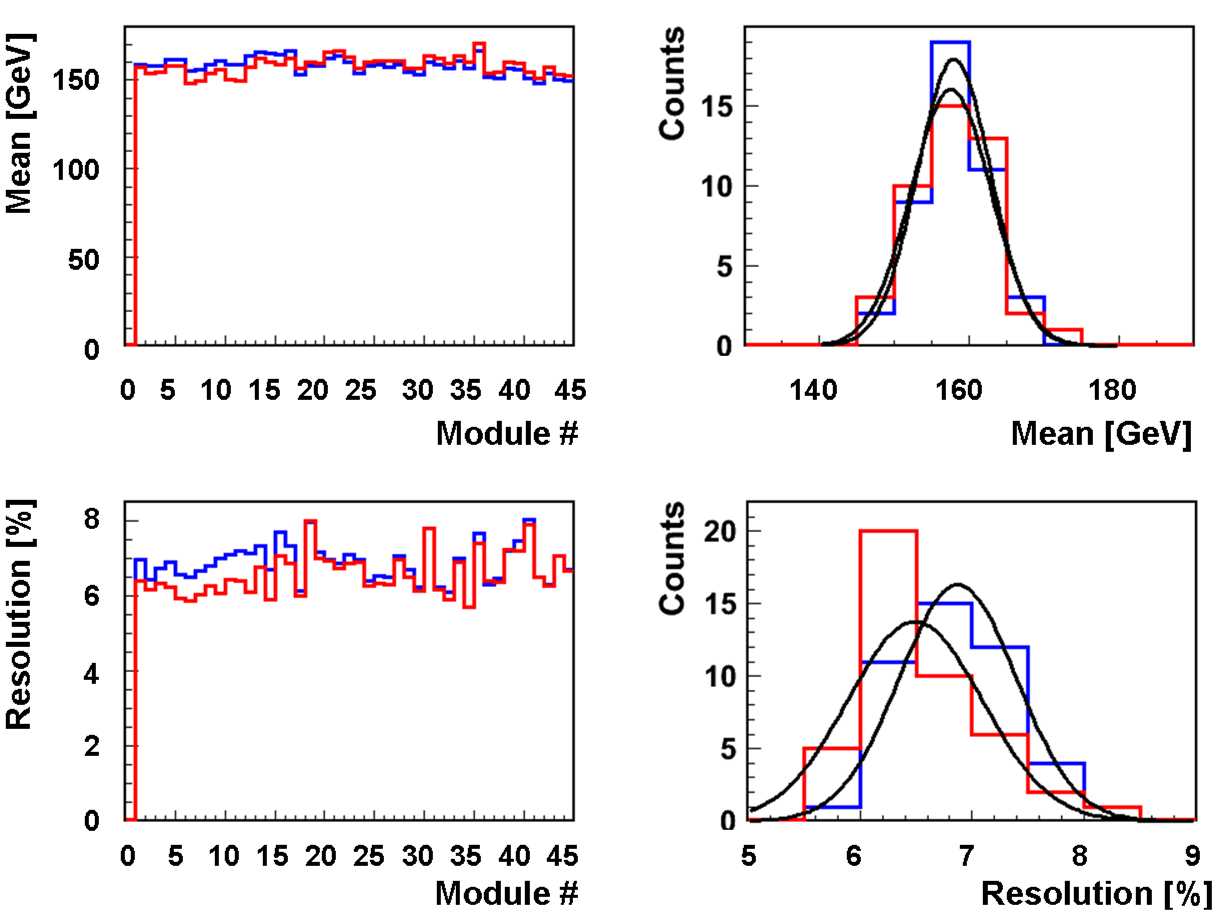}
\caption{ Mean values of reconstructed incident 158~\GeVc proton energy in
the PSD modules ($top~left$) and distributions of mean values for the 44 modules
($top~right$). Energy resolution for 158~\GeVc protons (bottom left)
and distributions of resolution for the 44 modules ($bottom~right$).}
\label{fig:psd12}
\end{figure}

Figure~\ref{fig:psd12} ($top$) shows the measured energy of
158~\GeVc protons as a function of the module number on which
the beam was centered.
Histograms and curves in Fig.~\ref{fig:psd12} ($top$) distinguish different energy
reconstruction methods by color. Red corresponds to the energy sum of all PSD
modules, while blue is for the sum of the corresponding clusters of modules
around the beam spot. As seen, the mean values of reconstructed energies
are  close to the real beam energy in both cases.

Figure~\ref{fig:psd12} ($bottom$) presents the energy resolution obtained for
each of the 44 modules from the scan with 158~\GeVc protons. The small
variations in the energy resolution can be explained by the precision of
the energy calibration. These results  are
consistent with the  prototype tests and with the MC simulations.

\section{\large Targets and LMPD}
\label{target}

The targets used by NA61/SHINE are positioned upstream
of the VTPC-1 and centered at $z \approx$ = -581~cm.
In data taking on p+Pb interactions the target was sourounded
by the LMPD detector.

\subsection{Targets}
\label{target:targets}

For data taking on p+p interactions
the liquid hydrogen target (LHT) of 20.29~cm length
(2.8\%~interaction length) and 3~cm diameter
was placed 88.4~cm upstream of the \mbox{VTPC-1}.
The target was filled with para-hydrogen obtained
in a closed-loop liquefaction system
which was operated at 75~mbar overpressure with respect to the atmosphere.
At the atmospheric pressure of 965~mbar the liquid hydrogen density
is $\rho_{LH} = 0.07$~g/cm$^3$.
The boiling rate in the liquid hydrogen was not monitored during the data taking
and thus the liquid hydrogen density is known only approximately.
Data taking with inserted and removed liquid hydrogen in the LHT
was alternated in order to calculate a data-based correction for
interactions with material surrounding the liquid hydrogen.
The density of gaseous hydrogen $\rho_{GH}$ present in the target after
removal of liquid hydrogen was estimated
from the ratio of high multiplicity events observed in a small
fiducial volume around the target center for data taken with inserted and
removed liquid hydrogen.
The density ratio $\rho_{GH}/\rho_{LH}$ varied within the range 0.4-0.6\%.
This indicates that the operational conditions of the LHT varied during the
data-taking period.

For data taking on $^7$Be+$^9$Be collisions in the period 2011-2013
two beryllium targets were used. Their density was $\rho$ = 1.85~g/cm$^3$
at 20$^o$C and dimensions were $2.5(W) \times 2.5(H) \times 1.2(L)$~cm$^3$
and $2.5(W) \times 2.5(H) \times 0.3(L)$~cm$^3$.
The targets consisted of more than 99.4\% $^9$Be.
The most abundant contaminant were oxygen nuclei (about~0.4\%).
The Be targets were placed in an aluminium container filled with
atmospheric pressure helium gas.

For data taking on p+Pb collisions a special thin Pb target disc was used
in order to reduce in-target absorption of slow protons which
mre easured by the LMPD and are used for
event centrality tagging. Two different target thicknesses, 0.5~mm and 1~mm,
were used to allow an experimental study of in-target absorption of slow protons.
The transverse shape of the target was
circular, with a diameter of 1~cm. The target material purity was
99.98\% lead with natural isotope composition
(52.4\% $\;^{208}$Pb, 22.1\% $\;^{207}$Pb, 24.1\% $\;^{206}$Pb, 1.4\% $\;^{204}$Pb).
The target density was 11.34~g/cm$^3$ at 20$^o$C with
a molar mass of 207.2~g/mol.
The target disc was placed in a Tedlar~\cite{Tedlar} foil container filled
with atmospheric pressure helium gas in order to reduce background of off-target
collisions in the vicinity of the target. The target was removable
from the beamline using a pneumatic piston for reference data taking
with the target removed needed
to estimate background due to off-target interactions.

For $\pi + C$ measurements at 158 and 350~\GeVc a  thin graphite
target of dimensions $2.5(W) \times 2.5(H) \times 2(L)$~cm$^3$ and
density $\rho$ = 1.84~g/cm$^3$, placed in an aluminium container filled with
helium gas, was employed by NA61/SHINE.
The same target was used  for hadro-production measurements
of interest for the T2K neutrino oscillation experiment in Japan. In these measurements
information about primary interactions of protons on carbon was extracted.
The thickness of this target along the beam axis was equivalent to about 4\% of a
nuclear interaction length $(\lambda_{I})$.

\begin{figure}[htb]
\begin{center}
\includegraphics[scale=.50]{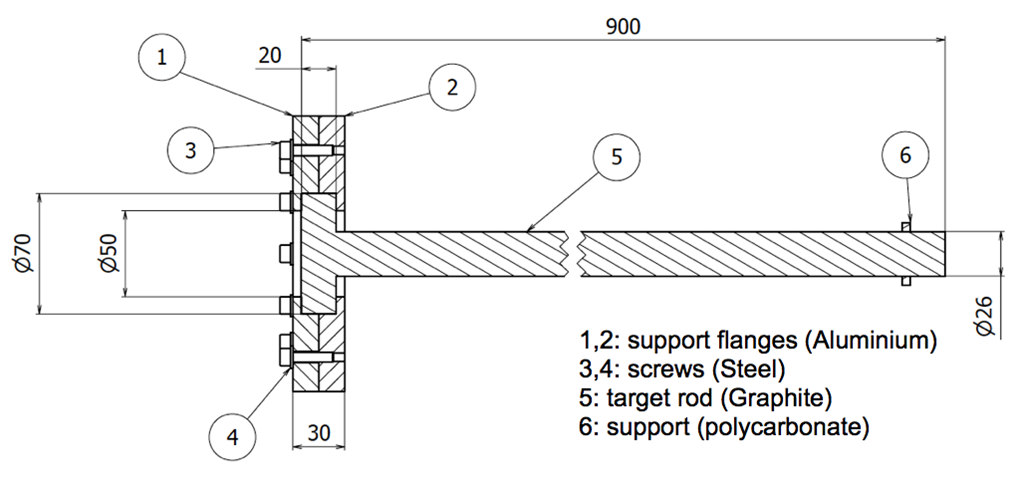}
\includegraphics[scale=.50]{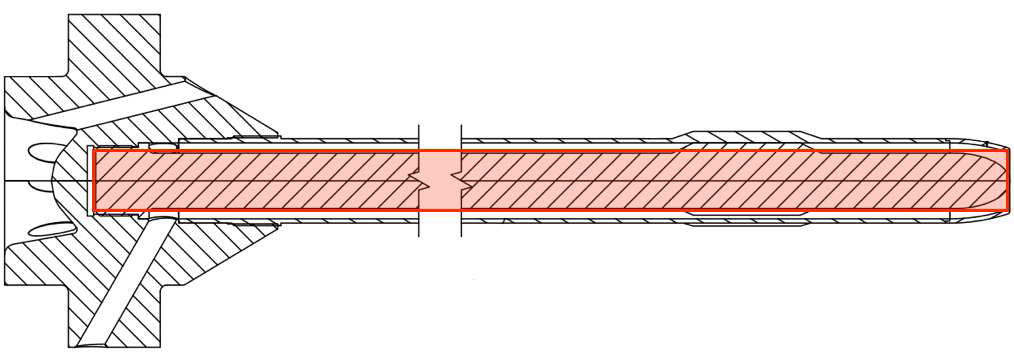}
\end{center}
\caption{
$Top:$ Technical drawing (side view) of
the replica target used during the NA61/SHINE data
taking consisting of a 90~cm long
graphite rod and aluminium support flanges.
$Bottom:$
Schematic layout of the complete geometry of the
T2K target. The overlaid red rectangle
represents the simplified geometry of
the replica target.
}
\label{fig:lt1}
\end{figure}

For hadro-production measurements also a thick graphite target, a replica of the T2K target, was used.
It consists of a 90~cm long rod with a radius of 1.3~cm and a density of
$\rho$~=~1.83~g/cm$^3$. The replica and the actual target of T2K are shown
in Fig.~\ref{fig:lt1}.
The upstream part of the graphite target is surrounded by aluminium flanges
which are inserted into a target holder.
Three screws in the target holder allow to align the target parallel
to the beam axis.
The target thickness along the beam axis is equivalent to about 1.9
interaction lengths.
The downstream face of the replica target was placed at around 50~cm from
\mbox{VTPC-1}.
Hadro-production measurements taken with the T2K replica target allow to
constrain the contributions from primary and secondary interactions
within the target (for details see ~\cite{T2Kreplica}).

\subsection{Low Momentum Particle Detector}
\label{target:lmpd}

In hadron-nucleus interactions the collision centrality can be deduced
from the number
of emitted low momentum protons (the so-called grey protons). For this
purpose a special
detector, the Low Momentum Particle Detector (LMPD), was constructed \cite{Marton:2014nima}. The
detector
consists of two small size TPC chambers on the two sides of the target with
a vertical drift
field, and a sequence of detection layers picking up the ionization signal
of radially emitted
particles. Between the detection layers plastic absorber layers are
inserted. Therefore the
range of particles in the detector material can be determined. The range
and energy loss by ionization
of a particle depends on its energy and type, and therefore
event-by-event
counting of the number of emitted low energy protons becomes
possible. Prototypes of the
detector were tested in 2009 and 2010.

\begin{figure}[ht]
\begin{center}
\begin{minipage}[b]{0.95\linewidth}
\includegraphics[width=0.44\linewidth]{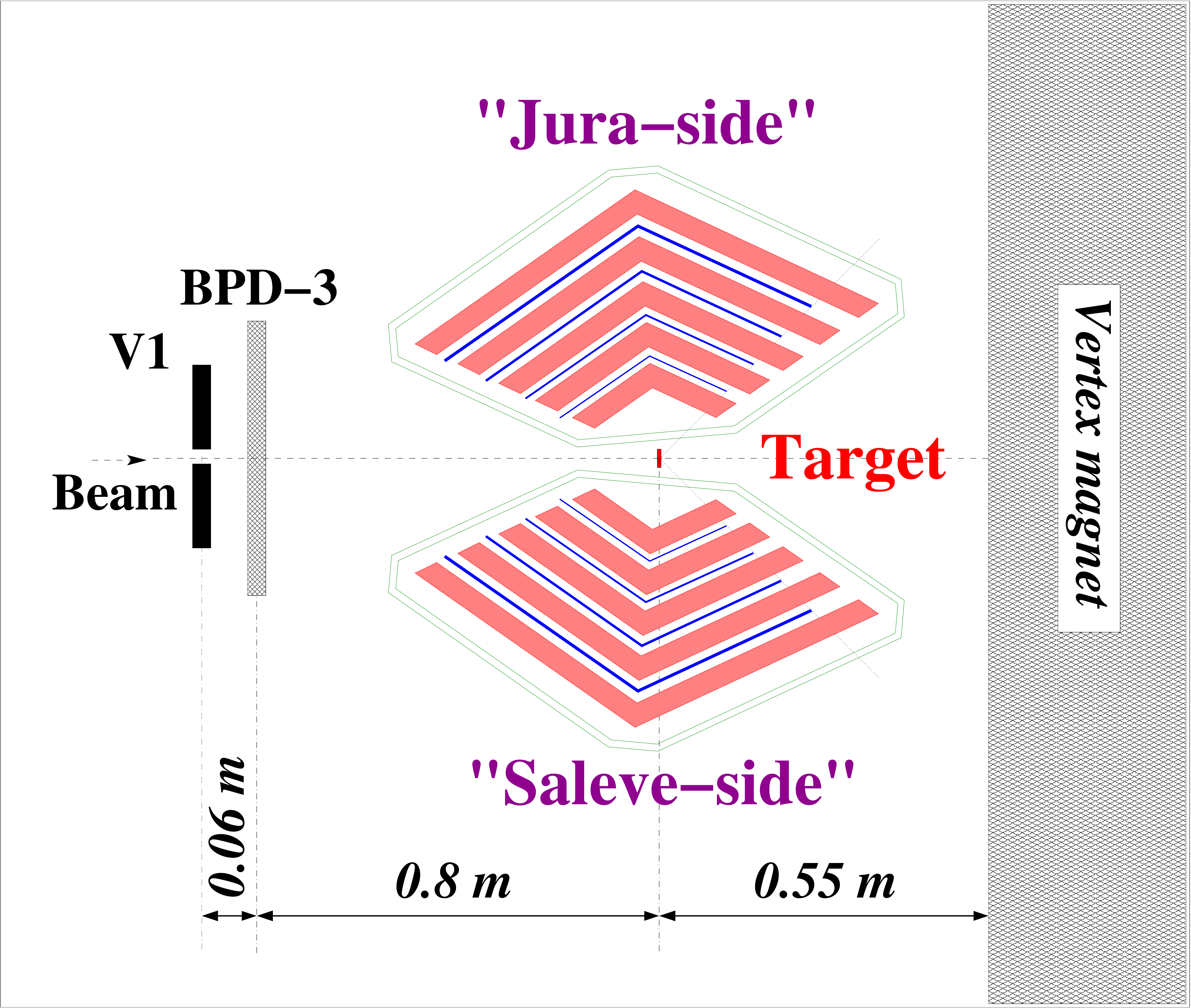}
\qquad
\includegraphics[width=0.5\linewidth]{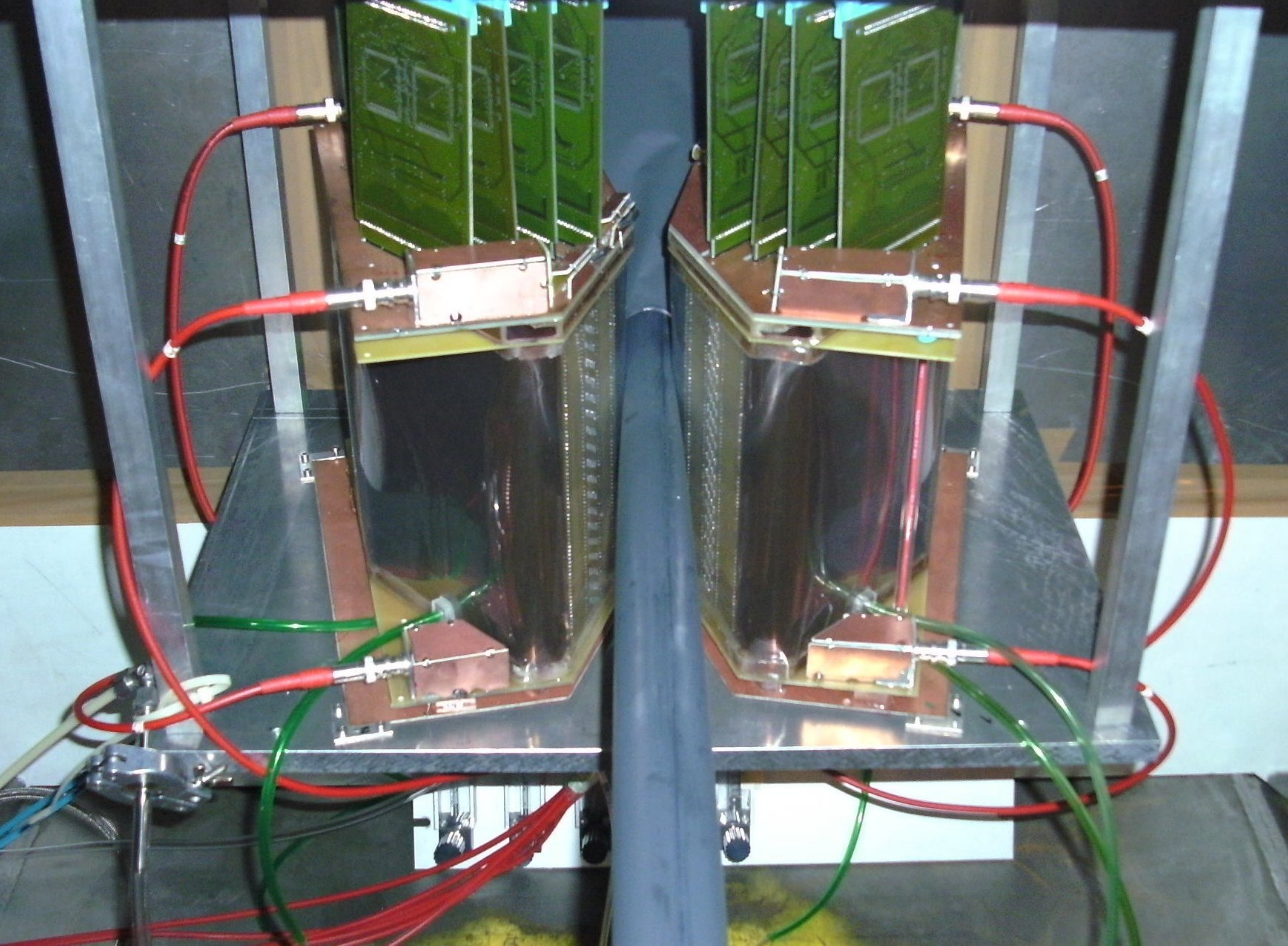}
\end{minipage}
\end{center}
\caption{
{\it Left:} Schematic top view of the LMPD and the associated beamline.
The internal structure of LMPD is also shown: thin absorber layers
reside between sensitive layers depicted by coloured bands. The particular
setting can be used to determine whether a
given particle from the target crossed the absorber layer or stopped inside
the absorber.
	{\it Right:}
	Front view photograph of the LMPD detector from the location upstream of
  the LMPD.
The Pb target is located in the center within the
thin Tedlar foil He container tube surrounding the
target disc.
The field cage strips (copper strips on capton) of the LMPD are also visible.
Standard NA61/SHINE TPC front end electronics were used for readout (on top).
}
\label{fig:lmpd1}
\end{figure}

\begin{figure}[ht]
\begin{center}
\hspace*{1.3cm}{
\begin{minipage}[b]{0.90\linewidth}
\raisebox{0.3\height}{\includegraphics[width=0.4\linewidth]{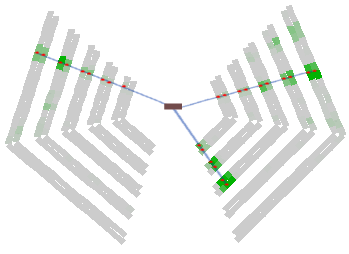}}
\qquad
\includegraphics[width=0.5\linewidth]{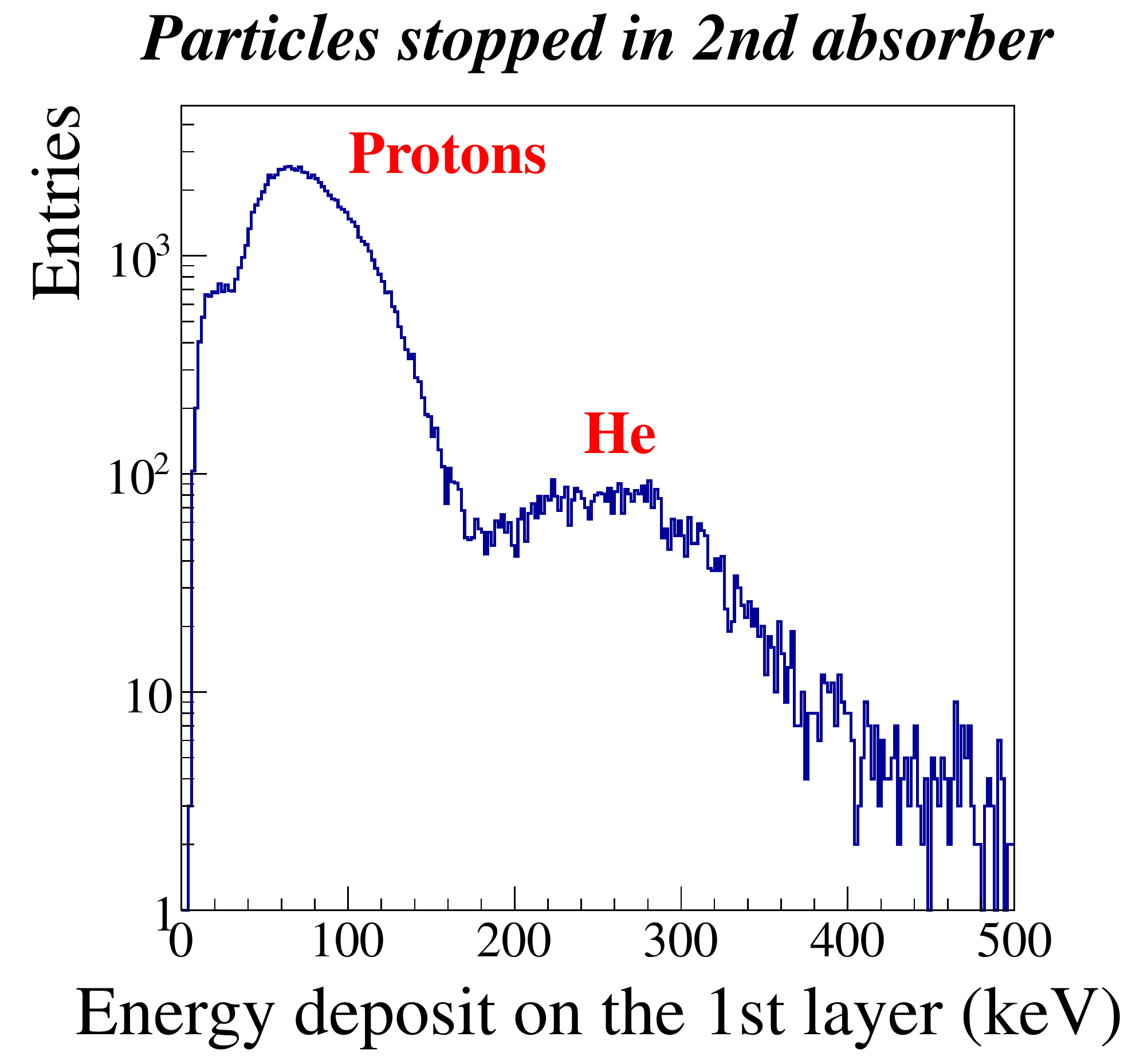}
\end{minipage}
}
\begin{minipage}[b]{0.95\linewidth}
\includegraphics[width=\linewidth]{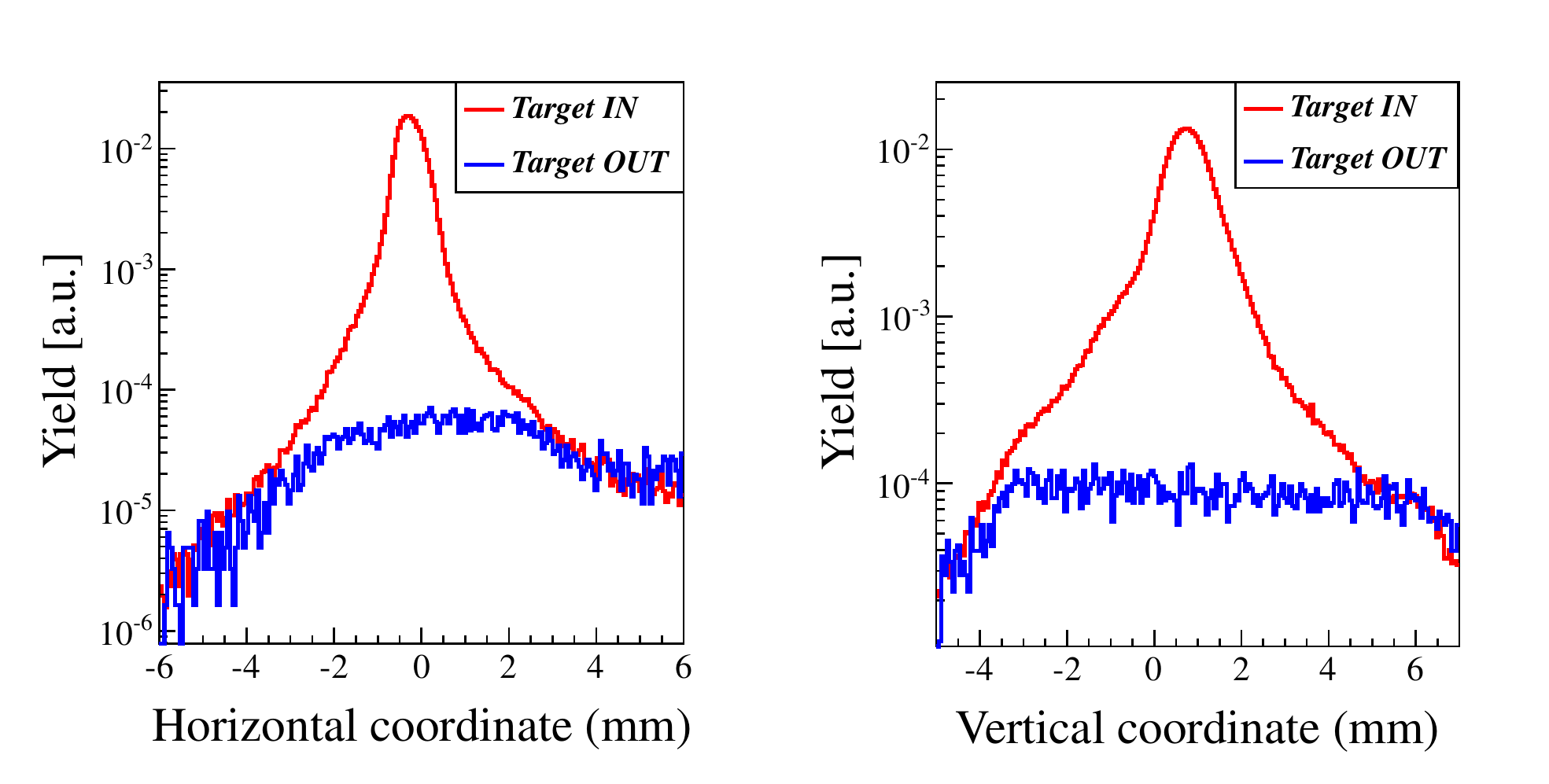}
\end{minipage}
\end{center}
\caption{
{\it Top left:} Event display of the LMPD for a typical 158~\GeVc p+Pb collision,
overlayed with the reconstructed clusters and tracks. Green color indicates
the energy deposit (ADC),  red points indicate the position of the
reconstructed clusters, blue lines depict the straight
particle tracks fitted to the clusters. The particle tracks point to a common
production vertex within the target.
	{\it Top right:}
Energy deposit of particles traversing the first absorber, but
stopped within the second absorber layer. The proton energy deposit along
with that of other nuclei are clearly seen.
	{\it Bottom:}
Distribution of reconstructed interaction coordinate for
data taking with target inserted and removed.
The off-target background is small.
}
\label{fig:lmpd2}
\end{figure}

The final version of the LMPD, see Fig.~\ref{fig:lmpd1} along with Table~\ref{table:lmpdparams},
was manufactured in 2011, and was first
tested in the NA61/SHINE experimental area, downstream of the NA61/SHINE detector,
in parasitic
mode. The tests showed the expected performance of the detector, namely its
capability of
proton identification and of counting the low energy (grey) protons. A
typical raw event overlayed with the reconstructed
clusters and tracks is seen in Fig.~\ref{fig:lmpd2}\lef.
The ionization produced by
particles with fixed range is shown in Fig.~\ref{fig:lmpd2}\rig.
The signal of protons with a selected penetration range is clearly visible along
with other fragment species such as helium nuclei. The cluster reconstruction
uses a closest neighbor search algorithm, while the track reconstruction uses
the Hough transform procedure combined with the maximum likelihood principle for pattern
recognition. Figure~\ref{fig:lmpd2}~(\textit{bottom}) shows the distributions of
interaction vertex coordinates reconstructed from LMPD tracks
for events recorded with inserted (red histograms) and removed target (blue histograms).
The signal of events from interactions in the Pb target is clearly visible and
the contribution from non-target interactions is small.

\begin{table}
\begin{center}
\vglue0.2cm
\begin{tabular}{|l||c|}
\hline
	& LMPD \\
\hline
\hline
size (L$\times$W$\times$H)~[cm] & 29 x 15.5 x 22.5 \\
\hline
No. of pads/sector & 140 \\
\hline
Pad size [mm] & (4-9) x 6 \\
\hline
Drift length [cm]& 20.5 \\
\hline
Drift velocity [cm/$\mu$s] & 0.9 \\
\hline
Drift field [V/cm] & 195 \\
\hline
Drift voltage [kV] & 4 \\
\hline
drift gas & Ar/CO$_2$ (85/15) \\
\hline
\# of sectors & 2 $\times$ 2 \\
\hline
\# of padrows & 10 \\
\hline
\# of pads/padrow & 8,8, 14,14, 16,16, 16,16, 16,16 \\
\hline
Absorber thickness [mm] & 0.5, 1.0, 2.0, 2.5 \\
\hline
\end{tabular}
\caption{Parameters of the LMPD.}
\label{table:lmpdparams}
\end{center}
\end{table}

The operational  parameters of the LMPD were optimized during the 2011 test.
In particular, a gradual decrease of amplification in the layers
close to the
target was introduced. This is necessary in order to adjust the dynamic
range for the
expected highly ionizing particles. After optimization large statistics
physics-quality data
were recorded in parasitic mode with the LMPD located  downstream of MTPCs.
They allow a
systematic study of gray proton production and absorption in the
target material.
During the last 3 days of the 2011 data taking period with proton beams,
the LMPD
was mounted
at the nominal position, and preliminary physics-quality
p+Pb data were
recorded  for the final test of the operational parameters.
Then, in 2012 large statistics 158~\GeVc p+Pb physics
data were taken with the LMPD sourounding the target.
This allows detection
and tagging of grey protons and thus characterizing the centrality of the p+Pb
collisions.

\section{\large Data acquisition and detector control systems}
\label{daq}

\subsection{Readout electronics and DAQ}
\label{daq:daq}

\begin{figure}[ht]
\begin{center}
\includegraphics[width=\linewidth]{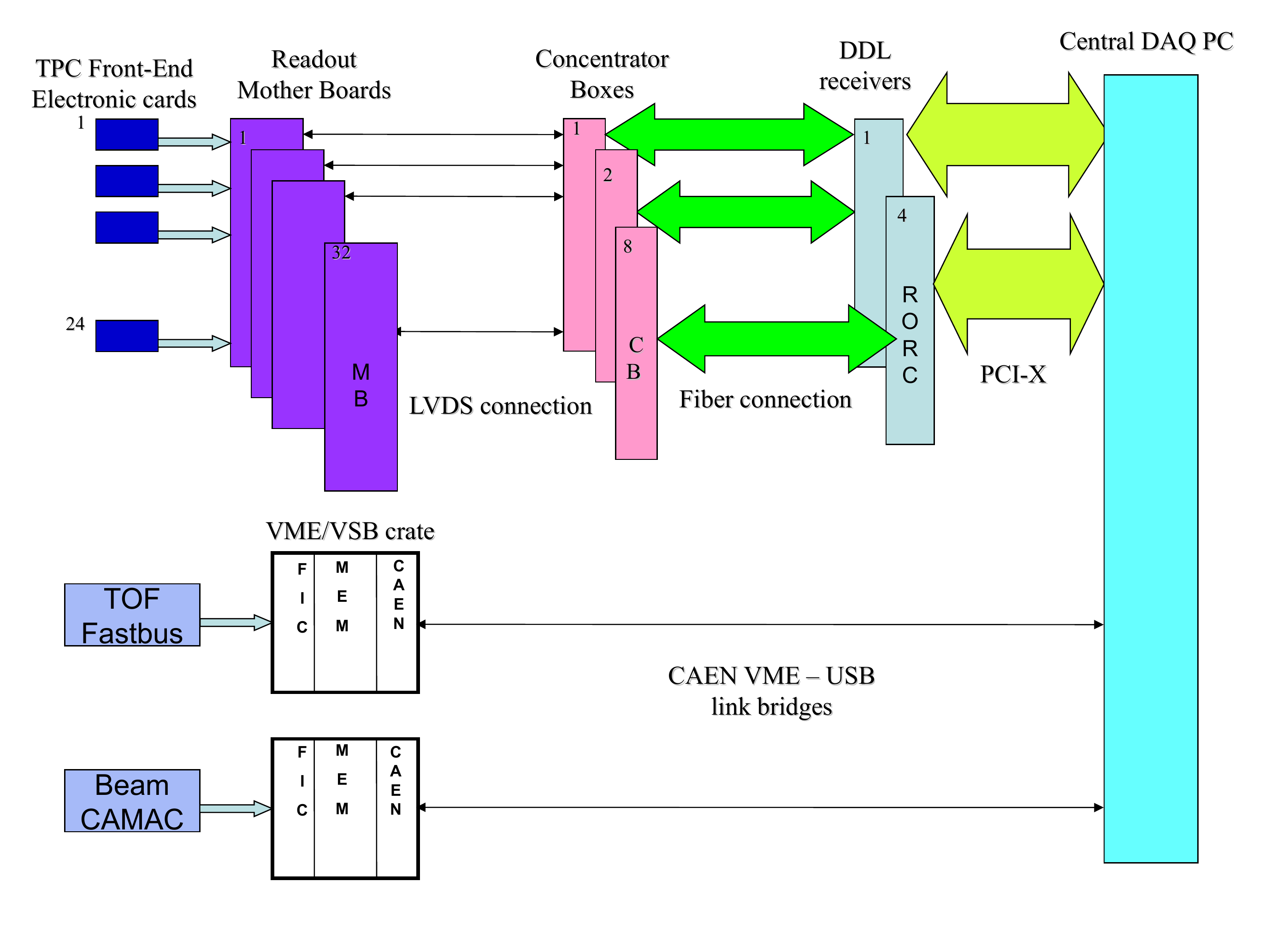}
\end{center}
\caption{
Block diagram of the three main parts of the NA61/SHINE  readout system:
TPC readout, ToF readout and CAMAC readout of beam detectors.
}
\label{fig:readout}
\end{figure}

The readout electronics \cite{Denes:2014nima} consists of three main parts: the electronics
responsible for reading out the TPC FEEs, the electronics for reading out
the FASTBUS based ToF system, and the electronics for reading out the CAMAC
based beam detectors (see Fig.~\ref{fig:readout}).

In case of the TPC system,
the FEE cards~\cite{Kleinfelder:1990,Bieser:1997} only host
pre-amplifiers, shapers, and time-sampling
capacitor arrays for 32 channels, along with ADCs. The necessary command
logic is managed by the readout Mother Boards, which transmit the necessary
clock signals to steer the time sampling, ADC conversion and data transmission
process. The FEE cards produce a 9 bit ADC value for each time slice of each
TPC pad. These are pedestal subtracted, noise suppressed and zero compressed
on the Mother Boards. The 250 Mother Boards, which are each capable of serving 24
FEE cards,
were implemented using Cyclone II FPGA (EP2C35F672) arrays and SRAM
units for pedestal table storage. Upon an event trigger, which is fanned out
to all Mother Boards, the time sampling with subsequent digitization begins on
the FEE cards, after which the ADC values are read out and processed on the fly
by the Mother Boards with polling. The processed data stream is serialized
onto a ground-independent LVDS connection line toward the Concentrator Boxes,
which can receive data from 32 Mother Boards. These boxes act like further
serializers and are
implemented on Cyclone II FPGA (EP2C20F484) arrays. In order to ensure galvanic ground
independence for the long-distance transfer,
the DDL optical transfer line~\cite{Rubin:1999,Carena:2004}
is used from this point to the Central DAQ PC. The data reaches the Central
DAQ in push-data mode, i.e. the data stream is triggered by the lower level
electronics and not via the Central DAQ. Upon the arrival of the first data
headers at the Concentrator Box level, a Busy signal is issued as a feedback
to the trigger electronics.

The ToF system uses legacy front end electronics based on FASTBUS
technology. The readout of the beam-related detectors (Beam Position Detectors,
beam counter pattern units, scalers, ADCs and TDCs) is based on
traditional CAMAC units. A FASTBUS-to-VME bridge and a CAMAC-to-VME bridge,
makes connection to VME crates dedicated to ToF and CAMAC readout,
respectively.
In each of these two VME crates a FIC8234 processor-based controller runs a
low level OS9 based software as low level DAQ. As the processors are capable
of receiving external signals as interrupt, the latter are used to initiate
low level
data taking. The measurement units are triggered by the experiment's
accurate pre-trigger signal, but the measurement data are only read out if
this is confirmed by the main-trigger signal within the time-out limit. In that case an
interrupt is passed to the FIC8234 controllers to initiate readout.
The data are transfered via the bridges to a MM6390 memory unit in the VME crate.
The received data are stored in a ring buffer overwriting older data and a trigger
counter is incremented. Meanwhile the Central DAQ polls for the incrementation
of the trigger counter, and the new events are drained via a CAEN V1718
VME-to-USB bridge to the Central DAQ's ring buffer.
Feedback to and control of the Busy logic is performed via RCB8047 CORBO register
units in the pertinent VME crates.

The readout of  the PSD, is designed along the
principles of
the TPC readout, and acts similarly to
the Mother Boards in the TPC readout system.

Due to the push-data mode method, the event data in the different hardware
channels arrive at the Central DAQ software asynchronously. The synchronization
is performed via trigger counters in the sub-events on the final
event-building level. For periodic checks the data stream is halted each minute,
the data pipelines are drained and the trigger counter synchronicity is verified.
In addition, the Busy signal of each hardware channel is monitored via a custom made
galvanically isolated TTL to RS232 register in order to make
sure that the event stream is never halted due to a stuck Busy signal caused by a
hardware failure.

The Central DAQ software runs on a single Central DAQ PC
(X7DB8-X motherboard, 64 bit, total 8 cores of Intel Xeon CPU @ 2~GHz, 8~GB
memory, 10 PCI-X slots, 4 USB ports, and a serial port).
The  main components of the software and their functions are
as follows:
\begin{enumerate}[(i)]
\setlength{\itemsep}{1pt}
 \item The core software of the Central DAQ written in C.
       Its user interface is a command line interpreter written using the GNU
       libreadline library.
 \item The Central DAQ GUI script written in Tcl/Tk language.
 \item The event server,
       created upon Central DAQ start for event monitoring.
       This is a fork-server serving up to 16 monitoring clients.
 \item The following processes created upon run start:
  \begin{enumerate}[(i)]
  \setlength{\itemsep}{1pt}
  \item The logger and monitoring / consistency checking process.
  \item The communication unit with the trigger system for summary information.
	This periodically gets scaler status information from the trigger server.
  \item The communication unit with the DCS system for summary information.
	This periodically gets information from the DCS server.
  \item The recorder process building the event, writing to disk and forwarding
  to the above event server upon request.
	This starts the following two processes for efficient parallelization
	of data receiving.
  \begin{enumerate}[(i)]
  \setlength{\itemsep}{1pt}
   \item The receiver process for the DDL channels. This looks for data on
   any DDL channel on a first-come first-served basis.
   \item The receiver process for the VME channels. This looks for data on
   any VME channel on a first-come first-served basis.
   \end{enumerate}
  \end{enumerate}
 \end{enumerate}
The GUI and the command line interpreter communicate via standard Unix pipes
and log files. The processes of the forked C program communicate via Linux
shared memory. The data itself gets read into a special fast shared memory,
called Physmem, outside the reach of the Linux kernel. Physmem is mapped
by a special Linux kernel module shipped with the DDL libraries.

The central DAQ software is completed by a system of failure-tolerant
scripts with checksum and size verification, which move the recorded data
after consistency and Quality Assessment check onto the tape system of CASTOR.

\subsection{Detector Control System}
\label{daq:dcs}

The NA61/SHINE Detector Control System (DCS) is responsible for online
monitoring and controlling of the working conditions of the detectors.

The system monitors parameters of the gas mixture in the TPCs (temperatures,
pressure, flow, water and oxygen content, drift velocity, amplification, etc.).
It also sets and monitors parameters of the high voltage in the sub-detectors: LMPD,
TPCs, BPDs and the beam counters. The system also controls low voltage power
supplies of the front end electronics and enables its cooling.

\begin{figure}[ht]
	\begin{center}
		\includegraphics[width=.6\linewidth]{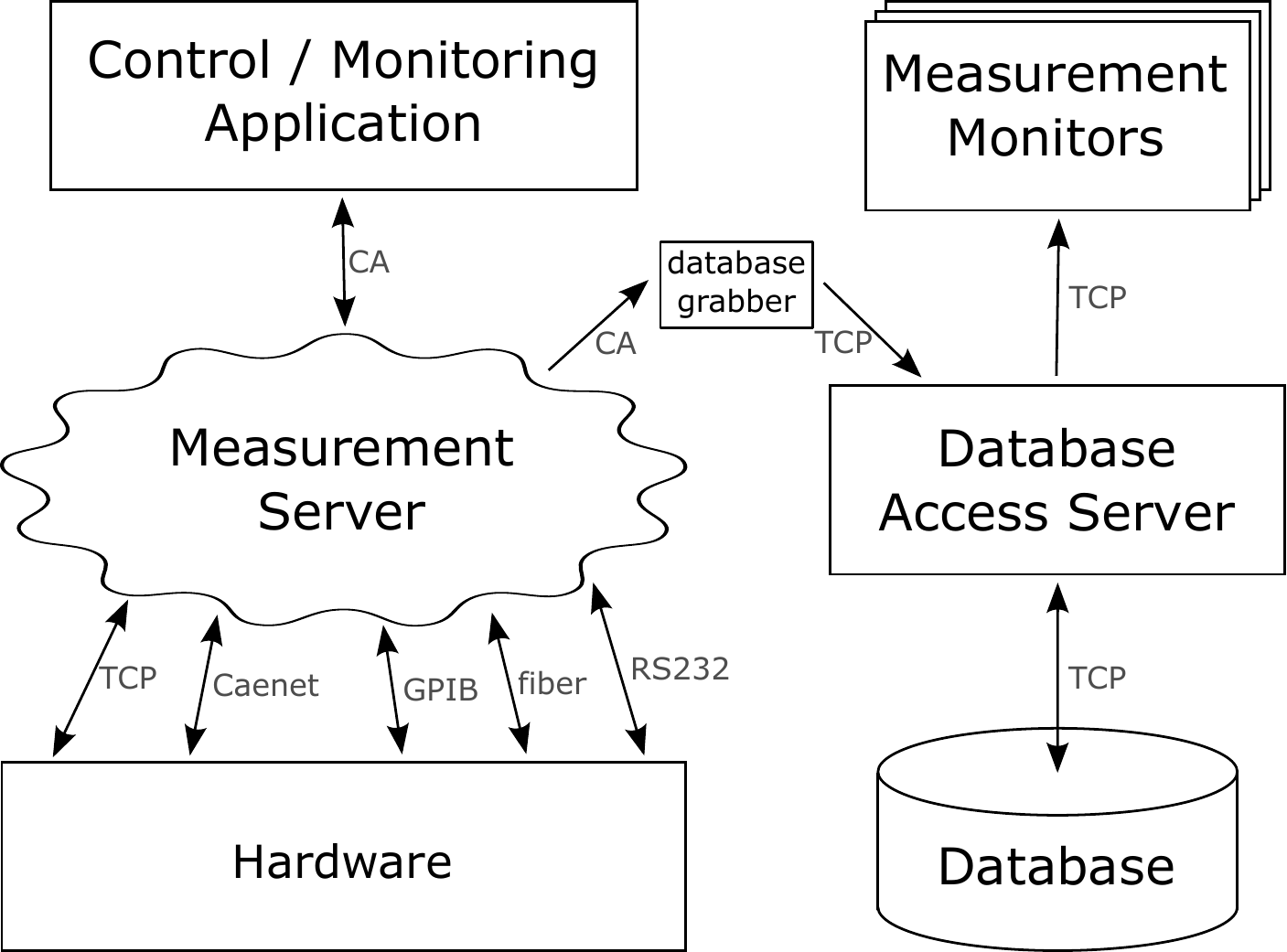}
	\end{center}
	\caption{Block diagram of the NA61/SHINE Detector Control System.}
	\label{fig:dcs}
\end{figure}

The block diagram of the DCS is shown in Fig.~\ref{fig:dcs}.
Although the system logically consists of the following subsystems:
\begin{enumerate}[(i)]
  \setlength{\itemsep}{1pt}
  \item Gas subsystem,
  \item High Voltage subsystem,
  \item Low Voltage subsystem,
  \item PSD subsystem,
\end{enumerate}
the main part of the system is a single EPICS-based \cite{epics_website}
measurement server distributed over a few PCs. It communicates with various
hardware (CAMAC, VME, PLC, etc.) via various interfaces (RS232, Caenet, TCP/IP,
GPIB, etc.), performs all the measurements and makes the results accessible for
the clients via the Channel Access protocol. The server runs constantly
regardless of the presence of the clients or the database availability.

All measured data is stored in the relational database (PostgreSQL) by one of
the EPICS clients. It can be accessed only via the Database Access Server by
issuing ASCII commands.

There are two GUIs available. One of them, which controls the Measurement Server
and monitors its performance, is an EPICS client and can be run only in the
experimental area. The other one can be run anywhere and is graphically presenting
results of measurements retrieved from the database.

\section{\large Summary and outlook}
\label{summary}

NA61/SHINE (SPS Heavy Ion and Neutrino Experiment)
is a multi-purpose facility for the
study of hadron production in hadron-proton, hadron-nucleus
and nucleus-nucleus collisions at
the CERN Super Proton Synchrotron.

NA61/SHINE has greatly profited from the long development of the
CERN proton and ion sources, the accelerator chain, as well as
the H2 beamline of the CERN North Area. The latter
has recently been modified to also serve as a fragment separator
as needed to produce the Be beams for NA61/SHINE.
Numerous components of the NA61/SHINE setup were inherited from
its predecessors, in particular, the last one, the NA49 experiment.

This paper describes the facility - the beams
and the detector system - as used up to March 2013, the start of
the CERN Long Shutdown I.
Special attention was paid to the presentation of the
components which were constructed for NA61/SHINE.
These are: the Projectile Spectator Detector, the Forward-ToF wall,
the Low Momentum Particle Detector,
the Z- and A-detectors, the Beam Position Detectors,
the digital part of the TPC readout and the DAQ system, the Trigger System,
the Detector Control System and the He-beam pipe.
Moreover the upgraded CERN accelerator chain and the H2 beamline
modified to serve as a fragment separator are described.
The components inherited from the past experiments and
described elsewhere are presented only briefly.
These are the Time Projection Chambers and their gas system,
the two super-conducting magnets, the ToF-L/R walls as well as
the beam and trigger counters.

Upgrades of the facility are continuing and the physics goals
are being expanded~\cite{NA61_future}.
Among  NA61/SHINE upgrades under preparation are:
construction of Forward-TPCs and of a Silicon Vertex Detector,
upgrade of the ToF and PSD readout systems, upgrade of
the ToF HV system and extension of the gas system.

This paper will be followed by further publications
presenting more details on the components which were
newly constructed for the NA61/SHINE facility.

\vspace*{1.5cm}

\noindent
Acknowledgements: This work was supported by
the Hungarian Scientific Research Fund (grants OTKA 68506 and 71989),
the Polish Ministry of Science and Higher Education (grants
667/N-CERN/2010/0, NN 202 48 4339 and  NN 202 23 1837),
the National Science Center of Poland (grant UMO-2012/04/M/ST2/00816),
the Federal Agency of Education of the Ministry of Education and Science
of the Russian Federation (grant RNP 2.2.2.2.1547), the Russian Academy of
Science and
the Russian Foundation for Basic Research (grants 08-02-00018, 09-02-00664,
and 12-02-91503-CERN),
the Ministry of Education, Culture, Sports, Science and Technology,
Japan, Grant-in-Aid for Scientific Research (grants 18071005, 19034011,
19740162, 20740160 and 20039012),
the German Research Foundation (grants GA 1480/2-1, GA 1480/2-2),
Bulgarian National Scientific Fondation (grant DDVU 02/19/ 2010),
Ministry of Education and Science of the Republic of Serbia (grant OI171002),
Swiss Nationalfonds Foundation (grant 200020-117913/1)
and ETH Research Grant TH-01 07-3.

\bibliographystyle{model1-num-names}


\end{document}